%% file: paper.tex
\RequirePackage{fix-cm}
\documentclass[smallextended]{svjour3}
\smartqed

\input{header.tex}
\usepackage{tikz}
\usepackage{marvosym}
\usepackage{graphicx}
\usepackage{wrapfig}
\usepackage{xurl}
\usepackage{soul}

\definecolor{mygreen}{rgb}{0.0, 0.5, 0.0}

\journalname{}

\begin{document}
\title{Evaluating Large Language Models for Multilingual Vulnerability Detection at Dual Granularities}

\titlerunning{Large Language Models for Multilingual Vulnerability Detection} 

\author{Honglin Shu \and
        Michael Fu \and 
        Junji Yu \and 
        Dong Wang \Letter \and 
        Chakkrit Tantithamthavorn \and 
        Junjie Chen \and 
        Yasutaka Kamei
}

\authorrunning{W. Dong et al.} 

\institute{ Honglin Shu \at
              Tianjin University, Tianjin, China, and Kyushu University, Fukuoka, Japan  \\
              \email{shu.honglin.167@s.kyushu-u.ac.jp}  \\
            \\
            Michael Fu \at
                The University of Melbourne, Melbourne, Australia \\
              \email{michael.fu@unimelb.edu.au}  \\
            \\ 
            Junji Yu,  Dong Wang, Junjie Chen\at
              School of Computer Software, Tianjin University, Tianjin, China \\
              \email{\{junjiyu, dong\_w, junjiechen\}@tju.edu.cn}\\
            \\
            Chakkrit Tantithamthavorn \at
              Monash University, Clayton, Australia \\
              \email{chakkrit@monash.edu} \\
            \\
            \\
            Yasutaka Kamei \at
              Kyushu University, Fukuoka, Japan \\
              \email{kamei@ait.kyushu-u.ac.jp} \\      
}

\date{Received: date / Accepted: date}

\maketitle

\begin{abstract} 
Various deep learning-based approaches utilizing pre-trained language models (PLMs) have been proposed for automated vulnerability detection. With recent advancements in large language models (LLMs), several studies have begun exploring their application to vulnerability detection tasks. However, existing studies primarily focus on specific programming languages (e.g., C/C++) and function-level detection, leaving the strengths and weaknesses of PLMs and LLMs in multilingual and multi-granularity scenarios largely unexplored. 
To bridge this gap, we conduct a comprehensive fine-grained empirical study evaluating the effectiveness of state-of-the-art PLMs and LLMs for multilingual vulnerability detection. 
Using over 30,000 real-world vulnerability-fixing patches across seven programming languages, we systematically assess model performance at both the function-level and line-level. Our key findings indicate that GPT-4o, enhanced through instruction tuning and few-shot prompting, significantly outperforms all other evaluated models, including CodeT5P. Furthermore, the LLM-based approach demonstrates superior capability in detecting unique multilingual vulnerabilities, particularly excelling in identifying the most dangerous and high-severity vulnerabilities. These results underscore the promising potential of adopting LLMs for multilingual vulnerability detection at function-level and line-level, revealing their complementary strengths and substantial improvements over PLM approaches. 
This empirical evaluation of PLMs and LLMs for multilingual vulnerability detection highlights LLMs' value in addressing real-world software security challenges.

\end{abstract}

\input{introduction.tex}
\input{evaluation.tex}
\input{results.tex}
\input{discussion.tex}
\input{threats.tex}
\input{related_work.tex}

\section{Conclusions}
\label{sec:conclusion}
This work systematically evaluated the performance of pre-trained language models (PLMs) and large language models (LLMs) across seven programming languages for multilingual vulnerability detection at both function and line granularity. 
Our findings highlight that LLMs, particularly GPT-4o enhanced with instruction tuning and few-shot prompting, significantly outperform existing PLMs like CodeT5P. 
Specifically, GPT-4o achieved superior accuracy in function-level detection and markedly higher precision and F1-scores in line-level detection tasks, demonstrating its effectiveness and versatility in multilingual contexts. 
Additionally, orthogonality analysis revealed that GPT-4o not only detects vulnerabilities uniquely missed by other models but also excels in identifying high-severity vulnerabilities, emphasizing its practical value in real-world security applications.
Despite these promising results, our study also reveals that the size and reasoning capabilities of LLMs do not necessarily correlate directly with improved vulnerability detection performance. 

Meanwhile, our study opens up several potential future research directions, including: extending the scale of multilingual vulnerability benchmark, introducing different intermediate representations for unifying the code representation, and refining the PLM/LLM learning strategies.
For instance, beyond supervised fine-tuning and instruction tuning, strategies like multi-task learning or meta-learning could better align PLMs and LLMs with multilingual domains by reducing the performance gap between high-resource and low-resource programming languages.
In general, our work provides critical insight and foundational knowledge that can guide the advancement of robust multilingual vulnerability detection tools, addressing key limitations identified in existing approaches.

\input{declaration.tex}

\bibliographystyle{spbasic}
\bibliography{additionalref}

\end{document}

%% file: header.tex
\usepackage{cite}
\usepackage[dvipsnames]{xcolor}
\usepackage{natbib}
\usepackage{verbatim}
\usepackage{caption}
\usepackage{graphicx}
\usepackage{booktabs}
\usepackage{relsize}
\usepackage{colortbl}
\usepackage{enumerate}
\usepackage{amsmath}
\usepackage{tikz}
\usepackage{pgf}
\usepackage{pbox}
\usepackage{rotating}
\usepackage{multirow}
\usepackage{balance}
\usepackage{tablefootnote}
\usepackage{flushend}
\usepackage{natbib}
\usepackage{hyperref}
\usepackage[T1]{fontenc}
\usepackage{enumitem}
\usepackage{amsmath,amsfonts}
\usepackage{algorithmic}
\usepackage{array}
\usepackage{textcomp}
\usepackage{stfloats}
\usepackage{url}
\usepackage{verbatim}
\usepackage{graphicx}
\usepackage{balance}
\usepackage{adjustbox}
\usepackage{booktabs}
\usepackage{multirow}
\usepackage{threeparttable}
\usepackage{arydshln}
\usepackage{tcolorbox}
\usepackage{soul}
\usepackage{fontawesome}
\usepackage{makecell}
\usepackage{natbib}
\usepackage{colortbl}
\usepackage{siunitx}
\usepackage{microtype}
\usepackage{tikz}
\usepackage[T1]{fontenc}
\usepackage{aecompl}
\usepackage[utf8]{inputenc}

\usepackage{hyperref}
\usepackage{graphicx}
\usepackage{subcaption}

\usepackage{soulutf8}

\renewcommand{\labelitemi}{--}
\newcommand{\startlist}{\begin{list}{\labelitemi}{\leftmargin=1em}\setlength{\itemsep}{-1mm}}
\newcommand{\stoplist}{\end{list}}

\usepackage{xcolor}
\definecolor{highlightcolor}{rgb}{1,1,0.7} 
\sethlcolor{highlightcolor}

%% file: introduction.tex
\section{Introduction}
\label{sec:introduction}
Software vulnerabilities are flaws in code that pose significant risks to software systems, potentially leading to compromised sensitive information and system failures.  
To address these risks, researchers have developed various automated vulnerability detection (AVD) techniques. 
These techniques typically fall into three main categories: rule-based techniques, machine learning (ML)-based techniques, and deep learning (DL)-based techniques.
Rule-based techniques~\citep{10.1145/3576915.3624401, kim2019rvfuzzer} and ML-based techniques~\citep{grieco2016toward, scandariato2014predicting} require significant human expertise to design rules and features, yet often struggle to define effective patterns for identifying vulnerabilities.
DL-based techniques often rely on Recurrent Neural Networks (RNNs)~\citep{li2021sysevr,li2021vuldeelocator,li2018vuldeepecker}, Graph Neural Networks (GNNs)~\citep{chakraborty2021deep, zhou2019devign, nguyen2022regvd, steenhoek2024dataflow}, and Pre-trained Language Models (PLMs)~\citep{hanif2022vulberta,fu2022linevul,liu2024pre}. These methods are relatively effective since they can learn vulnerability patterns in an end-to-end manner.
Particularly, leveraging pre-trained knowledge, PLMs formulate vulnerability detection as a binary classification on code embedding, proven superior to RNNs and GNNs at detecting vulnerabilities~\citep{steenhoek2023empirical}.
Although PLMs show promise to some extent, these methods still face limitations in fully understanding code semantics for extracting representative or unseen vulnerability patterns~\citep{ding2024traced}.

Large Language Models (LLMs) have recently garnered widespread attention and shown strong potential across a variety of code-related tasks~\citep{hou2023large}.
Several empirical studies~\citep{10479409, zhou2024large, yin2024multitask, yin2024pros, zhou2024largeworkshop} have explored using LLMs for vulnerability detection.
For instance, \citet{10479409} found that GPT-3.5-Turbo and GPT-4 fail to detect single-language (i.e., C/C++) vulnerabilities at both function-level and line-level.
~\citet{yin2024pros} revealed that ChatGPT, when using prompts alone, can be easily swayed to change vulnerability classifications, indicating low confidence.
Furthermore, ~\citet{yin2024multitask} demonstrated that while fine-tuned LLMs can detect vulnerabilities, they perform more weakly than PLMs.
Moreover, some LLM-based AVD techniques have been proposed, including Vul-RAG~\citep{du2024vul}, MSIVD~\citep{yang2024security}, and GRACE~\citep{lu2024grace}.
These techniques demonstrate that LLMs are competitive with PLMs at detecting vulnerabilities in individual programming languages.
For example, MSIVD applied multitask learning with LoRA to fine-tune the LLM and incorporated fine-tuned LLM embedding and Abstract Syntax Tree (AST) embedding.
Vul-RAG checks code vulnerability by reasoning about vulnerability causes and fixing solutions from retrieved vulnerability knowledge, proving LLMs' potential in detecting code vulnerabilities.

Despite significant advances in automated vulnerability detection using LLMs and PLMs, we identify three critical limitations in current approaches:
\textbf{(I) Existing research predominantly focuses on single-language vulnerability detection.}
Most existing studies limited their vulnerability detection capabilities to C and C++ using datasets like CVEfixes~\citep{bhandari2021cvefixes} and Big-Vul~\citep{fan2020ac}.
This narrow focus fails to address the reality of modern development environments, which frequently incorporate multiple programming languages~\citep{li2022vulnerability}. 
Studies have demonstrated the prevalence of vulnerabilities across diverse language ecosystems, including Python ~\citep{alfadel2023empirical} 
and Go~\citep{hu2024empirical}. 
For AVD to be practically valuable, research must expand beyond single-language approaches to address multilingual vulnerability detection.
In this context, we define multilingual detection not as analyzing mixed-language files simultaneously, but as the model's ability to robustly generalize detection performance across independent languages with varying syntax and semantics.
\textbf{(II) The generalizability of PLMs and LLMs across diverse language-specific vulnerability profiles remains insufficiently explored.}
While PLMs and LLMs are pre-trained on vast multilingual corpora, it is unclear if this linguistic exposure translates to an understanding of vulnerability semantics across different paradigms. 
The nature of software vulnerabilities is heavily influenced by the underlying language paradigm. 
Low-level languages like C and C++ frequently exhibit memory corruption issues (e.g., buffer overflows) due to manual memory management, whereas high-level languages like Python tend to manifest vulnerabilities related to business logic or input injection.
Existing study~\citep{steenhoek2024err} often assume that pre-training equates to detection capability, failing to empirically validate whether models can effectively generalize vulnerability detection rules across these distinct language ecosystems, rather than simply memorizing syntax. This knowledge gap hinders effective model selection for systems that must secure software across a diverse technology stack.
\textbf{(III) Current evaluation frameworks lack a comprehensive assessment across different granularity levels.}
Vulnerability detection operates primarily at two levels: function-level and line-level. Function-level detection identifies whether a function contains vulnerabilities but lacks precision, potentially impeding developers from efficiently addressing specific issues. Conversely, line-level detection aims to precisely identify the vulnerable lines within a function. 
Existing research~\citep{zhou2024large} has concentrated primarily on function-level detection in single-language contexts, leaving line-level multilingual vulnerability detection significantly underexplored. This gap is particularly problematic as multilingual contexts introduce greater challenges for line-level detection due to variations in syntax and semantics across programming languages, which complicate the effective generalization of language models.

In this work, we conduct an empirical study to systematically investigate the effectiveness of existing PLMs and examine the performance of advanced LLMs in multilingual vulnerability detection. 
Our hypothesis is that LLMs, with their remarkable semantic understanding and language-agnostic capabilities, combined with effective learning strategies, can enhance the effectiveness of detecting function-level and line-level multilingual vulnerability.
For our evaluation, we utilize REEF, a comprehensive multilingual vulnerability corpus containing 4,466 CVEs with 30,987 patches. 
This dataset spans seven major programming languages: C, C\#, C++, Go, JavaScript, Java, and Python.
To structure our investigation, we propose the following three research questions:
\\
\noindent
\textbf{- RQ1: How effective are PLMs and LLMs in detecting multilingual vulnerabilities at the function level?}
\\ 
\noindent
\textbf{- RQ2: How effective are PLMs and LLMs in detecting multilingual vulnerabilities at the line level?} 
\\
\noindent
\textbf{- RQ3: What are the strengths and weaknesses of the PLMs and LLMs in multilingual vulnerability detection?}

Our key findings are: 
(1) An appropriate LLM with instruction tuning and few-shot prompting demonstrates promising multilingual vulnerability detection performance. Specifically, for both function-level and line-level multilingual vulnerability detection, GPT-4o with instruction tuning and few-shot prompting demonstrates superior performance, achieving the highest accuracy (0.7196) at function-level compared to the best-performing CodeT5P (0.6037), and the best F1-score (0.6641) at line-level compared to other studied PLMs and LLMs.
(2) Across different programming languages, the best-performing LLM achieves its highest function-level accuracy (0.8082) with Go and its lowest (0.6626) with Python. At the line-level, GPT-4o attains its highest F1-score (0.7815) on JavaScript, while its lowest F1-score (0.4348) is observed on C\#.
(3) Orthogonality analysis reveals that the best-performing LLM excels not only in uniquely correct detections but also shows higher accuracy in identifying the top-25 most dangerous CWE-IDs compared to PLMs.
Base on CVSS severity, the top LLM also surpasses other studied PLMs and LLMs in detecting high-severity vulnerabilities (e.g., Critical and High level severity).
(4) Furthermore, we investigate the impact of LLM size and compare effectiveness between reasoning and non-reasoning LLMs.
The results show that the size of LLMs is not the decisive factor affecting performance on multilingual vulnerability detection, and the use of reasoning LLMs does not lead to significant improvements.

\smallskip
\noindent
\textbf{Contributions.}
To sum up, the contributions of this study are:
\begin{enumerate}
    \item This study systematically evaluates PLMs and LLMs for function- and line-level multilingual vulnerability detection across seven programming languages. It also compares various popular LLM strategies, including zero-shot prompting, retrieval-based few-shot prompting, and instruction tuning.
    \item The empirical results confirm the promising role of LLMs in multilingual vulnerability detection at function-level and line-level perspectives, particularly when instruction-tuning LLMs with few-shot prompting strategies.
    \item Through fine-grained analysis, we provide valuable insights into the capabilities and limitations of LLMs for multilingual vulnerability detection, offering essential guidance for future research aimed at advancing LLM-based vulnerability detection techniques.
    \item We further discuss the impact of reasoning capabilities and larger-scale LLMs on both function-level and line-level multilingual vulnerabilities, and analyze the deployment costs associated with LLMs and PLMs.
    \item We have made the replication package publicly available on our homepage~\citep{replicationpackage}, including all used data, codes, and analysis details involved in our study. 
\end{enumerate}

\noindent
\textbf{Paper Extension.} 
This paper extends our prior work on a preliminary study~\citep{yu2025preliminarystudylargelanguage}, published as a research paper at the 1st International Workshop on Large Language Model Supply Chain Analysis, co-located with ISSTA'25.
The key differences can be summarized as follows: (I) \textit{A broader range of studied models and prompting strategies}. 
Specifically, we enhance PLMs by incorporating three OpenAI text-embedding models, and for LLMs, we explore two additional strategies: few-shot prompting and instruction tuning.
(II) \textit{Incorporation of line-level detection}.
To provide a more comprehensive perspective, we extend the research scope beyond function-level detection to include line-level detection, enabling more detailed analysis of vulnerability locations.
(III) \textit{Finer-grained analyses of detection performance}.
We conduct an in-depth analysis focusing on unique correct and incorrect detections, prediction tendencies for the Top-25 most dangerous CWE-IDs, and the effectiveness of detecting vulnerabilities across different severity levels.
(IV) \textit{Exploration of effects of model size and architecture}. 
We further investigate how the size of open-source LLMs and the use of reasoning-capable LLMs impact performance in multilingual vulnerability detection.

%% file: evaluation.tex
\section{Study Design}
\label{sec:evaluation_design} 
This section introduces the overview of our study design.
Initially, based on the latest multilingual vulnerability dataset, we construct the training and test datasets for the detection of multilingual vulnerability at the function and line levels. 
We first investigate the effectiveness of existing PLM-based AVD approaches and LLMs in detecting function-level multilingual vulnerability (RQ1). 
After that, we further examine the performance of existing PLM-based AVD approaches and LLMs in detecting line-level multilingual vulnerability (RQ2).
Finally, we obtain an understanding of the strengths and weaknesses of the PLM-based AVD approaches and LLMs in multilingual vulnerability detection from different perspectives (RQ3).

\subsection{Dataset Preparation}
\label{subsec:dataset}

\textbf{Studied dataset}. 
To evaluate PLM-based AVD approaches and LLMs for multilingual vulnerability detection, we utilize the REEF dataset~\citep{wang2023reef}. 
This dataset contains 4,466 CVEs with 30,987 patches across seven programming languages, including comprehensive vulnerability information (CVE, CWE, CVSS, etc.) and project details (such as commit messages).
REEF is constructed from real-world vulnerabilities sourced from the National Vulnerability Database (NVD) and Mend's CVE list~\citep{whitesource2022mend}, an open-source vulnerability database covering 2016-2023.
At the function-level, the dataset consists of 6,957 functions written in C, 2,244 in C++, 1,529 in C\#, 3,187 in Go, 6,207 in Java, 5,066 in JavaScript, and 5,797 in Python.
Our study encompasses all seven programming languages supported by REEF: C, C++, C\#, Go, Java, JavaScript, and Python.

\textbf{Data pre-processing}. 
To adapt the REEF dataset for multilingual AVD tasks, our first step was to extract the vulnerable and non-vulnerable functions.
We received commit data in raw format with patches. To analyze both the original and modified versions of vulnerable functions, we processed these patches using the Linux patch command.\footnote{\url{https://www.man7.org/linux/man-pages/man1/patch.1.html}}
Following the commit data collection~\citep{fan2020ac}, we pre-processed the data by removing code comments to minimize bias.
Comments can contain misleading information, such as outdated or incorrect annotations. Additionally, they may include vulnerability-related details that pose a potential risk of data leakage.
We then extracted function definition code using Tree-sitter~\citep{brunsfeld2024tree}, an efficient parser generator and incremental parsing library capable of analyzing multiple programming languages.
To extract functions and their corresponding function-level and line-level labels, we iterated through function definitions in the pre-change file to match them with their post-change and pre-change function definitions.
Specifically, the function names are used as identifiers to match and compare corresponding functions between the pre-change and post-change versions of the files.
When multiple matching function definitions are found in the post-change file, we calculate the edit distance between the pre-change function and each post-change function. The pair with the smallest edit distance is then identified as the vulnerable and clean function pair.
In other words, we define the pre-change function as the vulnerable function and the post-change function as the non-vulnerable function.
Since many PLMs use the absolute position encoding, which is limited to the input length (e.g., 512), we filtered out functions that are greater than 512 in length.
Finally, we obtained a total of 20,165 functions and corresponding labels.
At the function-level, the dataset comprises 3,056 C, 1,792 C++, 427 C\#, 2,905 Go, 3,235 Java, 5,468 JavaScript, and 3,282 Python functions.
To obtain line-level labels, we first eliminated all non-vulnerable functions from the function-level dataset.
We then applied \texttt{DIFFLIB}~\citep{difflib} to match changed code lines by comparing each vulnerable function with its corresponding non-vulnerable version.
\texttt{DIFFLIB} identifies three matching cases: (1) vulnerable functions fixed by adding code lines only, (2) vulnerable functions fixed by removing code lines only, and (3) vulnerable functions fixed by both removing and adding code lines.
Since we need to locate specific line positions, only cases (2) and (3) can identify vulnerable line positions through deletion information. 
We use these two cases to obtain code line labels, i.e., line numbers and code lines, then filter out vulnerable functions where line labels cannot be extracted.
To ensure the integrity of our line-level dataset, we excluded 9,613 functions from the training set, 1,202 from the validation set, and 1,216 from the test set because their vulnerable line positions could not be identified. Consequently, these samples were filtered out during the construction of our line-level dataset.
Finally, we obtained a total of 8,134 vulnerable functions and the corresponding line labels.
Finally, we obtained a total of 8,134 lines and corresponding labels.
At the line-level, the dataset comprises 1,175 C, 651 C++, 151 C\#, 919 Go, 1,375 Java, 2,496 JavaScript, and 1,367 Python functions.

To validate the robustness of Tree-sitter's parsing capabilities, we conducted a manual validation on a statistically significant subset of 377 randomly selected samples (corresponding to a 95\% confidence level with a 5\% margin of error). 
To mitigate subjective bias, the first and third authors performed the verification independently. 
The first author validated 349 samples as correct (93\% correctness rate), while the third author confirmed 346 (92\% correctness rate). 
The resulting Cohen's Kappa coefficient of 0.9448 indicates near-perfect agreement, providing strong evidence for Tree-sitter's reliability. The validation followed a rigorous protocol: first, mapping code changes back to the original security patch hunks to isolate ground-truth vulnerable functions; second, cross-referencing these changes with pre- and post-patch function definitions to verify structural alignment and line-number consistency. Our inspection confirms that Tree-sitter provides a robust foundation for our parsing requirements.

\begin{table}[t]
\caption{Statistical summary of the function-level multilingual vulnerability dataset}
\label{tab:detection_dataset}
\centering
\begin{adjustbox}{width=0.9\textwidth,center}
\begin{tabular}{ccccccc}
\toprule
\textbf{Languages} & \textbf{\#Training} & \textbf{\#Validation} & \textbf{\#Test} & \textbf{\#Total} & \textbf{\#Vul} & \textbf{\#Non-Vul} \\ 
\midrule
C         & 2,444     & 305        & 307  & 3,056 & 1,541 & 1,515 \\
C++       & 1,432     & 179        & 181  & 1,792 & 911 & 881 \\
C\#       & 341     & 42         & 44   & 427  & 212 & 215 \\
Go        & 2,323    & 290        & 292  & 2,905  & 1,462 & 1,443 \\
Java      & 2,587     & 323        & 325  & 3,235 & 1,622 & 1,613 \\
JavaScript        & 4,374     & 546        & 548  & 5,468 & 2,743 & 2,725 \\
Python    & 2,625     & 328        & 329  & 3,282 & 1,642 & 1,640 \\ \midrule
Total     & 16,126     & 2,013       & 2,026 & 20,165 & 10,133 & 10,032\\ 
\bottomrule
\end{tabular}
\end{adjustbox}
\begin{tablenotes}
        \footnotesize
        \centering
        \tiny
        \item[*] \#Training, \#Validation, \#Test, \#Total, \#Vul and \#Non-Vul represent the number of training functions, validation functions, test functions, vulnerable functions, and non-vulnerable functions, respectively
\end{tablenotes}
\end{table}

\begin{table}[t]
\caption{Statistical summary of the line-level multilingual vulnerability dataset}
\label{tab:detection_dataset_line}
\centering
\begin{adjustbox}{width=0.9\textwidth,center}
\begin{tabular}{ccccccc}
\toprule
\textbf{Languages} & \textbf{\#Training} & \textbf{\#Validation} & \textbf{\#Test} & \textbf{\#Total} & \textbf{\#Vul} & \textbf{\#Non-Vul} \\  \midrule
C         & 937    & 118        & 120  & 1,175 & 3,764 & 32,191  \\
C++       & 528     & 63        & 60  & 651 & 2,223 & 14,380 \\
C\#       & 123      & 15         & 13   & 151  & 396 & 3,123\\
Go        & 735     & 92        & 92  & 919  & 3,123 & 19,958\\
Java      & 1,100     & 134        & 141  & 1,375  & 3,602 & 22,414\\
JavaScript        & 1,996     & 254   & 246  & 2,496  & 8,350 & 38,376\\
Python    & 1,094     & 135        & 138  & 1,367  & 3,702 & 21,165\\ 
\midrule
Total     & 6,513     & 811       & 810 & 8,134 & 25,161 & 150,607\\ 
\bottomrule
\end{tabular}
\end{adjustbox}
\begin{tablenotes}
        \footnotesize
        \centering
        \tiny
        \item[*] \#Training, \#Validation, \#Test, \#Total, \#Vul and \#Non-Vul represent the number of training functions, validation functions, test functions, vulnerable lines, and non-vulnerable lines, respectively
\end{tablenotes}
\end{table}

\textbf{Construction of training and test datasets}.
Similar to~\cite{fu2022linevul}, we divide the REEF dataset of function-level and line-level multilingual vulnerability into training, validation, and test sets at an 8:1:1 ratio.
Specifically, we employed stratified random sampling to generate the data splits. Rather than performing a single global split, which risks under-representing minority languages, we partitioned the dataset into seven distinct strata, one for each language. 
Within each stratum, we applied the predetermined 8:1:1 ratio to generate language-specific training, validation, and test subsets. This stratification strategy serves two purposes: first, it ensures that every language is consistently represented across all pipeline stages; second, it preserves the natural variation in data volume across languages. 
By maintaining this skew, our dataset reflects the real-world prevalence of vulnerabilities in open-source repositories, ensuring that the model is evaluated on a distribution that aligns with the actual development landscape.
As shown in Table~\ref{tab:detection_dataset}, the function-level multilingual vulnerability dataset contains 16,126, 2,013, and 2,026 functions and corresponding labels for training, validation, and testing over seven languages.
As shown in Table~\ref{tab:detection_dataset_line}, the line-level multilingual vulnerability dataset contains 6,513, 811, and 810 vulnerable functions and corresponding line labels for training, validation, and testing across the studied languages.

\begin{figure*}[!t]
    \centering
    \includegraphics[width=.7\linewidth]{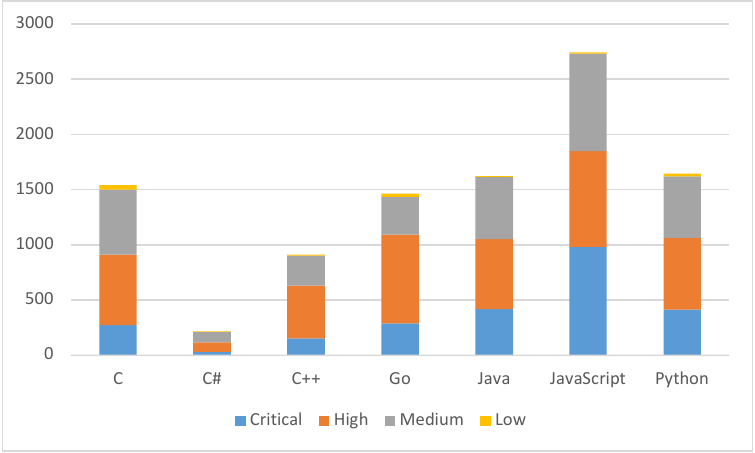}
    \caption{The severity distribution of multilingual vulnerability dataset}
    \label{fig:severity}
\end{figure*}

\textbf{Distribution of vulnerability severity.} 
To analyze the dataset quality, we examined the severity levels of both function-level and line-level datasets, as shown in Table~\ref{tab:detection_dataset} and~\ref{tab:detection_dataset_line}.
We used the Common Vulnerability Scoring System (CVSS), a standardized method for measuring vulnerability severity. 
Using CVSS v4.0 Ratings,\footnote{\url{https://nvd.nist.gov/vuln-metrics/cvss}} we classified the vulnerabilities into four levels: low, medium, high, and critical.
Figure~\ref{fig:severity} shows the severity distribution across the seven programming languages studied.
In this multilingual vulnerability dataset, only 1.2\% are rated as low severity, while 41.12\% are classified as high severity and 25.09\% as critical severity.
These findings demonstrate that the real-world vulnerabilities provided by REEF are high-quality and predominantly target severe vulnerabilities across both function-level and line-level multilingual vulnerability.

\subsection{Experimented Language Models}
\label{subsec:models}
In our study, we investigated the performance of two types of language models in detecting multilingual vulnerability at both the function level and the line level: pre-trained language models and large language models.
For the PLMs, we included six state-of-the-art PLMs for multilingual vulnerability detection, including:

\begin{itemize}
    \item \textbf{CodeBERT}~\citep{feng2020codebert} is a powerful pre-trained model that combines source code and natural language processing capabilities through its multilayer Transformer architecture.
    
    \item \textbf{CodeT5}~\citep{wang2021codet5} is a unified encoder-decoder model that integrates code token type information directly into its architecture. Building upon the T5 framework, it employs denoising sequence-to-sequence pre-training techniques.
    
    \item \textbf{CodeT5P}~\citep{wang2023codet5+} extends CodeT5's capabilities through an architecture featuring a shallow encoder paired with a deep decoder. The training process follows a multi-stage approach, beginning with unimodal training before progressing to bimodal data integration.

    \item \textbf{UniXCoder}~\citep{guo2022unixcoder} leverages a unified cross-modal pre-trained architecture that incorporates multiple information sources like abstract syntax trees (AST) and code comments to enhance code representation, all built on transformer-based foundations.

    \item \textbf{LineVul}~\citep{fu2022linevul} is a Transformer-based approach for predicting software vulnerabilities at the line level, aiming to enhance the granularity and accuracy of vulnerability detection. By leveraging the BERT architecture, LineVul captures long-term dependencies within code sequences, enabling precise identification of vulnerable lines.

    \item \textbf{Text-Embedding Models}~\citep{textemb2024} refer to a series of the latest generation of embedding models from OpenAI that offer enhanced representation capabilities for both text and code. Specifically, we utilized three text-embedding models: Text-Embedding-Ada-002, Text-Embedding-3-Small, and Text-Embedding-3-Large.
\end{itemize}
We focused exclusively on PLMs rather than employing GNNs or RNNs for multilingual vulnerability detection. 
This choice was motivated by PLMs' proven strengths in vulnerability detection compared to RNNs and GNNs~\citep{steenhoek2023empirical}. 
Moreover, alternative approaches~\citep{zhang2023vulnerability,liu2024pre,steenhoek2024dataflow} heavily rely on Abstract Syntax Trees (ASTs). Since ASTs are inherently language-specific structures, creating a unified AST representation across different programming languages is impractical due to varying syntax and semantic constructs.

For LLMs, we studied five state-of-the-art models, including three open-source LLMs and two closed-source LLMs, which have demonstrated promising performance in various code-related tasks~\citep{zhou2024large, hou2023large}. 
We strategically selected three open-source LLMs with parameter counts ranging from 6.7B to 8B.  This choice is driven by pragmatic considerations: many practitioners lack the high-end computational resources required to deploy or instruction-tune larger models.  
Furthermore, our selection is supported by recent findings~\citep{alizadeh2025language,zhong2025larger} indicating that increased model scale does not consistently guarantee superior performance on specialized tasks.  Consequently, these models represent an optimal balance between computational efficiency and competitive predictive capability.
The details of these LLMs are as follows:

\begin{itemize}  
    \item 
    \textbf{DeepSeek-Coder}~\citep{guo2024deepseek} trained on a massive dataset of 2 trillion tokens spanning 87 programming languages. With its large 16K context window and specialized training in code completion tasks, it achieves leading performance among open-source models and even outperforms some closed-source solutions like Code Llama and GPT-3.5-Turbo in certain areas.
    
    \item \textbf{Code Llama}~\citep{roziere2023code} derived from Llama 2, is a leading decoder-only language model specialized in code generation and infilling. The model achieves its capabilities through extensive fine-tuning on 500B tokens of code-focused data.

    \item \textbf{Llama 3}~\citep{dubey2024llama} represents an innovative suite of large-scale multilingual language models leveraging the Transformer architecture to enhance performance across diverse natural language understanding tasks. Through sophisticated optimizations in data quality, training methodology, and architectural design, Llama 3 demonstrates marked advancements in natural language processing capabilities.
    
    \item \textbf{GPT-3.5-Turbo}~\citep{chatgpt2022} is an improved version of GPT-3.5, optimized for conversations through reinforcement learning from human feedback. 
    As one of the core models powering ChatGPT, it has become among the most widely used general-purpose language models.

    \item \textbf{GPT-4o}~\citep{openai2024gpt4o}, introduced by OpenAI, is a multimodal AI model capable of processing and generating text, audio, and images in real time. 
    It integrates these modalities within a unified architecture, enabling seamless interaction across different data types.
\end{itemize}

To ensure a rigorous comparison against the state-of-the-art, we extend our evaluation to include GRACE~\citep{lu2024grace}, a representative framework that enhances LLMs by injecting structural program information derived from ASTs. Given that GRACE utilizes a zero-shot prompting strategy to elicit vulnerability insights without task-specific parameter updates, we categorize it as a zero-shot prompting baseline in our study.
We exclude MSIVD~\citep{du2024vul} because we lack attack complexity data and precise vulnerability explanations across multiple languages, which could lead to inconsistent instructional quality across languages. Similarly, we exclude Vul-RAG~\citep{yang2024security} to avoid knowledge availability bias, as we lack sufficient language-specific information to construct the vulnerability knowledge base.
As a result, we focused primarily on vanilla general and code-specific LLMs for multilingual vulnerability detection, following previous empirical studies~\citep{10479409,yin2024multitask, yin2024pros, zhou2024largeworkshop}.

Moreover, we also introduce two types of dummy classifiers at the function-level and line-level detection as the ordinary baselines.
The first type (${DummyClf}_{vul}$) predicts all functions or lines as vulnerable. 
The second type (${DummyClf}_{clean}$) predicts all functions or lines as clean. 
These baseline comparisons help us evaluate whether PLM and LLM's performance on multilingual vulnerability detection differs meaningfully from extreme predictions, allowing us to assess whether PLM and LLM predictions show any systematic patterns rather than being purely random.

\begin{figure}[]
    \centering
    \subfloat[\small A zero-shot prompting template for function-level multilingual vulnerability detection]{
        \includegraphics[width=.7\linewidth]{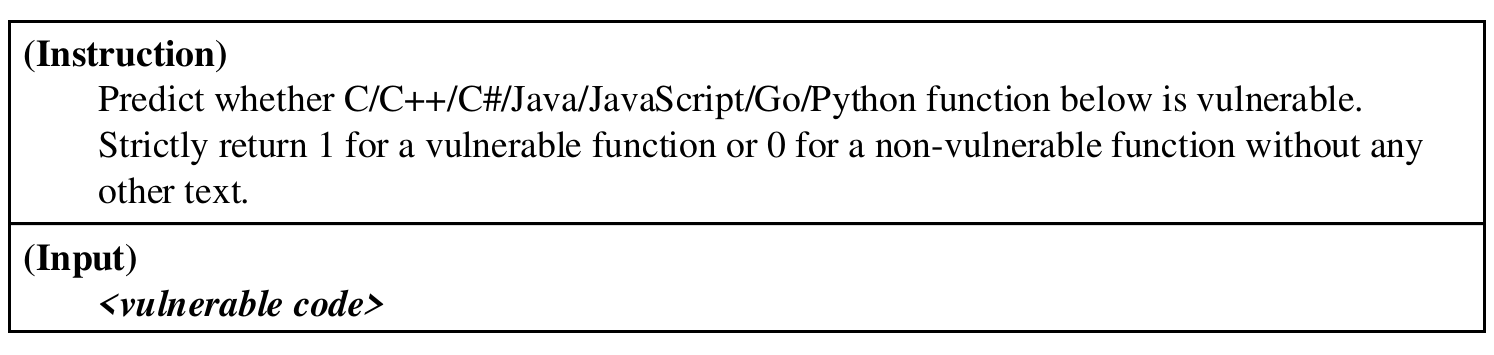}
        \label{fig:zsp_func}
    }\\
    \subfloat[\small A zero-shot prompting template for line-level multilingual vulnerability detection]{
        \includegraphics[width=.7\linewidth]{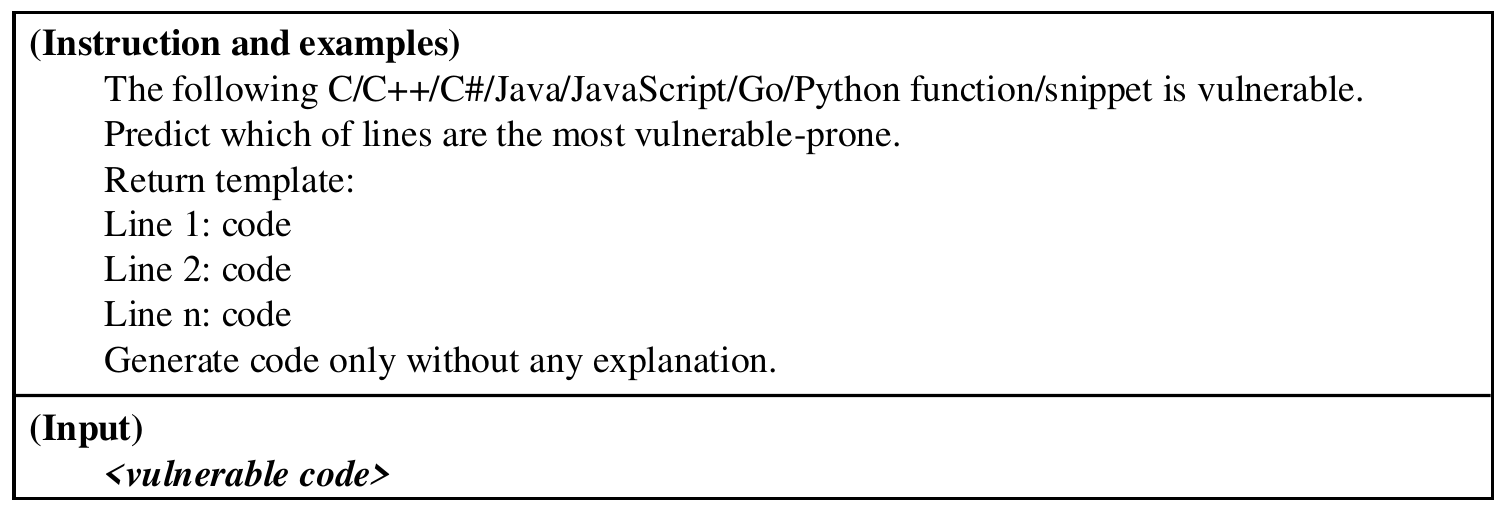}
        \label{fig:zsp_line}
    }
    \caption{Zero-shot prompt template for multilingual vulnerability detection}
    \label{fig:zsp}
\end{figure}

\subsection{Strategies for Large Language Models}
\label{subsec:strategy}
We also investigated the effectiveness of LLMs' various learning strategies for detecting multilingual vulnerability at both the function- and line-levels, since different learning strategies can significantly affect the LLMs' performance.
For instance, existing studies~\citep{tian2024large} have shown that fine-tuning strategies effectively enhance LLM performance by adapting general LLMs to specific tasks.
Additionally, diverse prompting strategies have been introduced as plug-and-play solutions to enhance LLM performance in code-related tasks~\citep{kojima2022large}.
Thus, we devised three LLM strategies:
\begin{itemize}[leftmargin=10pt]
   
    \item \textbf{Zero-shot prompting strategy}: 
    We developed a prompt structure with system and user roles, following prior works~\citep{zhou2024out, zhang2023pre}. We directly employed a structured instruction and a vulnerable function to prompt LLMs for multilingual vulnerability detection at both function and line levels, without using examples.
    Figure~\ref{fig:zsp} shows the zero-shot prompt templates for function-level and line-level multilingual vulnerability detection.
    For function-level detection, we specify that the input is a code function written in a particular programming language, then ask the LLM to determine whether the function is vulnerable or clean.
    For line-level detection, we specify that the input is a vulnerable function written in a particular programming language, then ask the LLM to identify the vulnerable line number and its corresponding code.

    \begin{figure}[t!]
    \centering
    \subfloat[A few-shot prompting template for function-level multilingual vulnerability detection]{
        \includegraphics[width=.8\linewidth]{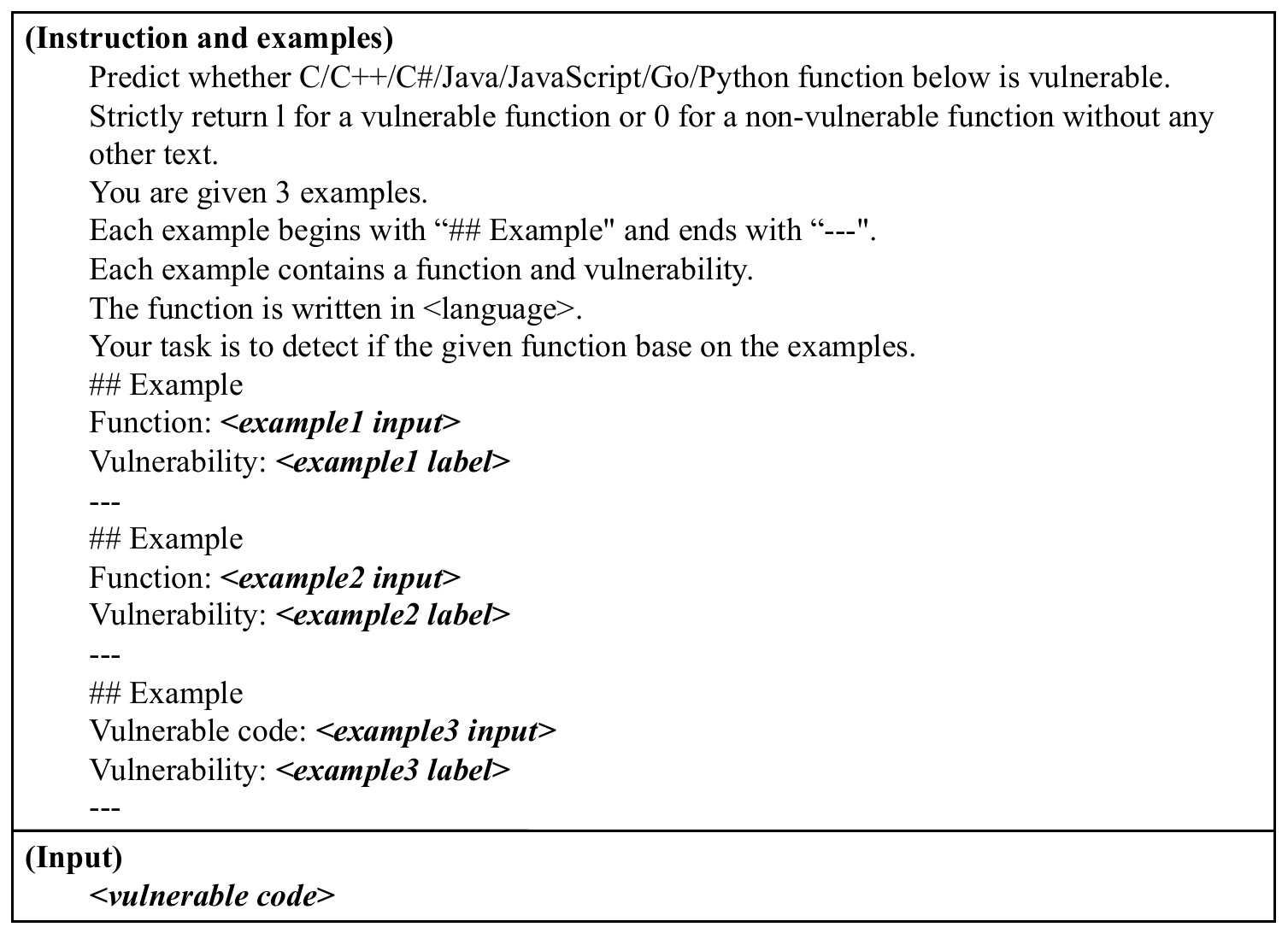}
        \label{fig:fsp_func}
    }\\
    \subfloat[A few-shot prompting template for line-level multilingual vulnerability detection]{
        \includegraphics[width=.8\linewidth]{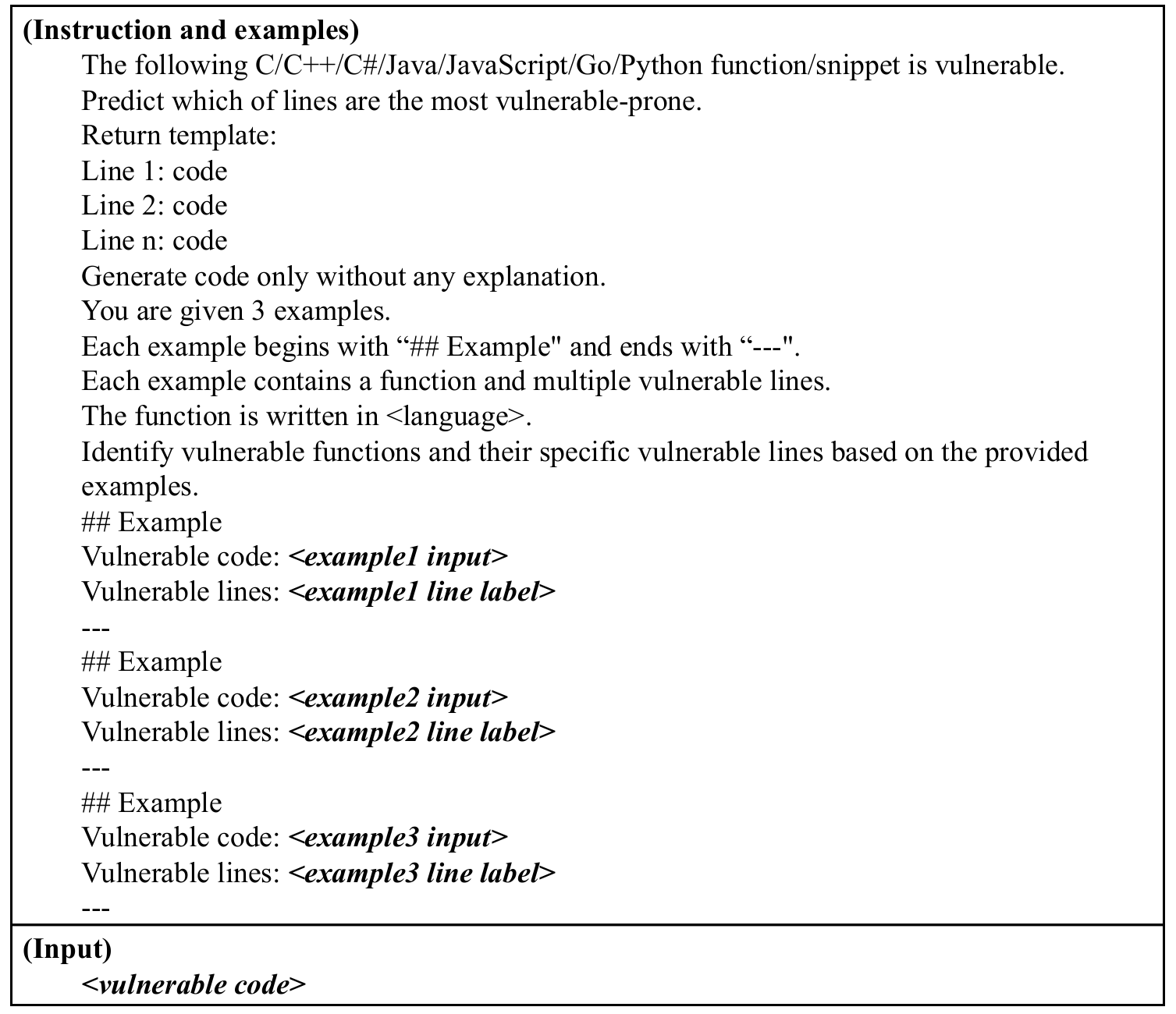}
        \label{fig:fsp_line}
    }
    \caption{Few-shot prompt template for multilingual vulnerability detection}
    \label{fig:fsp}
\end{figure}

    \item \textbf{Few-shot prompting strategy}: 
    It refers to providing LLMs with a small number of input-output examples in the prompt to help them understand the desired task before giving them a new input to complete.
    Specifically, it combines demonstration examples with a zero-shot prompt to form a new few-shot prompt, which is then fed to LLMs for detecting function-level or line-level multilingual vulnerability.
    Figure~\ref{fig:fsp} presents the few-shot prompt template for function-level and line-level multilingual vulnerability detection. 
    Based on prior research ~\citep{PORNPRASIT2024107523}, when analyzing a given function, we select three demonstration examples from the REEF training dataset using BM25~\citep{robertson2009probabilistic}. 
    These examples are used to prompt LLMs to detect function- and line-level multilingual vulnerabilities.
    BM25 was chosen as the sample selection method because previous studies~\citep{gao2023makes, yuan2023evaluating} demonstrate its superior performance over alternative approaches for software engineering tasks.
    
    \item \textbf{Instruction-tuning strategy}: 
    It enables LLMs to acquire task-specific knowledge and align with the desired response characteristics by supervised training on numerous instruction-filled function pairs.
    \citet{ouyang2022training} have shown that LLMs perform better on new, unfamiliar tasks when they are fine-tuned using diverse datasets that include natural language instructions.
    
    For function-level detection, an instruction-filled function pair contains three elements: a zero-shot prompt instruction (see Figure~\ref{fig:zsp_func}), a given function, and a label indicating whether that function is vulnerable or clean.
    Using all 16,126 instances from the function-level training set, we constructed a function-level instruction-filled fine-tuning set. We then fine-tuned the LLMs on this fine-tuning set to detect function-level multilingual vulnerability using either zero-shot or few-shot prompts.
    Similarly, for line-level detection, an instruction-filled function pair consists of a zero-shot prompt instruction (see Figure~\ref{fig:zsp_line}), a vulnerable function, and corresponding line labels. 
    Using all 6,513 instances from the line-level training set, we created a line-level instruction-filled fine-tuning set. Then, we fine-tuned the LLMs to detect line-level multilingual vulnerability using either zero-shot or few-shot prompts.

    In our study, we adopt Parameter-Efficient Fine-Tuning (PEFT) via Low-Rank Adaptation (LoRA)~\citep{hu2021lora} to perform instruction tuning on LLMs on both function- and line-level multilingual vulnerability detection. 
    LoRA reduces the number of trainable parameters by freezing the original model weights and injecting trainable low-rank matrices into each target weight matrix.
    Specifically, given an original weight matrix $W_0 \in \mathbb{R}^{d \times k}$, LoRA represents the weight update $\Delta W$ as the product of two low-rank matrices $A \in \mathbb{R}^{d \times r}$ and $B \in \mathbb{R}^{r \times k}$, i.e.,
    \begin{equation}
        \Delta W = A B, \quad r \ll \min(d, k),
    \end{equation}
    and the adapted weight is
    \begin{equation}
        W = W_0 + \alpha \cdot \Delta W,
    \end{equation}
    where $\alpha$ is a scaling factor. This formulation enables us to fine-tune large models for instruction-following tasks with significantly reduced computational cost while maintaining competitive performance. Moreover, its parameter efficiency allows us to adapt the same base model across multiple programming languages and domains without requiring full retraining, which is particularly advantageous for multilingual and cross-domain instruction tuning.
\end{itemize}

\subsection{Metrics}
\label{subsec:metrics}
Following existing work~\citep{fu2022linevul,steenhoek2024dataflow}, we adopted four widely-used metrics, \textit{Accuracy}, \textit{Precision}, \textit{Recall}, and \textit{F1-score}, to assess all our studied models' effectiveness in detecting function-level multilingual vulnerability as a binary classification task.
\textit{Accuracy} measures the proportion of correctly predicted instances out of all instances. \textit{Precision} quantifies how many of the predicted positive instances are actually correct, while \textit{Recall} assesses the proportion of actual positive instances correctly predicted by the model. 
The \textit{F1-score} combines precision and recall into a single metric, providing a balanced view especially useful when dealing with imbalanced datasets.
Specifically, in our study, we used binary-averaged \textit{Precision}, \textit{Recall}, and \textit{F1-score} because function-level vulnerability detection can be formulated as a binary classification task.
For line-level multilingual vulnerability detection, we formulated the line-level detection task as a binary classification problem for each code line by following the existing work~\citep{ding2024traced}.
Therefore, we also adopted \textit{Accuracy}, \textit{Precision}, \textit{Recall}, and \textit{F1-score} to assess all our studied models' effectiveness in detecting line-level multilingual vulnerability.

Furthermore, we adopted the \textit{False Positive Rate} (FPR) to measure how often a classifier incorrectly predicts the positive class when the actual class is negative, and \textit{False Negative Rate} (FNR) to quantify how often the model mistakenly classifies positive instances as negative.
In vulnerability detection, false positives (when clean code is flagged as vulnerable) cause developers to waste resources investigating secure code, while false negatives (when vulnerable code is flagged as clean) pose a fatal risk by leaving security flaws unaddressed.
The trade-off between FPR and FNR, achieving lower FNR at the cost of lower FPR, is critical for measuring the intrinsic effectiveness of PLMs and LLMs. 
Specifically, FPR is defined as $\frac{FP}{FP + TN}$ where FP represents false positives and TN represents true negatives. FNR is defined as $\frac{FN}{FN + TP}$ where FN represents false negatives and TP represents true positives.
A high FPR means many clean functions and lines remain unrecognized. A high FNR means many vulnerable functions or lines remain undetected.

Following the recommendations of~\citet{uddin2025deep}, we adopt the \textit{Matthews Correlation Coefficient} (MCC) and \textit{Area Under the Receiver Operating Characteristic Curve} (AUC) to provide a comprehensive evaluation of multilingual vulnerability detection at both the function and line levels. MCC serves as a robust measure of binary classification quality by accounting for all four quadrants of the confusion matrix, calculated as $\frac{TP \times TN - FP \times FN}{\sqrt{(TP + FP)(TP + FN)(TN + FP)(TN + FN)}}$. 
Complementing this, AUC is employed to evaluate the model's discriminative ability across all possible classification thresholds. Statistically, AUC represents the probability that a randomly selected vulnerable instance will be ranked higher than a benign one, making it invariant to the specific choice of threshold. We compute the AUC by using the trapezoidal rule.

To reduce ambiguity in calculating metrics across multiple programming languages, we use \textit{Accuracy} as an illustrative example.
Suppose our test set consists of only two languages: Python and Java. Java contains one sample and Python contains two samples. A model correctly predicts one Java sample and one Python sample.
Instead of calculating \textit{Accuracy} for each language separately (Python: 50\%, Java: 100\%) and then averaging those percentages (75\%), we aggregate all samples first. The Total Correct Predictions is two and the Total Samples is three. The global \textit{Accuracy} is calculated as $\text{Accuracy} = \frac{\text{Total Correct Predictions}}{\text{Total Samples}} = \frac{2}{3} \approx 66.7\%$.
This approach ensures that performance is weighted by the actual distribution of samples in each language, providing an objective measurement of the holistic performance of PLMs and LLMs.

\subsection{Implementation and Environment}
\label{subsec:implementation}
For pre-trained language models (PLMs), we replicated CodeBert, CodeT5, CodeT5P, UniXCoder, LineVul, and Text-Embedding-Models following existing works~\citep{steenhoek2024dataflow}. 
We obtained all open-source pre-trained checkpoints (i.e., CodeBert, CodeT5, CodeT5P, UniXCoder, and LineVul) from Huggingface~\citep{wolf2019huggingface}. 
Since Text-Embedding-Models are a series of closed-source PLMs, we accessed them through OpenAI's API.
For our experiments with closed-source LLMs, we accessed GPT-3.5-Turbo (model version gpt-3.5-turbo-0125) and GPT-4o (model version gpt-4o-2024-08-06) through OpenAI's APIs~\citep{openai2024}.
For open-source LLMs, we utilized pre-trained checkpoints from Huggingface~\citep{wolf2019huggingface} for DeepSeek-Coder (6.7B parameters), Code Llama (7B parameters), and Llama 3 (8B parameters). We implemented Low-Rank Adaptation during fine-tuning to optimize computational efficiency and prevent overfitting.
We conducted all the experiments on an Intel Xeon CPU Gold-6342 machine with 512 GB RAM, Ubuntu 20.04.6, and two A800 GPUs.
More implementation details can be found in our replication package. 
To facilitate future research, we have made all the used datasets and model execution scripts publicly available. 

%% file: results.tex
\section{Evaluation Results}
\label{sec:results_and_analysis}

\subsection{RQ1: How effective are PLMs and LLMs in detecting multilingual vulnerabilities at the function-level?}
\label{subsec:RQ1}

\noindent
\textbf{\emph{\underline{Approach.}}}
This research question provides a comparative analysis of performance between various PLMs and LLMs with different strategies in detecting function-level multilingual vulnerability.
Specifically, we investigated the effectiveness of eight PLMs and five LLMs.
The model details are summarized in Section~\ref{subsec:models}.
We now detail the training/inference process below.

Regarding PLMs, they are pre-trained on a corpus of natural language and source code snippets from various programming languages, but they have not been exposed to the task of function-level vulnerability detection, particularly across multilingual vulnerability.
Prior studies~\citep{fu2022linevul,steenhoek2023empirical,steenhoek2024dataflow} have demonstrated the effectiveness of fine-tuned PLMs for function-level vulnerability detection. 
Therefore, we fine-tune all studied open-source PLMs (i.e., CodeBert, CodeT5, CodeT5P, UniXCoder, and LineVul) on the function-level training set. 
The model architecture for function-level multilingual vulnerability detection employs a hybrid design that integrates a PLM encoder with a specialized binary classification head. 
Specifically, the binary classifier is implemented as a multi-layer perceptron (MLP) that operates on the aggregate sequence representation. To facilitate non-linear mapping and mitigate overfitting, the architecture incorporates a hyperbolic tangent ($\tanh$) activation function interleaved with stochastic dropout layers. 
The forward pass begins by extracting the representation of the leading sequence token (equivalent to the [CLS] or <s> token), which is then processed through a dense layer and non-linearities. 
The final output of this classification head is the raw logit, representing the model's unnormalized confidence score before being mapped to the binary prediction space.
We train all parameters of both the encoder and binary classifier through supervised learning on the function-level training set. 
The trained encoder outputs a code embedding representing the given input given code. 
This embedding is then fed into the trained binary classifier, which produces a binomial distribution indicating the probability of the given function being vulnerable.
For closed-source PLMs (i.e., Text-Embedding-Ada-002, Text-Embedding-3-Small, and Text-Embedding-3-Large), we follow the same training and inference approach. 
Since these PLMs do not allow access to their parameters, we treat them as frozen encoders and only train the binary classifier.

Regarding LLMs, they are decoder-only models with billions of parameters, pre-trained on vast data corpora of text and source code to process and generate human-like language.
Although LLMs have not been trained specifically for vulnerability detection, we can leverage their powerful in-context learning capabilities and pre-trained prior knowledge of various programming languages to perform function-level multilingual vulnerability detection with or without instruction tuning.
Without instruction tuning, we directly applied zero-shot or few-shot prompts to have LLMs predict either 0 or 1, where 1 indicates a vulnerable function and 0 indicates a non-vulnerable one.
With instruction tuning, we first trained the LLMs through supervised fine-tuning on the instruction-filled function-level training set. 
Then, we applied zero-shot or few-shot prompts to obtain the predictions of trained LLMs.

To evaluate the effectiveness of the studied PLMs and LLMs in detecting function-level multilingual vulnerabilities, we employed multiple metrics, including \textit{Accuracy}, \textit{Precision}, \textit{Recall}, \textit{F1-score}, FPR, FNR, MCC, and AUC as detailed in Section~\ref{subsec:metrics}. 
Given the balanced distribution between positive and negative labels, \textit{Accuracy} was used as the primary evaluation metric.

\begin{table}[ht]
    \caption{Performance of PLMs and LLMs for function-level multilingual vulnerability detection on balance scenario}
    \label{tab:rq1_func}
    \centering
    \tabcolsep=3.0mm
    \small
    \begin{adjustbox}{max width=1.0\textwidth, center}
    \begin{tabular}{lrrrrrrrr}
        \toprule
        \toprule
        \textbf{Techniques} & \textbf{Accuracy} & \textbf{Recall} & \textbf{Precision} & \textbf{F1-score} & \textbf{FPR} & \textbf{FNR} & \textbf{MCC} & \textbf{AUC} \\
        \midrule
        \rowcolor{gray!40}\multicolumn{9}{l}{\textbf{Dummy Classifiers}} \\ 
        \midrule
            ${DummyClf}_{vul}$ & 0.5030  & 1.0000 & 0.5030 & 0.6693 & 1.0000 & 0.0000 & 0.0000 & 0.5000\\
            ${DummyClf}_{clean}$ & 0.4970 & 0.0000 & 0.0000 & 0.0000 & 0.0000 & 1.0000 & 0.0000 & 0.5000\\
        \midrule
        \rowcolor{gray!40}\multicolumn{9}{l}{\textbf{PLMs}} \\ 
        \midrule
            Text-Embedding-3-Large & 0.5064 & 0.9696 & 0.5049 & 0.6640 & 0.9623 & \textbf{0.0304} & 0.0201 & 0.5037\\
            Text-Embedding-3-Small & 0.5044 & 0.9676 & 0.5038 & 0.6626 & 0.9643 & 0.0324 & 0.0093 & 0.5016\\
            Text-Embedding-Ada-002 & 0.5030 & \textbf{1.0000} & 0.5030 & 0.6693 & 1.0000 & 0.0000 & 0.0000 & 0.5000\\
            CodeBERT & 0.5173 & \textbf{1.0000} & 0.5103 & 0.6757 & 0.9712 & 0.0000 & 0.1212 & 0.5144\\
            LineVul & 0.5173 & \textbf{1.0000} & 0.5103 & 0.6757 & 0.9712 & 0.0000 & 0.1212 & 0.5144\\
            UniXcoder & 0.5814 & 0.8930 & 0.5518 & 0.6822 & 0.7339 & 0.1070 & 0.2045 & 0.5796\\
            CodeT5 & 0.5854 & 0.9342 & 0.5519 & 0.6939 & 0.7676 & 0.0658 & 0.2342 & 0.5833\\
            CodeT5P & 0.6037 & 0.9529 & 0.5626 & \textbf{0.7075} & 0.7496 & 0.0471 & 0.2860 & 0.6015\\
        \midrule
        \rowcolor{gray!40}\multicolumn{9}{l}{\textbf{LLMs (Zero-Shot Prompting)}} \\ 
        \midrule
            DeepSeek-Coder &0.5005 &0.1796 &0.5097 & 0.2656 & 0.1748 & 0.8204 & 0.0063 & 0.5024\\
            Code Llama & 0.4877 & 0.9156 & 0.4950 & 0.6426 & 0.9454 & 0.0844 & -0.0585 & 0.4851\\
            Llama 3 & 0.5005 & 0.5015 & 0.5034 & 0.5025 & 0.5005 & 0.4985 & 0.0010 & 0.5005\\
            GPT-3.5-Turbo & 0.4827 & 0.6183 & 0.4888 & 0.5459 & 0.6544 & 0.3817 & -0.0375 & 0.4819\\
            GPT-4o & 0.4985 & 0.3631 & 0.5020 & 0.4214 & 0.2632 & 0.7399 & -0.0035 & 0.4985\\
            GRACE & 0.5061 & 0.1673 & 0.5323 & 0.2546 & 0.1495 & 0.8327 & 0.0244 & 0.5089\\
        \midrule
        \rowcolor{gray!40}\multicolumn{9}{l}{\textbf{LLMs (Few-Shot Prompting)}} \\ 
        \midrule
            DeepSeek-Coder & 0.3870 & 0.5309 & 0.4146 & 0.4656 & 0.7587 & 0.4691 & -0.2378 & 0.3861\\
            Code Llama & 0.4763 & 0.6251 & 0.4840 & 0.5456 & 0.6743 & 0.3749 & -0.0515 & 0.4754\\
            Llama 3 & 0.4299 & 0.5260 & 0.4437 & 0.4814 & 0.6673 & 0.4740 & -0.1440 & 0.4293\\
            GPT-3.5-Turbo & 0.3618 & 0.4426 & 0.3835 & 0.4109 & 0.7200 & 0.5574 & -0.2810 & 0.3613 \\
            GPT-4o & 0.5035 & 0.4347 & 0.5074 & 0.4683 & 0.4270 & 0.5653 & 0.0078 & 0.5039\\
        \midrule
        \rowcolor{gray!40}\multicolumn{9}{l}{\textbf{LLMs (Instruction Tuning + Zero-Shot Prompting)}} \\ 
        \midrule
            DeepSeek-Coder & 0.5039 & 0.3023 & 0.5116 & 0.3800 & 0.2920 & 0.6977 & 0.0113 & 0.5051\\
            Code Llama & 0.4783 & 0.5103 & 0.4824 & 0.4959 & 0.5531 & 0.4897 & -0.0429 & 0.5091\\
            Llama 3 & 0.5192 & 0.5348 & 0.5215 & 0.5281 & 0.4965 & 0.4652 & 0.0383 & 0.5192\\
            GPT-3.5-Turbo & 0.5454 & 0.5132 & 0.5517 & 0.5318 & 0.4220 & 0.4868 & 0.0914 & 0.5456 \\
            GPT-4o &0.5874 &0.4622 & 0.6206 &  0.5298 & 0.2859 & 0.5378 & 0.1820 & 0.5881\\
        \midrule
        \rowcolor{gray!40}\multicolumn{9}{l}{\textbf{LLMs (Instruction Tuning + Few-Shot Prompting)}} \\ 
        \midrule
            DeepSeek-Coder & 0.4906  & 0.4789 & 0.4934 & 0.4861 & 0.4975 & 0.5201 & -0.0176 & 0.4912 \\
            Code Llama & 0.5094 & 0.6408 & 0.5098 & 0.5678 & 0.6226 & 0.3592 & 0.0188 & 0.5091\\
            Llama 3 & 0.4946 & 0.4848 & 0.4975 & 0.4911 & 0.4955 & 0.5152 & -0.0107 & 0.4946 \\
            GPT-3.5-Turbo & 0.5000 & 0.4524 & 0.5033 & 0.4765 & 0.4518 & 0.5475 & 0.0005 & 0.5003 \\
            GPT-4o & \textbf{0.7196} & 0.6722 & \textbf{0.7454} & 0.7069 & \textbf{0.2324} & 0.3278 & \textbf{0.4360} & \textbf{0.7169}\\
        \bottomrule
        \bottomrule
    \end{tabular}
    \end{adjustbox}
\end{table}

\begin{figure}[t!]
    \centering
    \subfloat[\small Accuracy of PLMs in detecting function-level vulnerability across seven language]{
        \includegraphics[width=.7\linewidth]{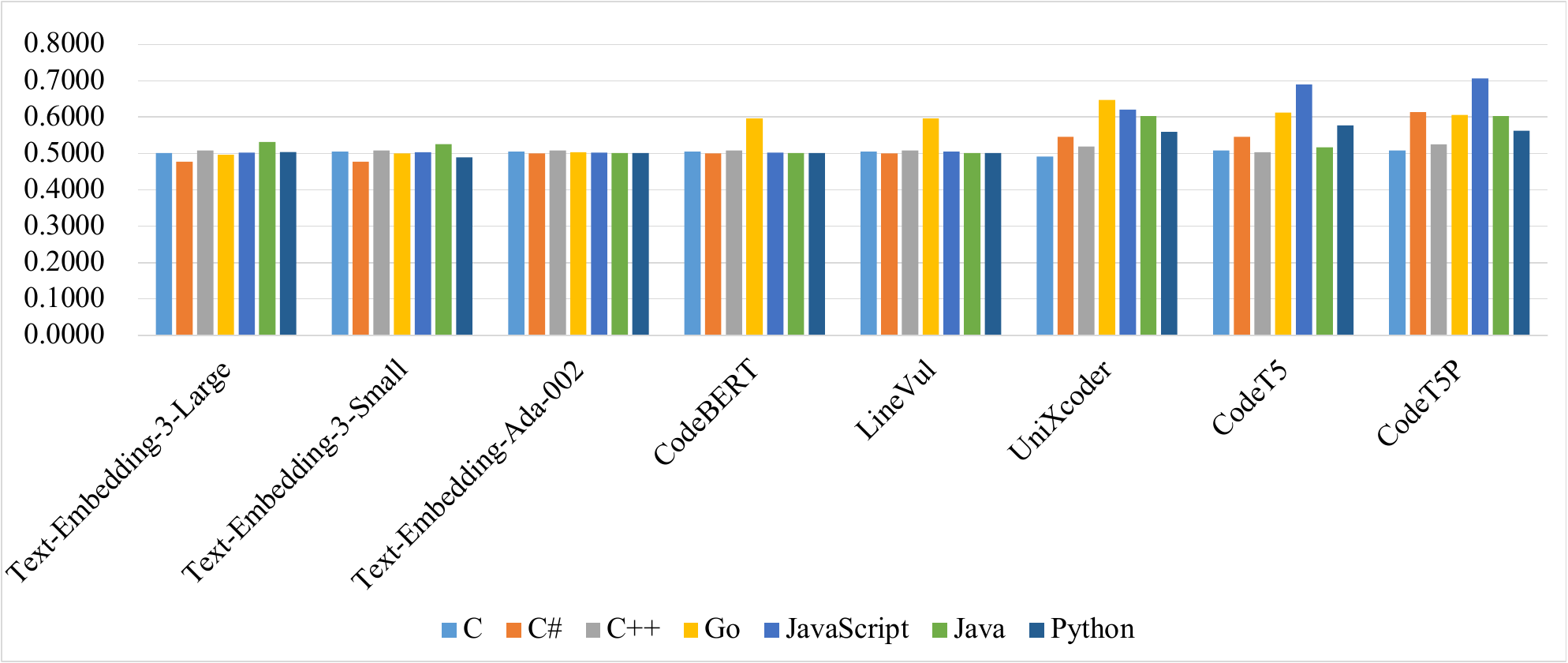}
        \label{fig:rq1_func_plm_lang}
    }\\
    \subfloat[\small Accuracy of LLMs (instruction tuning with few-shot prompting) in detecting function-level vulnerability across seven language]{
        \includegraphics[width=.7\linewidth]{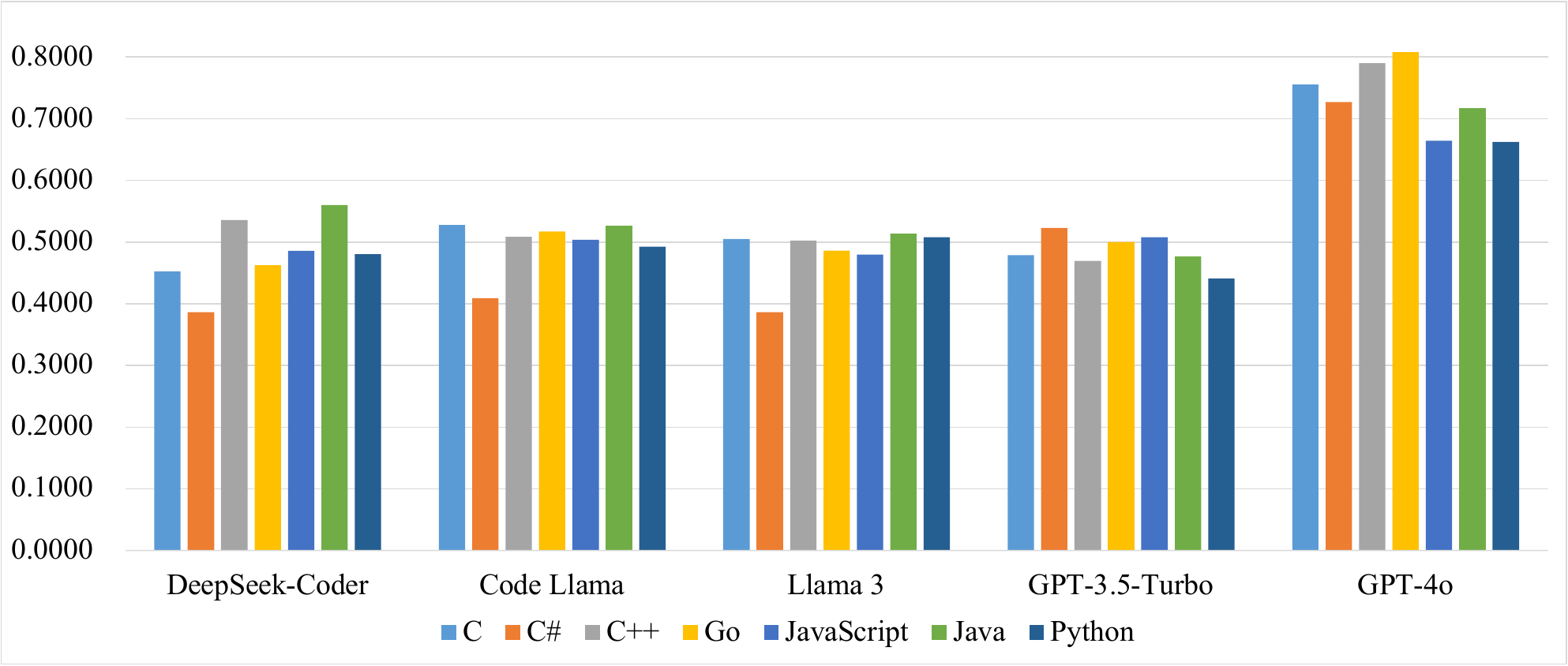}
        \label{fig:rq1_func_llm_lang}
    }
    \caption{Performance of PLMs and LLMs across seven programming languages in function-level multilingual vulnerability detection (y-axis: Accuracy)}
    \label{fig:rq1_func_lang}
\end{figure}

\noindent
\textbf{\emph{\underline{Results.}}}
Table~\ref{tab:rq1_func} shows the comparison results among PLMs and LLMs in terms of average Accuracy, Recall, Precision, F1-score, FPR, FNR, MCC, and AUC in detecting function-level multilingual vulnerability.
Within each metric, the values in bold indicate the model that exhibits the best performance among all PLMs and LLMs.
Figure~\ref{fig:rq1_func_lang} presents the performance of PLMs and LLMs with instruction tuning and few-shot prompting in seven programming languages in terms of Accuracy.

\textbf{Observation 1: Among PLMs, CodeT5P exhibits superior performance in function-level multilingual vulnerability detection.} 
In the PLMs' category, text-embedding models like Text-Embedding-3-Large, Text-Embedding-3-Small, and Text-Embedding-Ada-002 all hover around 0.5000 in Accuracy, indicating near-random performance on a balanced dataset.
Despite high Recall scores (all >0.9600), their low Precision (around 0.5000) results in modest F1-scores (around 0.6600). This suggests a bias toward over-identifying functions as vulnerable, in other words, a high FPR (all >0.9600).
The MCC scores for these models are consistently near zero (e.g., 0.0201 for Text-Embedding-3-Large), and AUC values hover around 0.5 (e.g., 0.5037), further confirming their inability to provide discriminative utility beyond a random guess on this balanced dataset.
Similarly, CodeBERT and LineVul achieve slightly higher Accuracy (0.5163 and 0.5173) and perfect Recall (1.0000) but suffer from low Precision (0.5098 and 0.5103). 
This leads to middling F1-scores (0.6753 and 0.6757) and limited practical utility, as they cannot reliably discriminate between vulnerable and non-vulnerable functions, yielding a low MCC of 0.1212 and an AUC of 0.5144.

More modern code-specific models show improved balance.
UniXcoder and CodeT5 both outperform general text-embedding models and simple code-specific models, with UniXcoder reaching 0.5814 Accuracy and CodeT5 achieving 0.5854.
These models strike a better trade-off between Recall and Precision, resulting in higher F1-scores (0.6822 and 0.6939, respectively) and improved discriminative power, as evidenced by their higher MCC values (0.2045 for UniXcoder and 0.2342 for CodeT5) and AUC scores (0.5796 and 0.5833, respectively).
Most notably, CodeT5P delivers the strongest performance across all PLMs, achieving the highest Accuracy of 0.6037 along with a solid F1-score of 0.7075. 
It also reaches the highest MCC (0.2860) and AUC (0.6015) in the PLM category.
Furthermore, the CodeT5P reaches FPR of 0.7496 and FNR of 0.0471, which implies that although CodeT5P can detect more vulnerable functions, it comes at the cost of wrongly detecting more clean functions.
This indicates that CodeT5P, which incorporates identifier-aware training objectives and enriched representation learning, is particularly effective at capturing the multilingual semantic nuances required for function-level multilingual vulnerability detection. 
Overall, the results demonstrate that while traditional PLMs struggle with precision and overall Accuracy, task-aware models like CodeT5P offer significant improvements, making them more suitable for multilingual vulnerability detection scenarios at the function-level.

\textbf{Observation 2: LLMs exhibit limitations in detecting function-level multilingual vulnerability through zero-shot or few-shot prompting strategies alone.}
In evaluating the effectiveness of LLMs for function-level multilingual vulnerability detection under zero-shot and few-shot prompting settings, Accuracy emerges as a critical metric due to the balanced nature of the dataset.
Under zero-shot prompting setting, the average Accuracies of LLMs across seven programming languages were 0.5005 for DeepSeek-Coder, 0.4877 for Code Llama, 0.5005 for Llama 3, 0.4758 for GPT-3.5-Turbo, and 0.4985 for GPT-4.
Specifically, DeepSeek-Coder achieves the highest Accuracy (0.5005), closely followed by Llama 3 (0.5005) and Code Llama (0.4877). 
Although DeepSeek-Coder exhibits the highest Accuracy, it achieves the lowest Recall (0.1796), leading to a poor F1-score (0.2656). This suggests a tendency to overpredict non-vulnerabilities.
In contrast, Code Llama stands out with the highest Recall (0.9156). However, its relatively lower Precision (0.4950) results in a moderate F1-score (0.6426), indicating a tendency to overpredict vulnerabilities, leading to a higher FPR (0.9454).
The MCC values for these models are either near zero or negative (e.g., -0.0585 for Code Llama and -0.0375 for GPT-3.5-Turbo), with AUC scores mostly below 0.5, indicating that they struggle to perform better than a random baseline in this setting.
GPT-3.5-Turbo and GPT-4o demonstrate lower Recall but more balanced Precision-Recall trade-offs, though with modest Accuracy scores (0.4827 and 0.4985, respectively).
Furthermore, we  examined GRACE, which integrates AST structural information through zero-shot prompting. As shown in Table 3, GRACE achieves the highest Accuracy (0.5061) and Precision (0.5323) among all zero-shot baselines, outperforming general-purpose LLMs.
However, this structural constraint induces a highly conservative prediction strategy. GRACE records the lowest Recall (0.1673) and F1-score (0.2546), along with the lowest FPR (0.1495). While explicit syntax information helps filter out benign code, reducing false alarms, it significantly hinders the model's ability to generalize and identify diverse vulnerability patterns. This results in a high FNR (0.8327) and minimal MCC (0.0244) and AUC (0.5089).

Under the few-shot prompting setting, this strategy does not consistently improve performance. 
In fact, most LLMs show decreased performance compared to zero-shot prompting, with GPT-4o being the only exception.
Specifically, DeepSeek-Coder drops sharply to 0.3870 Accuracy, Code Llama to 0.4763, Llama 3 to 0.4299, and GPT-3.5-Turbo to 0.3618, indicating that providing few examples may confuse rather than help the models.
In this case, GPT-4o is the only exception, slightly improving to 0.5035 Accuracy, though its Recall (0.4347) and F1-score (0.4683) remain relatively low, and its MCC (0.0078) and AUC (0.5039) show negligible improvement over the zero-shot baseline.
These results highlight that zero-shot prompting yields more stable Accuracy for LLMs, and few-shot prompting does not guarantee improvement. 
Overall, LLMs using only zero-shot or few-shot prompting show limited effectiveness in function-level multilingual vulnerability detection, emphasizing the need for further adaptation to the task domain.

\textbf{Observation 3: Instruction tuning is critical to improve the LLM's effectiveness in detecting function-level multilingual vulnerability.}
Instruction tuning is able to enhance the function-level multilingual vulnerability detection capabilities of LLMs, especially when paired with prompting strategies. 
In the instruction tuning with zero-shot prompting setting, GPT-4o achieves the highest Accuracy (0.5874), outperforming GPT-3.5-Turbo (0.5454), Code Llama (0.4783), Llama 3 (0.5192), and DeepSeek-Coder (0.5039). 
Notably, most of the instruction-tuned LLMs show improvement in  Accuracy, Precision, and F1-scores, compared to their non-instruction-tuned counterparts.
Specifically, GPT-4o improves its MCC to 0.1820 and AUC to 0.5881, demonstrating that instruction tuning effectively boosts the model's discriminative capabilities even without examples.
For instance, the most significant Accuracy improvements were seen in GPT-4o (17.83\%) and GPT-3.5-Turbo (14.63\%).
These results suggest that instruction tuning enables better generalization and contextual understanding of multilingual vulnerability patterns, even without in-context examples.

When few-shot prompting was combined with instruction tuning, most LLMs exhibited the same tendency as with few-shot prompting alone, i.e., LLMs' performance deteriorated, except for GPT-4o.
In this case, GPT-4o achieves a remarkable Accuracy of 0.7196, the highest across all LLMs with different strategies, along with the highest Precision (0.7454) and F1-score (0.7069). 
Additionally, GPT-4o achieves a relatively balanced ratio between its FNR of 0.3278 and FPR of 0.2324, indicating its ability to effectively detect vulnerable functions while maintaining accurate predictions for clean functions.
This configuration also produces the highest overall MCC (0.4360) and AUC (0.7169), signifying a significant leap in robustness over other approaches.
This suggests that GPT-4 has strong discriminative capabilities and robustness in detecting function-level multilingual vulnerability. 
Compared to the best-performing PLM, CodeT5P, GPT-4o exhibits a significant advantage in Accuracy and Precision, while maintaining a competitive F1-score. 
Although CodeT5P shows higher Recall (0.9529 vs. 0.6722), CodeT5P's FPR (0.7496) is also significantly higher than GPT-4o's FPR (0.2324).
In contrast, GPT-4o's balanced trade-off across all metrics, particularly its much lower harmonic mean between FPR and FNR ($\frac{0.2324 + 0.3278}{2}$ < $\frac{0.7496 + 0.0471}{2}$), making it more practical in real-world applications.
This comparison underscores the growing potential of instruction-tuned LLMs, especially GPT-4o, as powerful and adaptable techniques for function-level multilingual vulnerability detection when paired with minimal task-specific examples.

\textbf{Observation 4: The effectiveness of GPT-4o with instruction tuning and few-shot prompting is promising on seven different programming languages, and superior to other PLMs and LLMs.}
As shown in Figure~\ref{fig:rq1_func_plm_lang}, the three embedding models, Text-Embedding-3-Large, Text-Embedding-3-Small, and Text-Embedding-Ada-002, demonstrate consistent performance levels, maintaining an approximate Accuracy of 0.5000 across all seven programming languages evaluated.
The evaluation reveals that CodeBERT and LineVul demonstrate limited effectiveness, with an approximate Accuracy score of 0.5000 when detecting function-level vulnerability on C, C\#, C++, JavaScript, Java, and Python.
Notably, their performance improves in the context of the Go language, where both models achieve an Accuracy of 0.5959.
In contrast, UniXcoder, CodeT5, and CodeT5P consistently outperform other PLMs in Accuracy measurements across seven programming languages. 
Specifically, CodeT5P stands out as the best-performing PLM, demonstrating robust performance with an Accuracy of 0.5081 in C, 0.6136 in C\#, 0.5249 in C++, 0.6062 in Go, 0.7062 in JavaScript, 0.6031 in Java, and 0.5623 in Python.
UniXcoder and CodeT5 stand out as suboptimal PLMs, showing relatively weak performance with Accuracies of 0.4919 and 0.5081 in C, 0.5455 and 0.5455 in C\#, 0.5193 and 0.5028 in C++, 0.6473 and 0.6130 in Go, 0.6204 and 0.6898 in JavaScript, 0.6031 and 0.5169 in Java, and 0.5593 and 0.5775 in Python.

Figure~\ref{fig:rq1_func_llm_lang} shifts the focus to LLMs, revealing GPT-4o as the standout model in terms of Accuracy across all seven programming languages.
Specifically, GPT-4o achieves the strongest performance with an Accuracy of 0.7557 in C, 0.7273 in C\#, 0.7901 in C++, 0.8082 in Go, 0.6642 in JavaScript, 0.7169 in Java, and 0.6626 in Python.
Furthermore, GPT-4o not only surpasses CodeT5P in high-level languages but also exhibits higher and more stable Accuracy across traditionally challenging languages such as Go, C, and C++. 
In several cases (e.g., Python, JavaScript), GPT-4o exceeds 0.6500 Accuracy, reflecting its superior capacity to understand diverse syntactic structures and vulnerability patterns across languages. 
This comparison underscores the strength of instruction-tuned LLMs with few-shot prompting, positioning GPT-4o as a more versatile and reliable model for function-level multilingual vulnerability detection.

\begin{tcolorbox}\textbf{RQ1 Summary:}
GPT-4o (with instruction tuning and few-shot prompting) and CodeT5P achieved the highest performance among LLMs and PLMs, scoring 0.7196 and 0.6037 in average Accuracy across seven programming languages, respectively.
Furthermore, GPT-4o with instruction tuning and few-shot prompting outperformed all other PLMs and LLMs across all seven studied languages, showing improvements of 19.20\% and 85.94\% in average Accuracy.
In evaluating performance across multiple programming languages using instruction tuning and few-shot prompting, GPT-4o reached the highest Accuracy of 0.8082 with Go and the lowest Accuracy of 0.6626 with Python.
\end{tcolorbox}

\subsection{RQ2: How effective are PLMs and LLMs in detecting multilingual vulnerabilities at the line level?}
\label{subsec:RQ2}

\noindent
\textbf{\emph{\underline{Approach.}}}
This research question focuses on a comparative analysis of performance between different PLMs and LLMs with various strategies in detecting line-level multilingual vulnerability.
Specifically, we investigated the effectiveness of four PLMs and five LLMs. 
In this case, we excluded three Text-Embedding Models and LineVul. 
The Text-Embedding Models can only provide a single embedding for a given function, making them unsuitable for line-level detection. LineVul uses an unsupervised learning approach, which would be unfair to compare with the supervised learning method used in our experiment.
We detail the training and inference process below.

Regarding PLMs, we adopt a line-level embedding approach to align with the task of detecting vulnerabilities at the line-level.
Specifically, we represented each vulnerable function using up to \(n_{\text{lines}} = 113\) lines, with each line containing up to \(n_{\text{tokens}} = 31\) tokens.
This design is based on the observation that 95\% of functions in our dataset contain fewer than 113 lines.
Each input function is embedded into a tensor of shape \((n_{\text{lines}}, n_{\text{tokens}}, d_{\text{model}})\), where \(d_{\text{model}} = 768\) is the hidden dimension of the embedding.
To obtain line-level representations, we summarized each line by applying a single-layer Gated Recurrent Unit (GRU) over the token dimension (\(n_{\text{tokens}}\)), producing a tensor of shape \((n_{\text{lines}}, d_{\text{model}})\).
We formulated the task as a multi-label classification problem, aiming to identify which lines within a vulnerable function are vulnerable.
Each vulnerable function contains at least one, and potentially multiple, vulnerable lines.
We added a classification head for each PLM that outputs a single value per line, followed by a sigmoid activation function.
A line is classified as vulnerable if its output exceeds a threshold of 0.5.

Regarding LLMs, we expected to use their in-context learning capabilities to achieve line-level multilingual vulnerability detection with or without instruction tuning.
Line-level detection required LLMs to identify vulnerable code lines, but previous studies~\citep{fu2023chatgpt} and our initial experiments have shown that LLMs cannot accurately distinguish code lines based on line breaks.
To address this issue, we re-structured the functions by adding tags (i.e., \texttt{Line N:}) in front of each code line.
Without instruction tuning, we used zero-shot or few-shot prompts to guide LLMs in predicting vulnerable line numbers (i.e., tags of lines) and their corresponding code.
With instruction tuning, we first performed supervised fine-tuning on our instruction-filled line-level training set.
We then prompted these instruction-tuned LLMs to generate vulnerable line numbers and corresponding code using zero-shot or few-shot prompting.

To evaluate the effectiveness of the studied PLMs and LLMs in detecting line-level multilingual vulnerability, we used common classification metrics: \textit{Accuracy}, \textit{Precision}, \textit{Recall}, \textit{F1-score}, FPR, FNR, MCC, and AUC, as detailed in~\ref{subsec:metrics}.
Since our task involves classifying each line of code within a function, we faced a challenge with imbalanced data, i.e., most code lines are non-vulnerable, creating an uneven distribution in our line-level dataset.
For this reason, we chose \textit{F1-score} as our primary metric.

\begin{table}[htbp]
    \caption{Performance of PLMs and LLMs for line-level multilingual vulnerability detection}
    \label{tab:study_rq2_line}
    \centering
    \tabcolsep=3.0mm
    \small
    \begin{adjustbox}{max width=1.0\textwidth, center}
    \begin{tabular}{lrrrrrrrr}
        \toprule
        \toprule
        \textbf{Techniques} & \textbf{Accuracy} & \textbf{Recall} & \textbf{Precision} & \textbf{F1-score} & \textbf{FPR} & \textbf{FNR} & \textbf{MCC} & \textbf{AUC} \\
        \midrule
        \rowcolor{gray!40}\multicolumn{9}{l}{\textbf{Dummy Classifiers}} \\ 
        \midrule
            ${DummyClf}_{vul}$ & 0.0267  & 1.0000 & 0.0267 & 0.0520 & 1.0000 & 0.0000 & 0.0000 & 0.5000\\
            ${DummyClf}_{clean}$ & 0.9733 & 0.0000 & 0.0000 & 0.0000 & 0.0000 & 1.0000 & 0.0000 & 0.5000\\
        \midrule
        \rowcolor{gray!40}\multicolumn{9}{l}{\textbf{PLMs}} \\ 
        \midrule
            CodeBERT       & 0.9203         & 0.8822 & 0.2353         & 0.3715   & 0.0787  & 0.1177 & 0.4316 & \textbf{0.9018}   \\
            UniXcoder      & 0.8880          & \textbf{0.8879}         & 0.1787         & 0.2975  &   0.1120  & \textbf{0.1121} & 0.3688 & 0.8880 \\
            CodeT5         & 0.9472         & 0.8119        & 0.3123       & 0.4511  &   0.0491 & 0.1881  & 0.4838 & 0.8814   \\
            CodeT5P        & 0.9563         & 0.7673          & 0.3536 & 0.4841 & 0.0385  &  0.2327 & 0.5029 & 0.8644 \\
        \midrule
        \rowcolor{gray!40}\multicolumn{9}{l}{\textbf{LLMs (Zero-Shot Prompting)}} \\ 
        \midrule
            DeepSeek-Coder & 0.9536          & 0.3002          & 0.2246          & 0.2570    & 0.0284 & 0.6997 & 0.2362 & 0.6359    \\
            Code Llama     & \textbf{0.9589} & 0.2237          & 0.2272          & 0.2254  & 0.0208  & 0.7763 & 0.2043 & 0.6014     \\
            Llama 3        & 0.9307          & 0.4847          & 0.1891          & 0.2720 & 0.0570 & 0.5153  & 0.2730 & 0.7138       \\
            GPT-3.5-Turbo  & 0.9557          & 0.3215          & 0.2475          & 0.2797   & 0.0268 & 0.6785 & 0.2596 & 0.6473   \\
            GPT-4o         & 0.9537          & 0.4286          & 0.2700          & 0.3313  & 0.0318  & 0.5714 & 0.3175 & 0.6984     \\
        \midrule
        \rowcolor{gray!40}\multicolumn{9}{l}{\textbf{LLMs (Few-Shot Prompting)}} \\ 
        \midrule
            DeepSeek-Coder                          & 0.9521 & 0.3493 & 0.2344 & 0.2805 & 0.0313 & 0.6507 & 0.2622 & 0.6590 \\
            Code Llama                              & 0.9666 & 0.3018 & 0.3536 & 0.3257 & 0.0151 &  0.6982 & 0.3097 & 0.6433\\
            Llama 3                                 & 0.8980 & 0.7509 & 0.1738 & 0.2822 & 0.0980 & 0.2491 & 0.3295 & 0.8265 \\
            GPT-3.5-Turbo                           & 0.9774 & 0.4372 & 0.6057 & 0.5078 & 0.0078 & 0.5628 & 0.5035 & 0.7147 \\
            GPT-4o                                  & 0.9802 & 0.5689 & 0.6461 & 0.6050 & 0.0086 & 0.4311 & 0.5962 & 0.7802\\
        \midrule
        \rowcolor{gray!40}\multicolumn{9}{l}{\textbf{LLMs (Instruction Tuning + Zero-Shot Prompting)}} \\ 
        \midrule
            DeepSeek-Coder                          & 0.9404 & 0.2155 & 0.1297 & 0.1619 & 0.0397 & 0.7845 & 0.1376 & 0.5879 \\
            Code Llama                              & 0.8771 & 0.5902 & 0.1234 & 0.2042 & 0.1150 & 0.4098 & 0.2296 & 0.7376 \\
            Llama 3                                 & 0.9415 & 0.3403 & 0.1819 & 0.2371  & 0.0420 & 0.6597 & 0.2207 & 0.6491\\
            GPT-3.5-Turbo                           & 0.9578 & 0.5186 & 0.3211 & 0.3966 & 0.0301 & 0.4814 & 0.3877 & 0.7443 \\
            GPT-4o                                  & 0.9729 & 0.4241 & 0.4910 & 0.4551 & 0.0121 & 0.5759 & 0.4425 & 0.7060 \\
        \midrule
        \rowcolor{gray!40}\multicolumn{9}{l}{\textbf{LLMs (Instruction Tuning + Few-Shot Prompting)}} \\ 
        \midrule
            DeepSeek-Coder                          & 0.7149          & 0.2851 & 0.0278          & 0.0507 &  0.2733 & 0.7149 & 0.0042 & 0.5059     \\
            Code Llama                              & 0.8688          & 0.6221 & 0.1207          & 0.2022   & 0.1243   & 0.3779 & 0.2329 & 0.7488   \\
            Llama 3                                 & 0.9220          & 0.4720 & 0.1647          & 0.2442     & 0.0657  & 0.5280 & 0.2464 & 0.7031  \\
            GPT-3.5-Turbo                           & 0.9716          & 0.5804 & 0.4735          & 0.5215  & 0.0177 & 0.4196 & 0.5098 & 0.7813       \\
            GPT-4o                                  & 0.9830 & 0.6307 & \textbf{0.7012} & \textbf{0.6641} & \textbf{0.0074} & 0.3693 & \textbf{0.6563} & 0.8116 \\
        \bottomrule
        \bottomrule
    \end{tabular}
    \end{adjustbox}
\end{table}

\begin{figure}[t!]
    \centering
    \subfloat[\small F1-score of PLMs and LLMs (zero-shot prompting) in detecting line-level vulnerability across seven language]{
        \includegraphics[width=.7\linewidth]{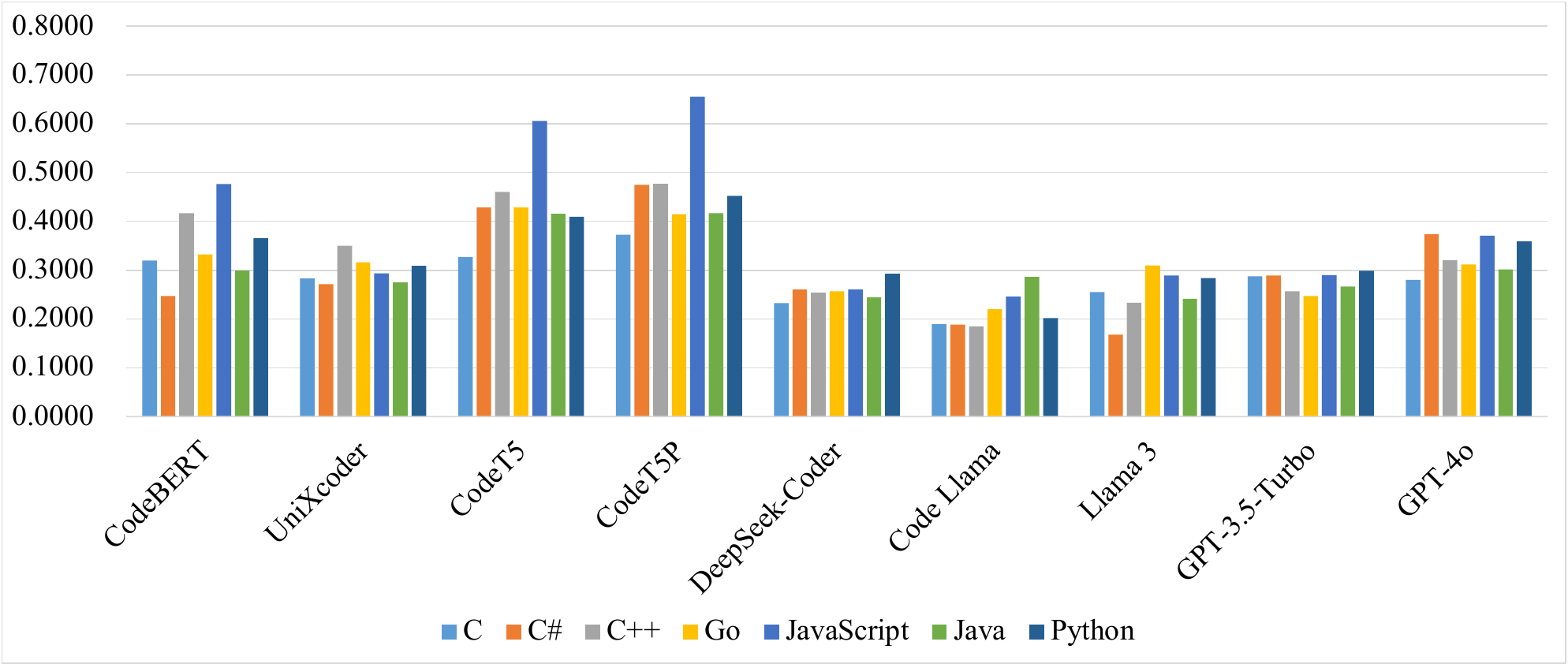}
        \label{fig:rq2_line_plm_lang}
    }\\
    \subfloat[\small F1-score of LLMs (instruction tuning with few-shot prompting) in detecting line-level vulnerability across seven language]{
        \includegraphics[width=.7\linewidth]{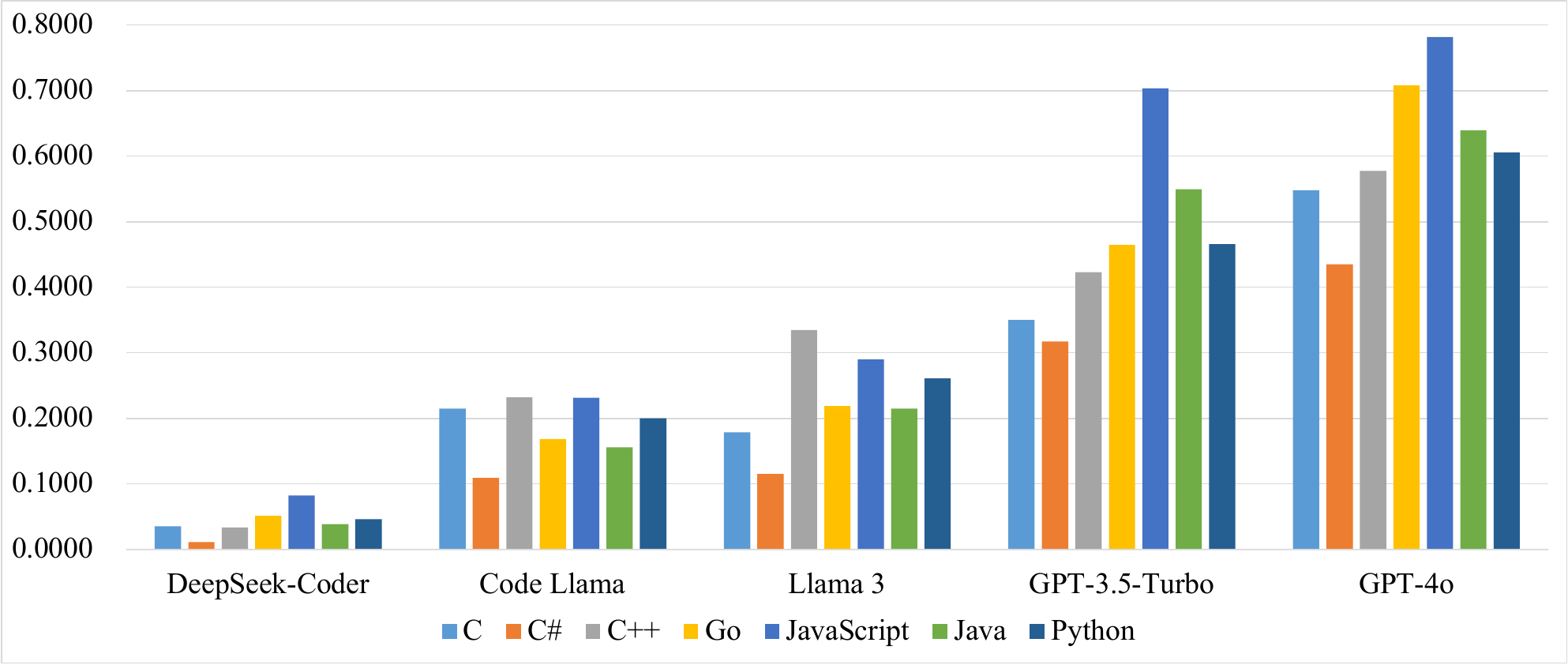}
        \label{fig:rq2_line_llm_lang}
    }
    \caption{Performance of PLMs and LLMs across seven programming languages in line-level multilingual vulnerability detection (y-axis: F1-score)}
    \label{fig:rq2_line_lang}
\end{figure}

\noindent
\textbf{\emph{\underline{Results.}}} 
Table~\ref{tab:study_rq2_line} shows the comparison results among PLMs and LLMs in terms of average Accuracy, Recall, Precision, F1-score, FPR, FNR, MCC, and AUC in detecting line-level multilingual vulnerability.
Within each metric, the values in bold indicate the model that exhibits the best performance among all PLMs and LLMs.
Figure~\ref{fig:rq2_line_lang} presents the performance of PLMs, LLMs with zero-shot prompting, and LLMs with instruction tuning and few-shot prompting in seven programming languages in terms of F1-score.

\textbf{Observation 5: CodeT5P is the best-performing model in identifying line-level multilingual vulnerability among PLMs.}
Among the evaluated PLMs, CodeT5P demonstrates the strongest overall performance in detecting line-level vulnerabilities, achieving the highest F1-score of 0.4841. 
This comparatively high F1-score primarily stems from its balanced approach, reflected in a moderate Recall of 0.7673 and Precision of 0.3536.
Furthermore, CodeT5P achieves the highest MCC (0.5029) in this category, indicating a superior correlation between predictions and ground truth compared to its peers. Interestingly, while CodeT5P leads in F1 and MCC, CodeBERT maintains the highest AUC (0.9018), suggesting a high potential for class discrimination despite its lower precision.
Additionally, CodeT5P balanced the trade-off between predicting false positives and false negatives, achieving the lowest FPR (0.0385) among studied PLMs.
Although UniXCoder achieves higher Recall (0.8879), indicating sensitivity in identifying vulnerable lines, its significantly lower Precision (0.1787) and higher FPR (0.1120) severely impacts its overall performance, yielding a much lower F1-score of 0.2975.
This is also reflected in its lower MCC of 0.3688 compared to CodeT5P.
CodeBert exhibits the modest performance among PLMs, with an F1-score of 0.3715, despite having relatively high Recall (0.8822). 
Thus, among traditional PLMs, CodeT5P emerges as the most viable choice for line-level vulnerability detection, effectively balancing FPR and FNR to mitigate the challenges of the imbalanced dataset.

\textbf{Observation 6: While using prompting alone, few-shot prompting can improve the effectiveness of LLMs in detecting line-level multilingual vulnerability.}
Under zero-shot prompting, all evaluated LLMs show limited practical effectiveness despite achieving notably high Accuracy scores. 
Code Llama attains the highest Accuracy (0.9589); however, it exhibits an extremely low F1-score (0.2254), reflecting significant weaknesses in Precision (0.2272), Recall (0.2237), and FNR (0.7763). 
The poor discriminative ability is further highlighted by its low MCC (0.2043) and AUC (0.6014).
Similarly, models like GPT-4o and GPT-3.5-Turbo display relatively high Accuracies (0.9537 and 0.9557, respectively) but similarly low F1-scores (0.3313 and 0.2797) and high FNR (0.5714 and 0.6785). 
Their zero-shot MCC values (0.3175 and 0.2596, respectively) and AUC scores (0.6984 and 0.6473) underscore the difficulty these models face in isolating vulnerable lines without specific guidance.
The consistently low Recall/Precision and high FNR across zero-shot prompting indicate a clear inability of LLMs to effectively detect vulnerable lines without further guidance or contextual clues, underscoring the insufficiency of using LLMs in isolation under zero-shot conditions for multilingual line-level vulnerability detection.

Few-shot prompting improves LLMs' performance through in-context examples. 
GPT-4o demonstrates the best performance, surpassing all other PLMs with a significantly higher F1-score of 0.6050, supported by strong Precision (0.6461) and moderate Recall (0.5689).
This improvement is validated by a substantial jump in MCC to 0.5962 and an AUC of 0.7802.
Compared to the zero-shot prompting setting, GPT-4o's FPR significantly decreases from 0.0318 to 0.0086, while its FNR also reduces from 0.5714 to 0.4311.
GPT-3.5-Turbo also shows notable improvement, with its F1-score rising to 0.5078. This improvement is primarily driven by enhanced Precision (0.6057), although Recall remains limited (0.4372).
Additionally, its MCC increases to 0.5035 and its AUC reaches 0.7147 under the few-shot setting.
These results demonstrate that few-shot prompting alone, without instruction tuning, can partially address the issue of imbalanced labels (where vulnerable lines are fewer than non-vulnerable lines) to improve LLMs' performance in detecting multilingual vulnerabilities at the line-level.

\textbf{Observation 7: Instruction tuning is also beneficial for improving LLMs in detecting multilingual vulnerabilities at the line-level.}
Finally, the addition of instruction tuning combined with few-shot prompting significantly enhances LLM performance, clearly outperforming all other prompting strategies. 
GPT-4o, in particular, achieves the highest overall performance with an impressive F1-score of 0.6641. 
This is further supported by the highest MCC (0.6563) and a strong AUC of 0.8116 among all tested LLM configurations.
This high F1-score results from a robust Precision of 0.7012 and substantial Recall of 0.6307, effectively handling the challenge posed by dataset imbalance.
Additionally, GPT-4o reaches the lowest FPR of 0.0074 and the lowest FNR of 0.3693 among the studied LLMs (except for Llama 3 with few-shot prompting). This indicates an excellent balance between correctly identifying vulnerable and clean lines.
GPT-3.5-Turbo also shows notable improvements under this condition, achieving an F1-score of 0.5215, along with an MCC of 0.5098 and an AUC of 0.7813, demonstrating the broader positive impact of instruction tuning combined with few-shot prompting.
While using instruction tuning with few-shot prompting, GPT-4o and GPT-3.5-Turbo significantly outperform the dummy classifier in F1-score and Precision. This demonstrates that their predictions are substantially more accurate than random guesses for line-level detection.

When comparing GPT-4o with the best-performing PLM, CodeT5P (F1-score: 0.4841), GPT-4o's superiority becomes even more evident. 
GPT-4o notably surpasses CodeT5P not only in F1-score (0.6641 vs. 0.4841) but also significantly in Precision (0.7012 vs. 0.3536). 
The superiority of GPT-4o is also reflected in the MCC comparison (0.6563 for GPT-4o vs. 0.5029 for CodeT5P). However, it is worth noting that CodeT5P maintains a higher overall AUC (0.8644) than GPT-4o (0.8116), indicating that while GPT-4o is more effective at the current threshold, the PLM may possess broader discriminative potential across different classification thresholds.
Although CodeT5P achieves slightly higher Recall (0.7673 vs. GPT-4o's 0.6307), its lower Precision substantially limits its practical utility, as it generates many more false-positive predictions. 
Supporting evidence comes from the FPR and FNR metrics. GPT-4o achieves a much lower FPR (0.0074 vs. CodeT5P's 0.0385), though it has a higher FNR (0.3693 vs. CodeT5P's 0.2327). 

In terms of F1-score on each programming language, GPT-4o from Figure~\ref{fig:rq2_line_lang} demonstrates a clear superiority in detecting line-level vulnerabilities. 
GPT-4o consistently achieves higher F1-scores across all languages, particularly excelling in Go and JavaScript, whereas CodeT5P shows more moderate performance with notable gaps in languages like C, Go, and Java.
Specifically, GPT-4o demonstrates the best performance with F1-scores of 0.5478 in C, 0.4348 in C\#, 0.5774 in C++, 0.7078 in Go, 0.7815 in JavaScript, 0.6391 in Java, and 0.6055 in Python.
These results are consistently superior to CodeT5P (except in C\#), which achieved F1-scores of 0.3727 in C, 0.4750 in C\#, 0.4768 in C++, 0.4146 in Go, 0.6556 in JavaScript, 0.4167 in Java, and 0.4519 in Python.
Consequently, these results indicate that GPT-4o’s instruction tuning combined with few-shot prompting substantially enhances its ability to generalize across diverse programming languages for line-level vulnerability detection, significantly surpassing the capability of traditional PLMs like CodeT5P.

\begin{tcolorbox}\textbf{RQ2 Summary:}
In a comparison of LLMs and PLMs, GPT-4o with instruction tuning and few-shot prompting emerged as the top performer, achieving an F1-score of 0.6641 across seven programming languages, followed by CodeT5P at 0.4841.
Additionally, GPT-4o demonstrated superior performance across all languages, with F1-score improvements ranging from 9.77\% to 310.19\% compared to other PLMs and LLMs.
In evaluating performance across multiple programming languages using instruction tuning and few-shot prompting, GPT-4o achieves the highest F1-score of 0.7815 with JavaScript and the lowest F1-score of 0.4348 with C\#.
\end{tcolorbox}

\subsection{RQ3: What are the strengths and weaknesses of the PLMs and LLMs in multilingual vulnerability detection?}
\label{subsec:RQ3}
\noindent
\textbf{\emph{\underline{Approach.}}} 
This research question aims to deepen our understanding of the performance of various PLMs and LLMs with different strategies in detecting multilingual vulnerability at the function-level and line-level.
According to RQ1 and RQ2, we selected the best-performing PLMs and LLMs with various strategies as our representative models.
In our function-level multilingual vulnerability detection analysis, we selected models based on Accuracy metric. We chose CodeT5P for PLMs and evaluated four LLM variants: Llama 3 using zero-shot prompting (ZSP), and three GPT-4o configurations - few-shot prompting (FSP), instruction tuning with zero-shot prompting (ITZSP), and instruction tuning with few-shot prompting (ITFSP).
For line-level multilingual vulnerability detection, we based our model selection on F1-score performance. We maintained CodeT5P as our PLM representative and utilized the four GPT-4o variants (ZSP, FSP, ITZSP, and ITFSP) for LLM representatives.
Our orthogonality analysis consists of the following three perspectives:
\begin{itemize}
    \item \textbf{The unique correct and incorrect detection. } 
    To investigate the complementarity of PLMs and LLMs, we performed a set-membership analysis using Venn diagrams. We categorized the results based on the instance-level alignment between model predictions and ground-truth labels. Specifically, we constructed sets of Correct Detections (True Positives and True Negatives) and Incorrect Detections (False Positives and False Negatives) for each model. This approach allows us to quantify the orthogonality of the models—identifying whether LLMs succeed on the specific code instances where PLMs fail, and vice-versa. The Venn diagrams thus represent the distribution of unique and shared performance over the total population of test instances, rather than semantic code features.
    \item \textbf{The tendency in predicting the Top-25 dangerous CWE-IDs. }
    Since certain PLMs and LLMs tend to be more effective at detecting specific CWE types, we further evaluated the accuracy of detecting specific CWE types at both function-level and line-level.
    We used test data from the 2023 CWE Top 25 Most Dangerous Software Weaknesses, published by the CWE community.\footnote{\url{https://cwe.mitre.org/top25/archive/2023/2023_top25_list.html}} 
    Note that our test data encompasses all of the top 25 CWE-IDs.
    \item \textbf{The effectiveness across the different vulnerability severity. }
    For the last perspective, PLMs and LLMs tend to detect different vulnerability severity, which can affect the effectiveness of models in real-world scenarios. 
    Hence, we evaluated the number of correct detections at both function- and line-levels using test data categorized by CVSS v4.0 ratings.
\end{itemize}

\begin{figure*}[!t]
    \centering
    \includegraphics[width=.9\linewidth]{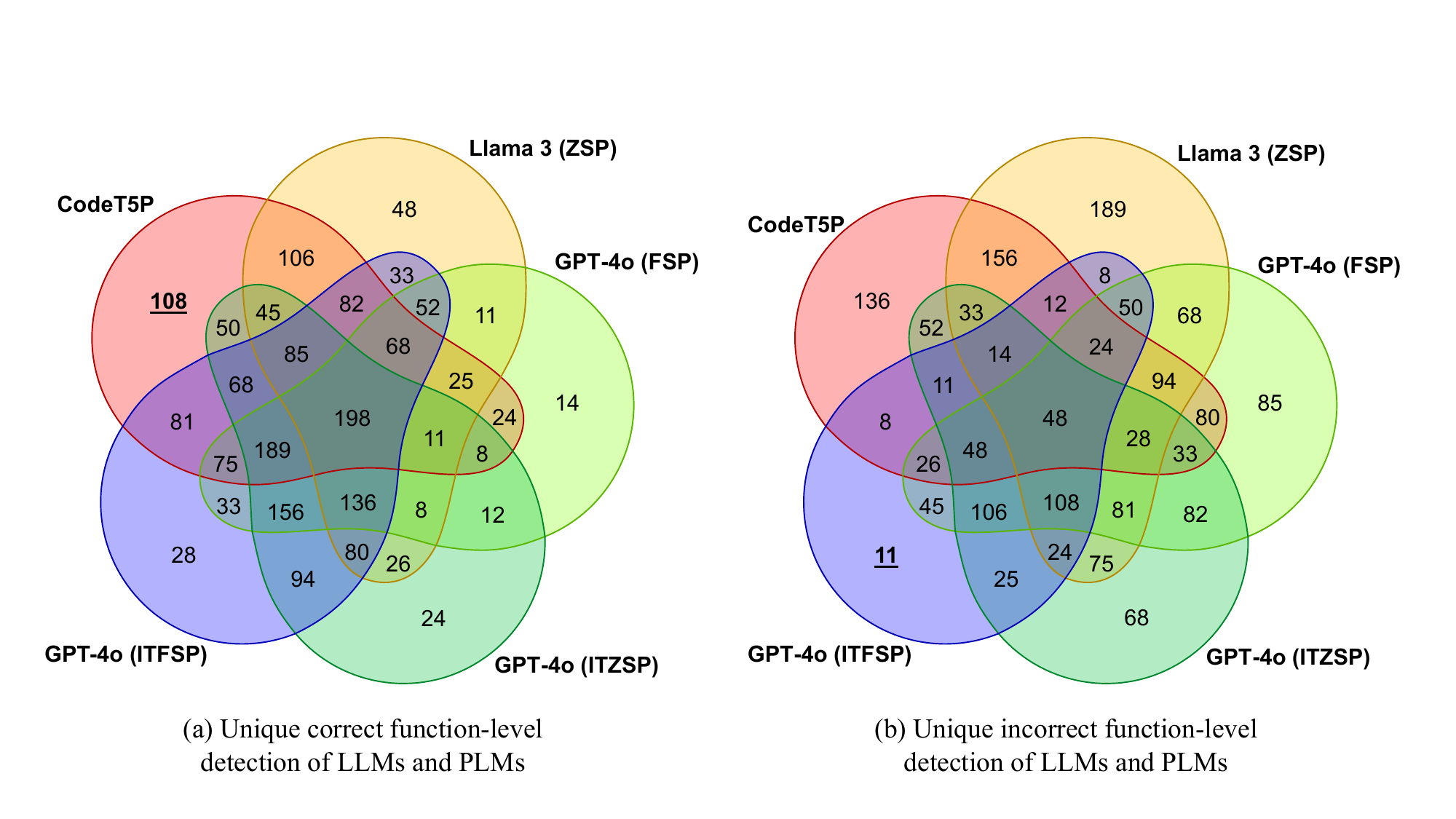}
    \caption{The unique correct/incorrect detection of function-level multilingual vulnerability in PLMs and LLMs}
    \label{fig:rq3_func_venn}
\end{figure*}

\begin{figure*}[!t]
    \centering
    \includegraphics[width=.9\linewidth]{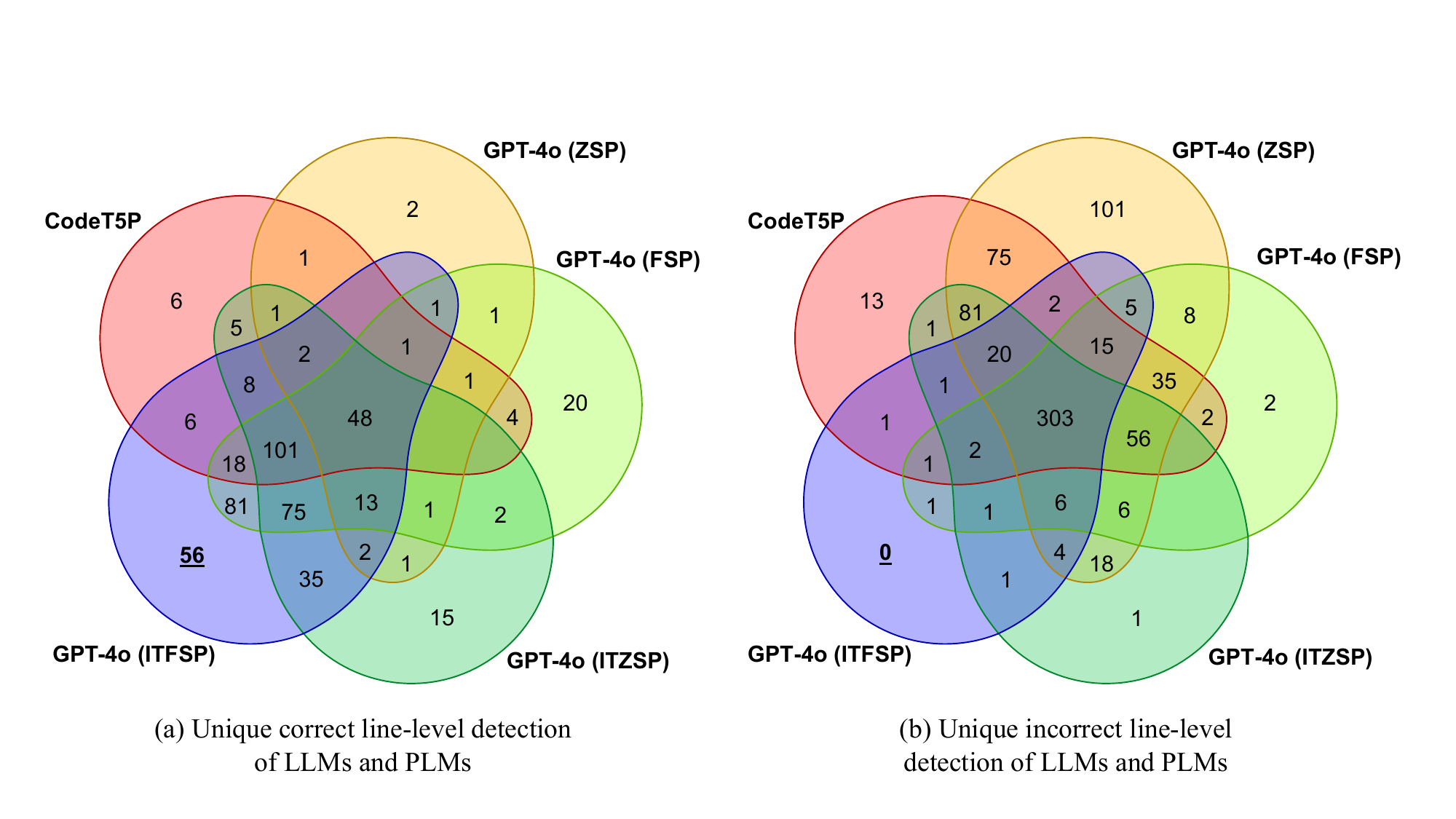}
    \caption{The unique correct/incorrect detection of line-level multilingual vulnerability in PLMs and LLMs}
    \label{fig:rq3_line_venn}
\end{figure*}

\begin{figure*}[!t]
    \centering
    \includegraphics[width=.7\linewidth]{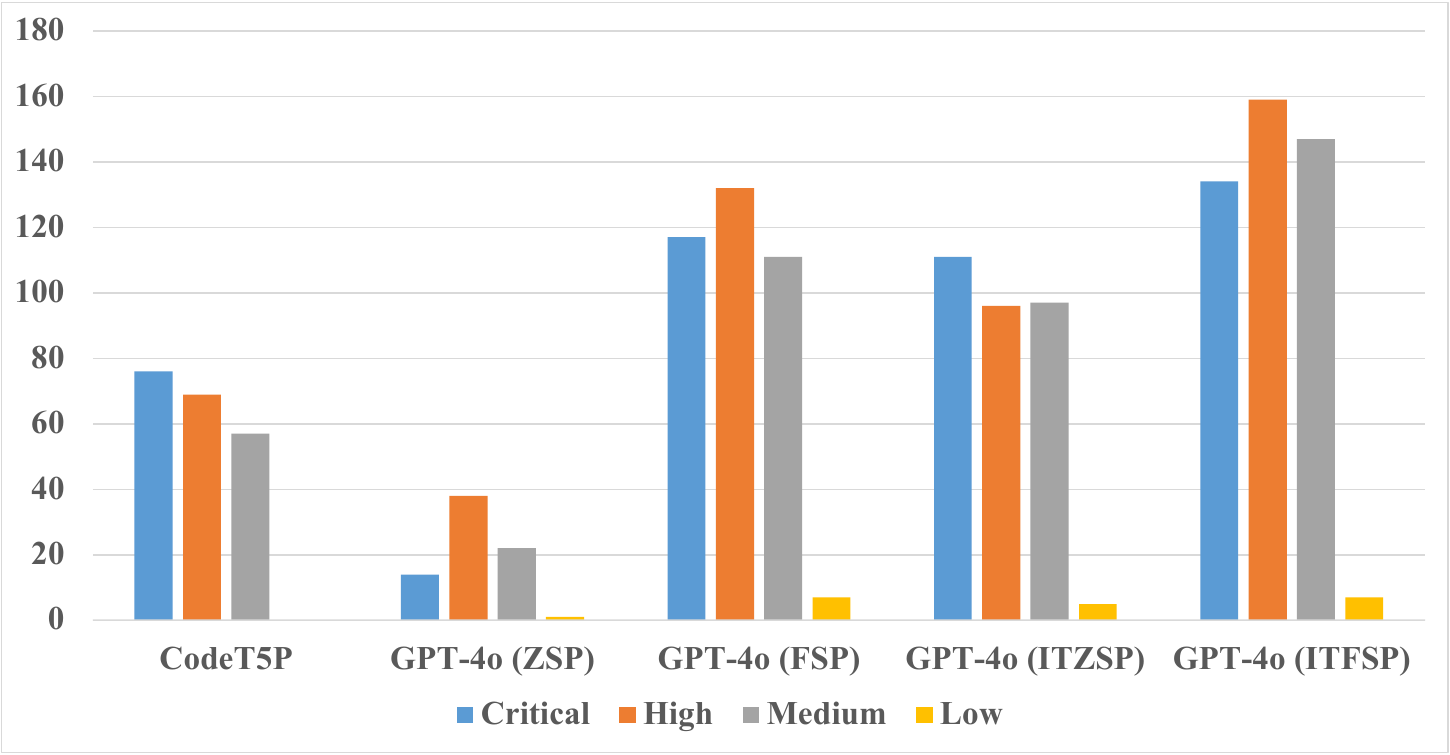}
    \caption{The number of correct detections based on CVSS severity at the function-level in PLMs and LLMs}
    \label{fig:rq3_func_cvss}
\end{figure*}

\begin{figure*}[!t]
    \centering
    \includegraphics[width=.7\linewidth]{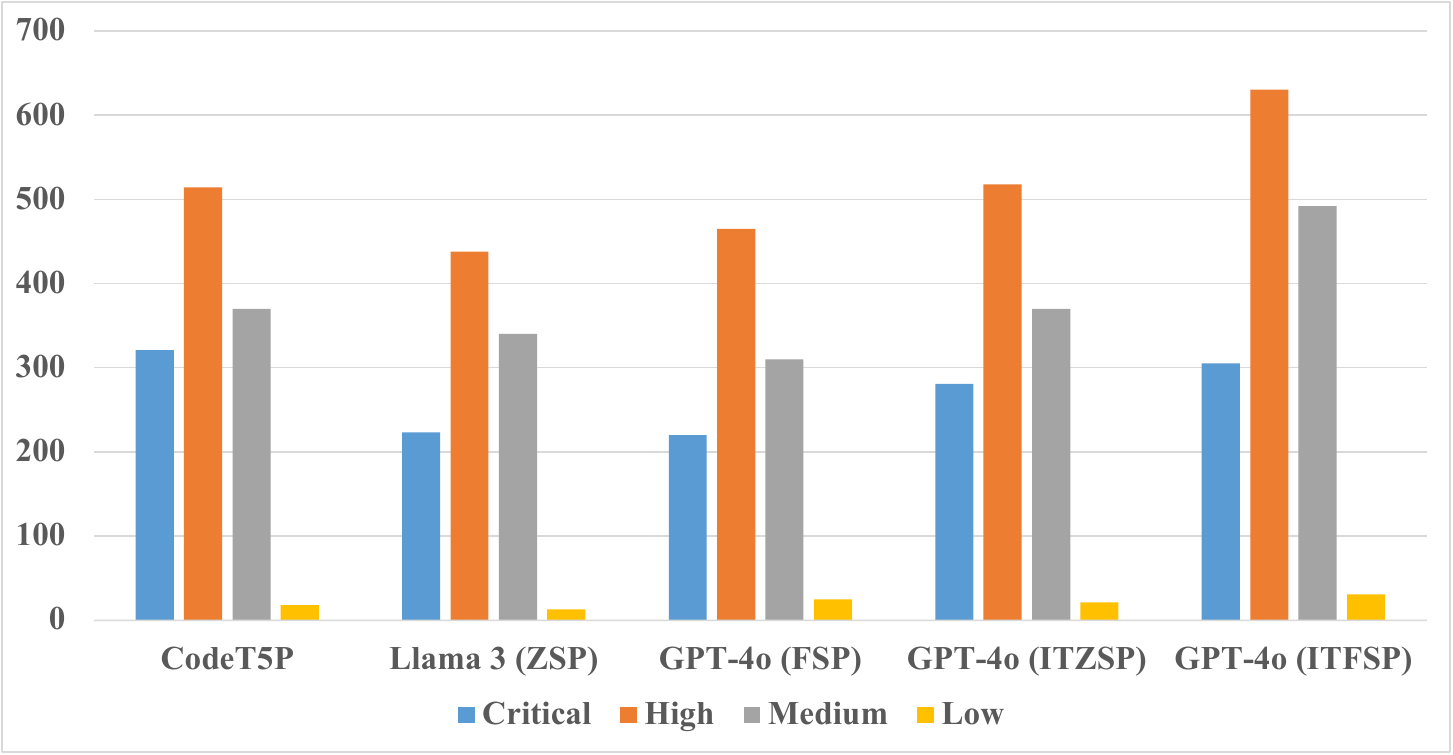}
    \caption{The number of correct detections based on CVSS severity at the line-level in PLMs and LLMs}
    \label{fig:rq3_line_cvss}
\end{figure*}

\noindent
\textbf{\emph{\underline{Results.}}}
Figures~\ref{fig:rq3_func_venn} and~\ref{fig:rq3_line_venn} show Venn diagrams depicting the intersection of correct and incorrect detections at both function-level and line-level among the studied PLMs and LLMs.
The overlapping regions indicate shared correct or incorrect detections, while the non-overlapping areas show detections unique correct or incorrect predictions to each model.
Figures~\ref{fig:rq3_func_cvss} and~\ref{fig:rq3_line_cvss} illustrate the performance comparison of multilingual vulnerability detection at both function- and line-level, with results categorized by vulnerability severity.
Tables~\ref{tab:func_top25} and~\ref{tab:line_top25} further present the detection performance of studied PLMs and LLMs for the top 25 most dangerous CWE-IDs in 2023 at function-level and line-level.
The values highlighted in bold denote the model or strategy that exhibits the best performance for the corresponding CWE-ID.

\textbf{Observation 8: In terms of unique correct and incorrect detection, GPT-4o with ITFSP outperforms the other studied models, achieving better results in both function-level detection (with fewer incorrect detections) and line-level detection (with more unique correct detections and zero incorrect ones).}
For function-level detection, Figure ~\ref{fig:rq3_func_venn}a demonstrates that CodeT5P achieves superior performance in unique correct detections compared to LLMs using the four studied strategies. CodeT5P identifies 108 unique correct detections, substantially outperforming Llama 3 with ZSP (48), GPT-4o with FSP (14), GPT-4o with ITZSP (24), and GPT-4o with ITFSP (28).
However, Figure ~\ref{fig:rq3_func_venn}b shows that GPT-4o with ITFSP has fewer unique incorrect detections at the function-level, only 11 compared to CodeT5P (136), Llama 3 with ZSP (189), GPT-4o with FSP (85), and GPT-4o with ITZSP (68).
For line-level detection, Figure ~\ref{fig:rq3_line_venn}a shows that GPT-4o with ITFSP outperforms all studied PLMs and LLMs, with 56 unique correct detections compared to CodeT5P (6), GPT-4o with ZSP (2), GPT-4o with FSP (20), and GPT-4o with ITZSP (15).
Figure~\ref{fig:rq3_line_venn}b demonstrates that GPT-4o with ITFSP achieves the best performance in unique incorrect detections.
GPT-4o with ITFSP has only 0 unique incorrect detections, substantially outperforming CodeT5P (13), GPT-4o with ZSP (101), GPT-4o with FSP (2), and GPT-4o with ITZSP (1).
These results further demonstrate the effectiveness of GPT-4o with ITFSP in both function- and line-level multilingual vulnerability detection, reinforcing our findings from RQ1 and RQ2.

\begin{table}[t!]
    \caption{The percentage of correct detection for function-level multilingual vulnerability detection using two representative techniques across the Top 25 most dangerous CWE-IDs in 2023}
    \label{tab:func_top25}
    \centering
    \begin{adjustbox}{max width=1.0\textwidth, center}
    \begin{threeparttable}
    \begin{tabular}{ccrrrrrc}
    \toprule
    \toprule
\textbf{Ranking} & \textbf{CWE-ID} & \multicolumn{1}{c}{\textbf{Llama 3 (ZSP)}} & \multicolumn{1}{c}{\textbf{GPT-4o (FSP)}} & \multicolumn{1}{c}{\textbf{GPT-4o (ITZSP)}} & \multicolumn{1}{c}{\textbf{GPT-4o (ITFSP)}} & \multicolumn{1}{c}{\textbf{CodeT5P}} & \textbf{Total} \\
\midrule
1 & CWE-787 (Out-of-bounds Write) & 19(46.34\%) & 21(51.22\%) & 22(53.66\%) & \textbf{34(82.93\%)} & 23(56.1\%) & 41 \\
2 & CWE-79 (Cross-site Scripting) & 97(48.99\%) & 115(58.08\%) & 126(63.64\%) & \textbf{155(78.28\%)} & 125(63.13\%) & 198 \\
3 & CWE-89 (SQL Injection) & 21(45.65\%) & 33(71.74\%) & 33(71.74\%) & \textbf{36(78.26\%)} & 27(58.7\%) & 46 \\
4 & CWE-416 (Use After Free) & 20(58.82\%) & 17(50.0\%) & 17(50.0\%) & \textbf{26(76.47\%)} & 19(55.88\%) & 34 \\
5 & CWE-78 (OS Command Injection) & 5(33.33\%) & 8(53.33\%) & 10(66.67\%) & \textbf{12(80.0\%)} & 5(33.33\%) & 15 \\
6 & CWE-20 (Improper Input Validation) & 43(46.74\%) & 46(50.0\%) & 47(51.09\%) & \textbf{62(67.39\%)} & 52(56.52\%) & 92 \\
7 & CWE-125 (Out-of-bounds Read) & 28(54.9\%) & 21(41.18\%) & 24(47.06\%) & \textbf{35(68.63\%)} & 31(60.78\%) & 51 \\
8 & CWE-22 (Path Traversal) & 45(59.21\%) & 43(56.58\%) & 56(73.68\%) & \textbf{60(78.95\%)} & 43(56.58\%) & 76 \\
9 & CWE-352 (Cross-Site Request Forgery) & 42(52.5\%) & 56(70.0\%) & 56(70.0\%) & \textbf{62(77.5\%)} & \textbf{62(77.5\%)} & 80 \\
10 & CWE-434 (Unrestricted Upload of File with Dangerous Type) & 2(40.0\%) & \textbf{3(60.0\%)} & \textbf{3(60.0\%)} & \textbf{3(60.0\%)} & \textbf{3(60.0\%)} & 5 \\
11 & CWE-862 (Missing Authorization) & \textbf{1(100.0\%)} & 0(0.0\%) & 0(0.0\%) & \textbf{1(100.0\%)} & \textbf{1(100.0\%)} & 1 \\
12 & CWE-476 (NULL Pointer Dereference) & 19(42.22\%) & 28(62.22\%) & 23(51.11\%) & \textbf{34(75.56\%)} & 19(42.22\%) & 45 \\
13 & CWE-287 (Improper Authentication) & 64(45.71\%) & 82(58.57\%) & 105(75.0\%) & 108(77.14\%) & \textbf{110(78.57\%)} & 140 \\
14 & CWE-190 (Integer Overflow or Wraparound) & 17(53.12\%) & 18(56.25\%) & 18(56.25\%) & \textbf{27(84.38\%)} & 18(56.25\%) & 32 \\
15 & CWE-502 (Deserialization of Untrusted Data) & 18(62.07\%) & 17(58.62\%) & 16(55.17\%) & \textbf{21(72.41\%)} & 18(62.07\%) & 29 \\
16 & CWE-77 (Command Injection) & 13(50.0\%) & 15(57.69\%) & \textbf{19(73.08\%)} & 18(69.23\%) & 14(53.85\%) & 26 \\
17 & CWE-119 (Improper Restriction of Operations within the Bounds of a Memory Buffer) & 27(43.55\%) & 28(45.16\%) & 33(53.23\%) & \textbf{49(79.03\%)} & 30(48.39\%) & 62 \\
18 & CWE-798 (Use of Hard-coded Credentials) & \textbf{2(66.67\%)} & 1(33.33\%) & 1(33.33\%) & \textbf{2(66.67\%)} & 1(33.33\%) & 3 \\
19 & CWE-918 (Server-Side Request Forgery) & 13(39.39\%) & 13(39.39\%) & 16(48.48\%) & \textbf{22(66.67\%)} & 20(60.61\%) & 33 \\
20 & CWE-306 (Missing Authentication for Critical Function) & 9(50.0\%) & \textbf{15(83.33\%)} & 13(72.22\%) & \textbf{15(83.33\%)} & 9(50.0\%) & 18 \\
21 & CWE-362 (Race Condition) & 7(53.85\%) & 6(46.15\%) & 7(53.85\%) & \textbf{10(76.92\%)} & 5(38.46\%) & 13 \\
22 & CWE-269 (Improper Privilege Management) & 7(63.64\%) & 1(9.09\%) & 3(27.27\%) & \textbf{8(72.73\%)} & \textbf{8(72.73\%)} & 11 \\
23 & CWE-94 (Code Injection) & 11(45.83\%) & 13(54.17\%) & 15(62.5\%) & 17(70.83\%) & \textbf{18(75.0\%)} & 24 \\
24 & CWE-863 (Incorrect Authorization) & 28(65.12\%) & 18(41.86\%) & 24(55.81\%) & \textbf{32(74.42\%)} & 22(51.16\%) & 43 \\
25 & CWE-276 (Incorrect Default Permissions) & \textbf{1(100.0\%)} & 1(100.0\%) & 0(0.0\%) & 0(0.0\%) & \textbf{1(100.0\%)} & 1 \\
\midrule
& Mean & 559(49.96\%) & 619(55.32\%) & 687(61.39\%) & \textbf{849(75.87\%)} & 684(61.13\%) & 1,119\\
\bottomrule
\bottomrule
\end{tabular}
 \begin{tablenotes}
        \small
        \item[*] The numbers in parentheses represent the percentage of correct prediction of the corresponding techniques, while the "Total" column represents the total amount of data corresponding to CWE-ID in the test dataset of this paper
    \end{tablenotes}
    \end{threeparttable}
    \end{adjustbox}
\end{table}

\begin{table}[t!]
    \caption{The percentage of correct detection for line-level multilingual vulnerability detection using two representative techniques across the Top 25 most dangerous CWE-IDs in 2023}
    \label{tab:line_top25}
    \centering
    \begin{adjustbox}{max width=1.0\textwidth, center}
    \begin{threeparttable}
\begin{tabular}{ccrrrrrc}
\toprule
\toprule
\textbf{Ranking} & \textbf{CWE-ID} & \multicolumn{1}{c}{\textbf{GPT-4o (ZSP)}} & \multicolumn{1}{c}{\textbf{GPT-4o (FSP)}} & \multicolumn{1}{c}{\textbf{GPT-4o (ITZSP)}} & \multicolumn{1}{c}{\textbf{GPT-4o (ITFSP)}} & \multicolumn{1}{c}{\textbf{CodeT5P}} & \textbf{Total} \\
\midrule
1 & CWE-787 (Out-of-bounds Write) & 0(0.0\%) & 0(0.0\%) & \textbf{1(6.25\%)} & \textbf{1(6.25\%)} & 0(0.0\%) & 16 \\
2 & CWE-79 (Cross-site Scripting) & 9(10.59\%) & 46(54.12\%) & 44(51.76\%) & \textbf{57(67.06\%)} & 26(30.59\%) & 85 \\
3 & CWE-89 (SQL Injection) & 5(27.78\%) & 5(27.78\%) & \textbf{9(50.0\%)} & \textbf{9(50.0\%)} & 5(27.78\%) & 18 \\
4 & CWE-416 (Use After Free) & 1(6.25\%) & 5(31.25\%) & 5(31.25\%) & \textbf{7(43.75\%)} & 2(12.5\%) & 16 \\
5 & CWE-78 (OS Command Injection) & 1(20.0\%) & 3(60.0\%) & 1(20.0\%) & \textbf{4(80.0\%)} & 2(40.0\%) & 5 \\
6 & CWE-20 (Improper Input Validation) & 2(4.88\%) & 16(39.02\%) & 10(24.39\%) & \textbf{17(41.46\%)} & 3(7.32\%) & 41 \\
7 & CWE-125 (Out-of-bounds Read) & 0(0.0\%) & 1(4.55\%) & 3(13.64\%) & \textbf{6(27.27\%)} & 1(4.55\%) & 22 \\
8 & CWE-22 (Path Traversal) & 0(0.0\%) & \textbf{9(32.14\%)} & 4(14.29\%) & \textbf{9(32.14\%)} & 2(7.14\%) & 28 \\
9 & CWE-352 (Cross-Site Request Forgery) & 13(33.33\%) & 23(58.97\%) & 20(51.28\%) & \textbf{24(61.54\%)} & 17(43.59\%) & 39 \\
10 & CWE-434 (Unrestricted Upload of File with Dangerous Type) & 0(0.0\%) & 0(0.0\%) & 0(0.0\%) & 0(0.0\%) & 0(0.0\%) & 1 \\
11 & CWE-862 (Missing Authorization) & 0(0.0\%) & 0(0.0\%) & 0(0.0\%) & 0(0.0\%) & 0(0.0\%) & 1 \\
12 & CWE-476 (NULL Pointer Dereference) & 2(14.29\%) & 6(42.86\%) & 3(21.43\%) & \textbf{9(64.29\%)} & 1(7.14\%) & 14 \\
13 & CWE-287 (Improper Authentication) & 0(0.0\%) & 19(51.35\%) & 19(51.35\%) & \textbf{22(59.46\%)} & 14(37.84\%) & 37 \\
14 & CWE-190 (Integer Overflow or Wraparound) & 1(10.0\%) & 3(30.0\%) & 3(30.0\%) & \textbf{5(50.0\%)} & 2(20.0\%) & 10 \\
15 & CWE-502 (Deserialization of Untrusted Data) & 2(15.38\%) & 6(46.15\%) & 7(53.85\%) & \textbf{8(61.54\%)} & 3(23.08\%) & 13 \\
16 & CWE-77 (Command Injection) & 2(16.67\%) & 7(58.33\%) & 7(58.33\%) & \textbf{10(83.33\%)} & 4(33.33\%) & 12 \\
17 & CWE-119 (Improper Restriction of Operations within the Bounds of a Memory Buffer) & 0(0.0\%) & \textbf{5(21.74\%)} & 1(4.35\%) & \textbf{5(21.74\%)} & 1(4.35\%) & 23 \\
18 & CWE-798 (Use of Hard-coded Credentials) & 0(0.0\%) & \textbf{1(100.0\%)} & 0(0.0\%) & \textbf{1(100.0\%)} & 0(0.0\%) & 1 \\
19 & CWE-918 (Server-Side Request Forgery) & 2(14.29\%) & \textbf{6(42.86\%)} & 4(28.57\%) & \textbf{6(42.86\%)} & 3(21.43\%) & 14 \\
20 & CWE-306 (Missing Authentication for Critical Function) & 0(0.0\%) & 5(55.56\%) & 2(22.22\%) & \textbf{6(66.67\%)} & 4(44.44\%) & 9 \\
21 & CWE-362 (Race Condition) & 0(0.0\%) & 0(0.0\%) & 0(0.0\%) & \textbf{2(50.0\%)} & 1(25.0\%) & 4 \\
22 & CWE-269 (Improper Privilege Management) & 0(0.0\%) & 1(25.0\%) & 0(0.0\%) & \textbf{1(25.0\%)} & 0(0.0\%) & 4 \\
23 & CWE-94 (Code Injection) & 2(16.67\%) & \textbf{7(58.33\%)} & 6(50.0\%) & \textbf{7(58.33\%)} & 3(25.0\%) & 12 \\
24 & CWE-863 (Incorrect Authorization) & 2(13.33\%) & 5(33.33\%) & 4(26.67\%) & \textbf{7(46.67\%)} & 2(13.33\%) & 15 \\
25 & CWE-276 (Incorrect Default Permissions) & 0(0.0\%) & 0(0.0\%) & 0(0.0\%) & \textbf{1(100.0\%)} & 0(0.0\%) & 1 \\
\midrule
0 & Mean & 44(9.98\%) & 179(40.59\%) & 153(34.69\%) & \textbf{224(50.79\%)} & 96(21.77\%) & 441 \\
\bottomrule
\bottomrule
\end{tabular}
 \begin{tablenotes}
        \small
        \item[*] The numbers in parentheses represent the percentage of correct prediction of the corresponding techniques, while the "Total" column represents the total amount of data corresponding to the CWE-ID in the test dataset of this paper
    \end{tablenotes}
    \end{threeparttable}
    \end{adjustbox}
\end{table}

\textbf{Observation 9: Among the Top 25 most dangerous vulnerabilities, such as CWE-89 (SQL Injection), GPT-4o with ITFSP demonstrates superior detection capabilities in both function-level and line-level detection, achieving remarkable detection rates of 75.87\% and 50.79\%, respectively.}
Regarding function-level detection, Table ~\ref{tab:func_top25} shows that GPT-4o with ITFSP outperforms all studied PLMs and LLMs.
Specifically, GPT-4o with ITFSP successfully detects 75.87\% of vulnerable and clean functions associated with the top 25 most dangerous CWE-IDs in 2023, substantially outperforming Llama 3 with ZSP (49.96\%), GPT-4o with FSP (55.32\%), GPT-4o with ITZSP (61.39\%), and CodeT5P (61.13\%).
GPT-4o with ITFSP achieves the best performance on 23 out of 25 CWE-IDs, while the second-best model, CodeT5P, leads in only 7.
Regarding line-level detection, Table ~\ref{tab:line_top25} shows the same trend as function-level detection, where GPT-4o with ITFSP is superior to other studied PLMs and LLMs.
GPT-4o with ITFSP successfully detects 50.79\% of vulnerable and clean functions for the top 25 most dangerous CWE-IDs in 2023, significantly outperforming GPT-4o with ZSP (9.98\%), GPT-4o with FSP (40.59\%), GPT-4o with ITZSP (34.69\%), and CodeT5P (21.77\%). 
These findings highlight the effectiveness of GPT-4o with ITFSP in addressing dangerous vulnerabilities at both function-level and line-level, reinforcing the results from RQ1 and RQ2.

\textbf{Observation 10: Most vulnerabilities are relatively easy to detect at the function-level but challenging to detect at the line-level.}
To compare the overlap between vulnerabilities detected at the function-level and line-level, we consider a CWE detectable by the language models only if all 5 techniques in Table 5 and Table 6 can detect it. Otherwise, we consider the CWE undetectable by the language models. For example, CWE-276 (Incorrect Default Permissions) cannot be detected by the language models since the percentage of correct detection is 0 on both GPT-4o (ITZSP) and GPT-4o (ITFSP).
In this case, we found 13 CWEs that can be detected at both function-level and line-level: CWE-79, CWE-89, CWE-416, CWE-78, CWE-20, CWE-352, CWE-476, CWE-190, CWE-502, CWE-77, CWE-918, CWE-94, and CWE-863.
Additionally, 11 CWEs can be detected at function-level but not at line-level. For example, CWE-787, CWE-125, and CWE-22, three of the most dangerous CWEs, can be effectively detected at function-level but fail to be detected at line-level.
Finally, there is 1 exception (CWE-276) that cannot be detected at either function-level or line-level.
These results demonstrate that language models exhibit higher detection capabilities at the function-level than at the line-level, with some of the most critical vulnerabilities being identifiable only through function-level analysis.

\textbf{Observation 11: Across all CVSS severity levels, GPT-4o with ITFSP demonstrates superior capability in identifying multilingual vulnerabilities, showing exceptional performance at both function and line-level detection compared to other studied models.}
For function-level detection, Figure~\ref{fig:rq3_func_cvss} demonstrates that GPT-4o with ITFSP performs better at detecting High, Medium, and Low severity vulnerabilities. However, GPT-4o with ITFSP performs slightly worse than CodeT5P when detecting Critical severity issues.
For example, GPT-4o with ITFSP exhibits 630 detections of High severity vulnerability, surpassing 514 by CodeT5P, 438 by Llama 3 with ZSP, 465 by GPT-4o with FSP, and 518 by GPT-4o with ITZSP.
GPT-4o with ITFSP exhibits 305 detections of Critical severity vulnerability, which is slightly lower than 321 by CodeT5P.
For line-level detection, Figure~\ref{fig:rq3_line_cvss} shows that GPT-4o with ITFSP outperforms all studied PLMs and LLMs at detecting Critical, High, Medium, and Low severity vulnerabilities.
For example, GPT-4o with ITFSP exhibits 134 detections of Critical severity vulnerability, significantly surpassing 76 by CodeT5P, 14 by GPT-4o with (ZSP), 117 by GPT-4o with FSP, and 111 by GPT-4o with ITZSP.
These findings underscore that GPT-4o with ITFSP consistently delivers superior performance in correctly detecting vulnerabilities at both function-level and line-level granularity across different CVSS severity categories, reflecting the advantage of incorporating instruction tuning and few-shot prompting in enhancing LLMs.

\begin{tcolorbox}\textbf{RQ3 Summary:}
For both function-level and line-level analysis, GPT-4o, with instruction tuning and few-shot prompting, outperforms the PLM representative (CodeT5P) and other studied LLM representatives. This superiority is shown in unique correct/incorrect multilingual detection, vulnerability detection in the Top-25 dangerous CWE-IDs, and vulnerability detection across various severity levels.
\end{tcolorbox}

%% file: discussion.tex
\section{Discussion}
\label{sec:discussion}

\subsection{A Fine-grained Analysis of Code Structural Properties}

\begin{figure*}[thp!]
    \centering
    \includegraphics[width=1.\linewidth]{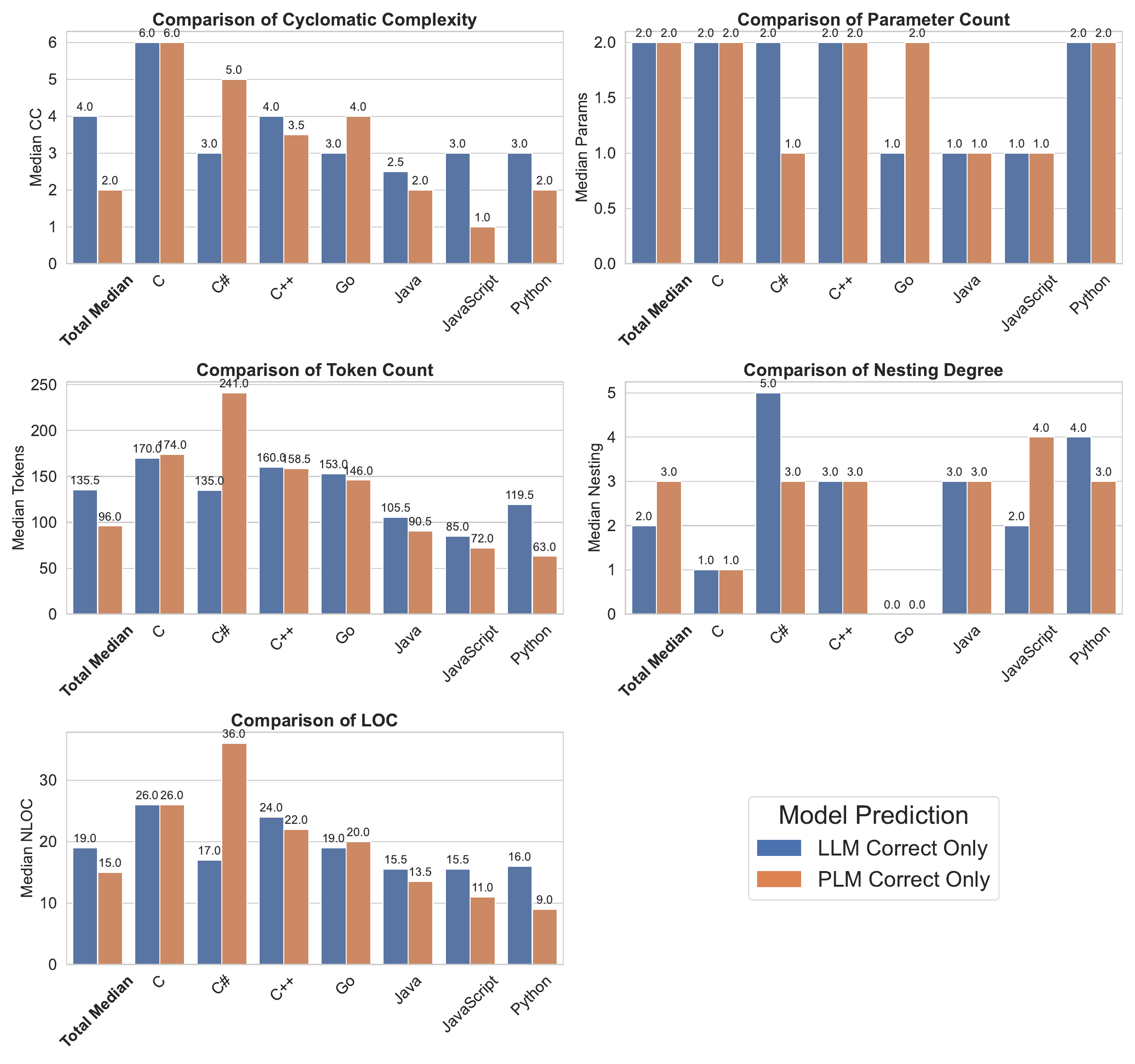}
    \caption{Qualitative analysis of code structural properties for samples where language models exhibit divergent performance. The bars represent the median values of five complexity metrics for samples correctly detected only by the LLM (Blue) versus only by the PLM (Orange).}
    \label{fig:qualitative}
\end{figure*}

To gain deeper insights into the performance differences between PLMs and LLMs, we conducted a granular qualitative analysis of code structural properties. We utilized CodeT5P as the representative PLM and instruction-tuned GPT-4o (with few-shot prompting) as the representative LLM, since they are the best-performing language model on each category. Our analysis focuses on samples where the models exhibited divergent performance. Specifically, we isolated 608 functions from the test set that were correctly classified only by the LLM (where the PLM failed) and 370 functions correctly classified only by the PLM (where the LLM failed). 

To understand ``why'' behind language model failures and success, we selected five metrics that serve as proxies for the cognitive and structural difficulty of code analysis~\citep{10.1109/ICSE43902.2021.00056, weissberg2025llm}:

\begin{itemize}
    \item \textbf{Cyclomatic Complexity (CC)}~\citep{mccabe1976complexity}. Measures the number of linearly independent paths through the code. High CC indicates intricate branching (loops, conditionals), which increases the difficulty of tracking control flow to identify logic-based vulnerabilities.
    \item \textbf{Nesting Degree}~\citep{harrison1981complexity}. Represents the maximum depth of nested structures (e.g., loops inside loops). Deep nesting obscures variable scope and lifetime, making it harder to detect data-flow anomalies or resource management errors.
    \item \textbf{Token Count \& Lines of Code (LOC)}~\citep{halstead1977elements} . These measure the volume of the code. In vulnerability detection, longer contexts dilute the model's attention, making it harder to isolate the specific ``needle in the haystack'', the vulnerable line, amidst benign code.
    \item \textbf{Parameter Count}~\citep{chidamber1994metrics}. Indicates the number of arguments in a function signature. High parameter counts often correlate with complex interfaces and state dependencies, complicating the detection of improper input validation or taint propagation.
\end{itemize}

The structural comparison, summarized in Figure~\ref{fig:qualitative}, reveals distinct operational profiles for each model architecture when analyzing median values. The most significant differentiator between the models is their ability to handle logical complexity. As evidenced by the Median CC metric, the total median for samples correctly detected only by the LLM is 4.0, compared to a median CC of 2.0 for samples detected only by the PLM. This disparity is particularly stark in languages like C\#, where the LLM-exclusive successes had a median CC of 3.0 while the PLM-exclusive successes reached 5.0. Overall, the lower total median for the PLM indicates that CodeT5P struggles to maintain coherence in code characterized by heavy branching, whereas LLM demonstrates superior reasoning capabilities in tracing complex execution paths.

A similar trend is evident when analyzing metrics related to code verbosity and length. The LLM exhibited a clear advantage in processing larger contexts, with the median for its exclusive correct predictions reaching 135.5 tokens and 19.0 lines of code (LOC). Conversely, the samples where the PLM uniquely succeeded were notably shorter, with median values of 96.0 tokens and 15.0 LOC. This data suggests that the PLM's performance degrades as the input volume increases, likely due to fixed context window constraints or limited attention capacity. LLM, however, remains more robust in verbose scenarios, identifying vulnerabilities where the PLM fails due to information overload.

While the LLM generally dominates in measures of scale and logical complexity, the analysis of Median Nesting Degree offers a nuanced counterpoint. The total median nesting degree for the PLM-exclusive successes was 3.0, which is higher than the 2.0 observed for the LLM's unique successes. 
This reversal, notably seen in JavaScript where the PLM handled a median nesting of 4.0 compared to the LLM's 2.0, implies that while CodeT5P falters with broad logical branching, it remains highly effective at capturing local structural patterns even when they are deeply nested. 
Regarding interface complexity, the Median Parameter Count also reflects this trend; the LLM correctly classified more complex functions in languages like C\# (median of 2.0 parameter count vs 1.0 for PLM), though the total median for both models sat at 2.0.

The data highlights a trade-off between complexity reasoning and pattern recognition. The PLM (CodeT5P) falters primarily when faced with higher logical complexity and extended contexts. In contrast, the LLM (GPT-4o) occasionally fails on samples that are structurally simpler and shorter but may contain specific local patterns that the PLM’s training objective captures more effectively.

\subsection{How Does the Model Size Affect the LLM's Strategy?}

\begin{table}[thp!]
    \caption{Performance of Larger LLMs in detecting function-level multilingual vulnerability}
    \label{tab:dis_func}
    \centering
    \tabcolsep=3.0mm
    \small
    \begin{adjustbox}{width=0.9\textwidth, center}
    \begin{tabular}{lrrrrrrrr}
        \toprule
        \toprule
        \textbf{Techniques} & \textbf{Accuracy} & \textbf{Recall} & \textbf{Precision} & \textbf{F1-Score} & \textbf{FPR} & \textbf{FNR} & \textbf{MCC} & \textbf{AUC}\\ 
        \midrule
        \rowcolor{gray!40}\multicolumn{9}{l}{\textbf{LLMs (Zero-Shot Prompting)}} \\ 
        \midrule
            Code Llama (7B) & 0.4877 & 0.9156 & 0.4950 & 0.6426 & 0.9454 & 0.0844 & -0.0585 & 0.4851\\
            Code Llama (70B) & 0.5089 & 0.7360 & 0.5081 & 0.6012 & 0.7210 & 0.2640 & 0.0169 & 0.5075 \\
            Llama 3 (8B) & 0.5005 & 0.5015 & 0.5034 & 0.5025 & 0.5005 & 0.4985 & 0.0010 & 0.5005\\
            Llama 3 (70B) & 0.4956 & 0.7733 & 0.4991 & 0.6066 & 0.7855 & 0.2267 & -0.0147 & 0.4939\\
        \midrule
        \rowcolor{gray!40}\multicolumn{9}{l}{\textbf{LLMs (Few-Shot Prompting)}} \\ 
        \midrule
            Code Llama (7B) & 0.4763 & 0.6251 & 0.4840 & 0.5456 & 0.6743 & 0.3749 & -0.0515 & 0.4754\\
            Code Llama (70B) & 0.4931 & 0.6133 & 0.4968 & 0.5490 & 0.6286 & 0.3867 & -0.0157 & 0.4924 \\
            Llama 3 (8B) & 0.4299 & 0.5260 & 0.4437 & 0.4814 & 0.6673 & 0.4740 & -0.1440 & 0.4293\\
            Llama 3 (70B) & 0.4659 & 0.6762 & 0.4781 & 0.5602 & 0.7468 & 0.3238 & -0.0779 & 0.4647\\
        \midrule
        \rowcolor{gray!40}\multicolumn{9}{l}{\textbf{LLMs (Instruction Tuning + Zero-Shot Prompting)}} \\ 
        \midrule
            Code Llama (7B) & 0.4783 & 0.5103 & 0.4824 & 0.4959 & 0.5531 & 0.4897 & -0.0429 & 0.5091\\
            Code Llama (70B) & 0.5035 & 0.5339 & 0.5060 & 0.5196 & 0.5273 & 0.4661 & 0.0066 & 0.5033\\
            Llama 3 (8B) & 0.5192 & 0.5348 & 0.5215 & 0.5281 & 0.4965 & 0.4652 & 0.0383 & 0.5192\\
            Llama 3 (70B) & 0.5079 & 0.6300 & 0.5087 & 0.5629 & 0.6157 & 0.3700 & 0.0148 & 0.5072\\
        \midrule
        \rowcolor{gray!40}\multicolumn{9}{l}{\textbf{LLMs (Instruction Tuning + Few-Shot Prompting)}} \\ 
        \midrule
            Code Llama (7B) & 0.5094 & 0.6408 & 0.5098 & 0.5678 & 0.6226 & 0.3592 & 0.0188 & 0.5091 \\
            Code Llama (70B) & 0.4941 & 0.5564 & 0.4973 & 0.5252 & 0.5690 & 0.4436 & -0.0126 & 0.4937\\
            Llama 3 (8B) & 0.4946 & 0.4848 & 0.4975 & 0.4911 & 0.4955 & 0.5152 & -0.0107 & 0.4946 \\
            Llama 3 (70B) & 0.5093 & 0.5996 & 0.5104 & 0.5514 & 0.5819 & 0.4004 & 0.0180 & 0.5088\\
        \bottomrule
        \bottomrule
    \end{tabular}
    \end{adjustbox}
\end{table}

\begin{table}[thp!]
    \caption{Performance of Larger LLMs in detecting line-level multilingual vulnerability}
    \label{tab:dis_line}
    \centering
    \tabcolsep=3.0mm
    \small
    \begin{adjustbox}{width=0.9\textwidth, center}
    \begin{tabular}{lrrrrrrrr}
    \toprule
    \toprule
        \textbf{Techniques} & \textbf{Accuracy} & \textbf{Recall} & \textbf{Precision} & \textbf{F1-Score} & \textbf{FPR} & \textbf{FNR} & \textbf{MCC} & \textbf{AUC} \\ 
        \midrule
        \rowcolor{gray!40}\multicolumn{9}{l}{\textbf{LLMs (Zero-Shot Prompting)}} \\ 
        \midrule
            Code Llama (7B)     & 0.9589 & 0.2237          & 0.2272          & 0.2254  & 0.0208  & 0.7763 & 0.2043 & 0.6014      \\
            Code Llama (70B)     & 0.9628 & 0.1321          & 0.2016          & 0.1596    & 0.0144 & 0.8679 & 0.1448 & 0.5589    \\
            Llama 3 (8B)        & 0.9307          & 0.4847          & 0.1891          & 0.2720 & 0.0570 & 0.5153  & 0.2730 & 0.7138          \\
            Llama 3 (70B)        & 0.9619          & 0.3092          & 0.2960          & 0.3025  & 0.0202 & 0.6908 & 0.2830 & 0.6445     \\
        \midrule
        \rowcolor{gray!40}\multicolumn{9}{l}{\textbf{LLMs (Few-Shot Prompting)}} \\ 
        \midrule
            Code Llama (7B) & 0.9666 & 0.3018 & 0.3536 & 0.3257 & 0.0151 &  0.6982 & 0.3097 & 0.6433 \\
            Code Llama (70B)     & 0.9565 & 0.2385          & 0.2161          & 0.2267 & 0.0237 &     0.7616 & 0.2047 & 0.6074    \\
            Llama 3 (8B)  & 0.8980 & 0.7509 & 0.1738 & 0.2822 & 0.0980 & 0.2491 & 0.3295 & 0.8265  \\
            Llama 3 (70B)        & 0.9745          & 0.5190          & 0.5237          & 0.5214 & 0.0130 & 0.4810  & 0.5083 & 0.7530       \\
        \midrule
        \rowcolor{gray!40}\multicolumn{9}{l}{\textbf{LLMs (Instruction Tuning + Zero-Shot Prompting)}} \\ 
        \midrule
            Code Llama (7B) & 0.8771 & 0.5902 & 0.1234 & 0.2042 & 0.1150 & 0.4098 & 0.2296 & 0.7376 \\
            Code Llama (70B)     & 0.9565 & 0.2568          & 0.2251          & 0.2400  & 0.0243 & 0.7431  & 0.2182 & 0.6163     \\
            Llama 3 (8B) & 0.9415 & 0.3403 & 0.1819 & 0.2371  & 0.0420 & 0.6597 & 0.2207 & 0.6491 \\
            Llama 3 (70B)        & 0.9611          & 0.2573          & 0.2654          & 0.2613 & 0.0195 & 0.7427 & 0.2413 & 0.6189        \\
        \midrule
        \rowcolor{gray!40}\multicolumn{9}{l}{\textbf{LLMs (Instruction Tuning + Few-Shot Prompting)}} \\ 
        \midrule
            Code Llama (7B) & 0.8688          & 0.6221 & 0.1207          & 0.2022   & 0.1243   & 0.3779 & 0.2329 & 0.7488    \\
            Code Llama (70B)     & 0.9496 & 0.3918          & 0.2345          & 0.2934   & 0.0351 &  0.6082 & 0.2786 & 0.6784   \\
            Llama 3 (8B) & 0.9220          & 0.4720 & 0.1647          & 0.2442     & 0.0657  & 0.5280 & 0.2464 & 0.7031   \\
            Llama 3 (70B)        & 0.9752          & 0.5092          & 0.5378          & 0.5231 & 0.0120 & 0.4908 & 0.5105 & 0.7485       \\
    \bottomrule
    \bottomrule
    \end{tabular}
    \end{adjustbox}
\end{table}

Our prior experiments demonstrate that combining instruction tuning with zero-shot/few-shot prompting does not effectively improve all LLMs.
This limitation likely stems from the LLMs' basic capabilities, which typically improve with larger model sizes.
Based on performance results from RQ1 and RQ2, we selected Code Llama (70B) and Llama 3 (70B) to investigate how model size influences different LLM strategies and to minimize potential bias in our findings.
The implementation details for both Code Llama (70B) and Llama 3 (70B) remain consistent with Section 3.
As in RQ1 and RQ2, due to differences in positive and negative sample distribution, we evaluate using Accuracy for the function-level task and F1-Score for the line-level task.

As shown in Table~\ref{tab:dis_func}, Code Llama (70B) and Llama 3 (70B) achieve Accuracies of 0.5089 and 0.4956, respectively, in detecting function-level vulnerability using zero-shot prompting. 
Code Llama (70B) shows slight improvement with a positive MCC of 0.0169 and AUC of 0.5075 compared to Code Llama (7B), while Llama 3 (70B) performs slightly worse than Llama 3 (8B), yielding a negative MCC.
With instruction tuning and zero-shot prompting, both Code Llama (70B) and Llama 3 (70B) demonstrate similar performance patterns.
When using few-shot prompting, Code Llama (70B) and Llama 3 (70B) achieve Accuracies of 0.4931 and 0.4659, exceeding the performance of their smaller counterparts, Code Llama (7B) at 0.4763 and Llama 3 (8B) at 0.4299.
However, the MCC values across few-shot settings generally remain negative, indicating that detecting function-level vulnerabilities remains challenging for these larger LLMs.
When combining instruction tuning with few-shot prompting, Code Llama (70B) and Llama 3 (70B) perform similarly to their smaller versions, achieving Accuracies of 0.4941 and 0.5093 compared to 0.5094 and 0.4946, respectively.

As shown in Table~\ref{tab:dis_line}, for line-level detection, Code Llama (70B) reaches F1-Scores of 0.1596 and 0.2267 when using zero-shot prompting and few-shot prompting, which is degraded from 0.2254 and 0.3257 by Code Llama (7B).
This is further corroborated by the MCC and AUC metrics. For example, in zero-shot prompting, Code Llama (70B) drops to an MCC of 0.1448 and AUC of 0.5589, compared to the 7B model's MCC of 0.2043 and AUC of 0.6014.
Llama 3 (70B) is superior to its smaller version while using zero-shot and few-shot prompting, with F1-Score of 0.3025 and 0.5214 compared to 0.2720 and 0.2822, respectively.
Notably, in the few-shot setting, Llama 3 (70B) achieves strong performance with an MCC of 0.5083 and an AUC of 0.7530.
Instruction tuning can better improve the larger LLMs at detecting line-level vulnerability. 
For example, when using instruction tuning with few-shot prompting, Code Llama (70B) and Llama 3 (70B) reach F1-Scores of 0.2934 and 0.5231, significantly superior to 0.2022 by Code Llama (7B) and 0.2442 by Llama 3 (8B).
This improvement is reflected in the robustness of their discrimination, with Llama 3 (70B) maintaining a relative strong MCC of 0.5105 in this configuration.

These results indicate that larger LLMs are not the decisive factor in function-level detection performance. While few-shot prompting works better with larger models, instruction tuning can lead to under-fitting problems when there are too many parameters. Consequently, when using a larger model for function-level detection, careful instruction tuning and appropriate hyper-parameter (e.g. number of adapters in LoRA) selection are essential.
Larger LLMs have a more significant impact on line-level vulnerability detection, especially when using Llama 3, likely due to their more concentrated training sets and more informative labels. The differences between these models, despite having the same parameters, can be attributed to the models' inherent characteristics. 
For example, Llama 3 differs from Code Llama by using a new tokenizer and training on a larger token dataset, which can significantly affect the LLMs' fundamental ability and then further affect the effectiveness of instruction tuning.

\subsection{Comparison between Reasoning LLMs and Non-Reasoning LLMs}
\begin{table}[thp!]
    \caption{Comparison between reasoning and non-reasoning LLMs.}
    \label{tab:dis_reasoning}
    \centering
    \small
    \begin{adjustbox}{width=0.9\textwidth, center}
    \begin{tabular}{lrrrrrrrr}
    \toprule
    \toprule
        \textbf{Techniques} & \textbf{Accuracy} & \textbf{Recall} & \textbf{Precision} & \textbf{F1-Score} & \textbf{FPR} & \textbf{FNR} & \textbf{MCC} & \textbf{AUC}  \\ 
        \midrule
        \rowcolor{gray!40}\multicolumn{9}{l}{\textbf{Non-reasoning LLMs (Function-level)}} \\ 
        \midrule
            DeepSeek-Coder & 0.5005 &0.1796 &0.5097 & 0.2656 & 0.1748 & 0.8204 & 0.0063 & 0.5024\\
            Code Llama & 0.4877 & 0.9156 & 0.4950 & 0.6426 & 0.9454 & 0.0844 & -0.0585 & 0.4851\\
            Llama 3 & 0.5005 & 0.5015 & 0.5034 & 0.5025 & 0.5005 & 0.4985 & 0.0010 & 0.5005\\
            GPT-3.5-Turbo & 0.4827 & 0.6183 & 0.4888 & 0.5459 & 0.6544 & 0.3817 & -0.0375 & 0.4819\\
            GPT-4o & 0.4985 & 0.3631 & 0.5020 & 0.4214 & 0.2632 & 0.7399 & -0.0035 & 0.4985\\
            GRACE & 0.5061 & 0.1673 & 0.5323 & 0.2546 & 0.1495 & 0.8327 & 0.0244 & 0.5089\\
        \midrule
        \rowcolor{gray!40}\multicolumn{9}{l}{\textbf{Reasoning LLMs (Function-level)}} \\ 
        \midrule
            DeepSeek-R1	 & 0.5217  & 	0.6585  & 	0.5193  & 	0.5807 & 0.6167 & 0.3415 & 0.0435 & 0.5209 \\
            QwQ-plus & 	0.5346  & 	0.4838  & 	0.5418  & 	0.5111 & 0.4141 & 0.5162 & 0.0701 & 0.5349 \\
        \midrule
        \midrule
        \rowcolor{gray!40}\multicolumn{9}{l}{\textbf{Non-reasoning LLMs (Line-level)}} \\ 
        \midrule
             DeepSeek-Coder & 0.9536          & 0.3002          & 0.2246          & 0.2570    & 0.0284 & 0.6997 & 0.2362 & 0.6359    \\
            Code Llama     & 0.9589 & 0.2237          & 0.2272          & 0.2254  & 0.0208  & 0.7763 & 0.2043 & 0.6014     \\
            Llama 3        & 0.9307          & 0.4847          & 0.1891          & 0.2720 & 0.0570 & 0.5153  & 0.2730 & 0.7138       \\
            GPT-3.5-Turbo  & 0.9557          & 0.3215          & 0.2475          & 0.2797   & 0.0268 & 0.6785 & 0.2596 & 0.6473   \\
            GPT-4o         & 0.9537          & 0.4286          & 0.2700          & 0.3313  & 0.0318  & 0.5714 & 0.3175 & 0.6984     \\
        \midrule
        \rowcolor{gray!40}\multicolumn{9}{l}{\textbf{Reasoning LLMs (Line-level)}} \\ 
        \midrule
            DeepSeek-R1	 & 0.9692  & 	0.2184  & 	0.3698  & 	0.2746 & 0.0102 & 0.7816 & 0.2694 & 0.6041 \\
            QwQ-plus & 	0.9655  & 	0.2953  & 	0.3344  & 	0.3136 & 0.0161 & 0.7047 & 0.2966 & 0.6396\\
    \bottomrule
    \bottomrule
    \end{tabular}
    \end{adjustbox}
\end{table} 

With the development of LLMs, a new category called reasoning LLMs~\citep{treude2025interacting} has emerged to handle more complex tasks.
The distinction between reasoning and non-reasoning LLMs primarily lies in their approach to problem-solving and the complexity of tasks they can handle.
Reasoning LLMs can perform step-by-step processing and tackle multi-step tasks like code-related challenges. For example, DeepSeek-R1~\citep{guo2025deepseek} and QwQ-plus~\citep{aliyun_qwq_2024} use reinforcement learning techniques (e.g., GRPO) to enhance their reasoning abilities.
Non-reasoning LLMs, such as GPT-4o, generate responses based on learned patterns without explicit intermediate reasoning steps.
To further validate our findings, we selected DeepSeek-R1 and QwQ-plus as our studied reasoning LLMs to evaluate their effectiveness in detecting function-level and line-level multilingual vulnerabilities.
To ensure experimental fairness and maximize the capabilities of reasoning LLMs, we selected the DeepSeek-R1 with 671B parameters and the default QwQ-plus model, whose parameters are not publicly available.
To minimize prompt-related variability, we used the same zero-shot prompt described in Section 3 for both reasoning and non-reasoning LLMs.

Table~\ref{tab:dis_reasoning} compares how reasoning and non-reasoning LLMs perform at detecting multilingual vulnerabilities at both function-level and line-level.
For function-level detection, reasoning LLMs outperform all studied non-reasoning LLMs. DeepSeek-R1 and QwQ-plus achieve Accuracies of 0.5217 and 0.5346, respectively, significantly higher than DeepSeek-Coder (0.5005), Code Llama (0.4877), Llama 3 (0.5005), GPT-3.5-Turbo (0.4758), and GPT-4o (0.4985).
The MCC and AUC results further emphasize this performance gap. Non-reasoning models show MCCs near zero or negative (e.g., $-$0.0585 for Code Llama) and AUCs around 0.5, indicating near-random performance. In contrast, reasoning LLMs demonstrate relatively better discriminative power, with QwQ-plus reaching an MCC of 0.0701 and an AUC of 0.5349.
For line-level detection, reasoning LLMs perform similarly to non-reasoning LLMs in terms of F1-score. For instance, DeepSeek-R1 achieves 0.2746 and QwQ-plus reaches 0.3136, comparable to GPT-3.5-Turbo's 0.2797 and GPT-4o's 0.3313.
The MCC and AUC metrics reflect this comparable performance. QwQ-plus achieves an MCC of 0.2966 and AUC of 0.6396, rivaling GPT-4o's MCC of 0.3175 but falling short of Llama 3's AUC of 0.7138.

Although reasoning LLMs are superior to non-reasoning LLMs at detecting function-level vulnerabilities, their effectiveness is only slightly better than random guessing, and their performance is not significantly better than non-reasoning LLMs.
When using reasoning LLMs to detect line-level vulnerabilities, the improvements are minimal and match the performance of non-reasoning LLMs.
Since reasoning LLMs require longer inference times but offer minimal or no improvements, implementing them for multilingual vulnerability detection requires careful balancing of efficiency and effectiveness.
While we have explored the feasibility of reasoning LLMs for multilingual vulnerability detection, the effectiveness of fine-tuning these models using reinforcement learning still needs to be explored.

\subsection{Data Leakage Analysis}

Given that LLMs are pre-trained on extensive open-source corpora, there is a substantial risk of data leakage where standard test samples may already exist in the model's pre-training data. To mitigate this concern, we employ the REEF data collection methodology to curate a dataset of recent vulnerabilities from 2024 to 2025. We maintain experimental consistency by applying the identical preprocessing pipeline and test split ratio (10\%) described in Section 2.1. This yields 3,138 function-level samples (balanced 1:1 between vulnerable and benign functions) and 1,328 line-level samples. Finally, we evaluate GPT-4, the best-performing model trained on the original dataset, on these temporally splitting test sets to assess its ability to generalize to future data. The results of this evaluation are presented in Table 10.

For function-level detection, the combination of instruction tuning with few-shot prompting demonstrates superior generalization capabilities compared to the zero-shot baseline. 
Specifically, the few-shot configuration achieves an F1-score of 0.7804 and an MCC of 0.5767, representing a substantial improvement over the zero-shot performance (F1-score of 0.4396 and MCC of 0.1056). 
In the case of line-level detection, we observe that the model retains discriminative power on unseen data. 
The few-shot configuration yields the highest stability with an F1-score of 0.3450 and an MCC of 0.3301. While the absolute values are lower than those at the function level, reflecting the inherent difficulty of precise localization, the AUC scores for both zero-shot (0.6541) and few-shot (0.6480) settings remain well above the random baseline (0.5). 

Overall, these results confirm that the LLMs does not merely memorize training patterns but retains robust predictive power on temporally distinct data, thereby validating the effectiveness of our methodology against potential data leakage issues.

\begin{table}[thp!]
    \caption{Performance of LLMs for function-level multilingual vulnerability detection on time-split data.}
    \label{tab:discussion_timesplit}
    \centering
    \tabcolsep=3.0mm
    \small
    \begin{adjustbox}{max width=1.0\textwidth, center}
    \begin{tabular}{lrrrrrrrr}
        \toprule
        \toprule
        \textbf{Techniques} & \textbf{Accuracy} & \textbf{Recall} & \textbf{Precision} & \textbf{F1-score} & \textbf{FPR} & \textbf{FNR} & \textbf{MCC} & \textbf{AUC} \\
        \midrule
        \rowcolor{gray!40}\multicolumn{9}{l}{\textbf{Function-level Detection}} \\ 
        \midrule
            GPT-4o (Instruction-tuning + Zero-shot prompting) & 0.5475 & 0.3530 & 0.5826 & 0.4396 & 0.2558 & 0.6470 & 0.1056 & 0.5486\\
            GPT-4o (Instruction-tuning + Few-shot prompting) & 0.7874 & 0.7510 & 0.8122 & 0.7804 & 0.1756 & 0.2490 & 0.5767 & 0.7877\\
        \midrule
        \rowcolor{gray!40}\multicolumn{9}{l}{\textbf{Line-level Detection}} \\ 
        \midrule
            GPT-4o (Instruction-tuning + Zero-shot prompting) & 0.9642 & 0.3251 & 0.3613 & 0.3423 & 0.0169 & 0.6749 & 0.3244 & 0.6541\\
            GPT-4o (Instruction-tuning + Few-shot prompting) & 0.9663 & 0.3104 & 0.3883 & 0.3450 & 0.0144 & 0.6896 & 0.3301 & 0.6480\\
        \bottomrule
        \bottomrule
    \end{tabular}
    \end{adjustbox}
\end{table}

\subsection{How Does the Impact of the Imbalance Scenario on Language Models?}
\begin{table}[ht!]
    \caption{Performance of PLMs and LLMs for function-level multilingual vulnerability detection on imbalance scenario}
    \label{tab:discuss_imbalanced}
    \centering
    \tabcolsep=3.0mm
    \small
    \begin{adjustbox}{max width=1.0\textwidth, center}
    \begin{tabular}{lrrrrrrrr}
        \toprule
        \toprule
        \textbf{Techniques} & \textbf{Accuracy} & \textbf{Recall} & \textbf{Precision} & \textbf{F1-score} & \textbf{FPR} & \textbf{FNR} & \textbf{MCC} & \textbf{AUC} \\
        \midrule
        \rowcolor{gray!40}\multicolumn{9}{l}{\textbf{Dummy Classifiers}} \\ 
        \midrule
            ${DummyClf}_{vul}$ & 0.1684 & 1.0000 & 0.1684 & 0.2883 & 1.0000 & 0.0000 & 0.0000 & 0.5000\\
            ${DummyClf}_{clean}$ & 0.8316 & 0.0000 & 0.0000 & 0.0000 & 0.0000 & 1.0000 & 0.0000 & 0.5000\\
        \midrule
        \rowcolor{gray!40}\multicolumn{9}{l}{\textbf{PLMs}} \\ 
        \midrule
            Text-Embedding-3-Large & 0.6614 & 0.7036 & 0.291 & 0.4117 & 0.3472 & 0.2964 & 0.2715 & 0.6782 \\
            Text-Embedding-3-Small & 0.6477 & 0.7105 & 0.2827 & 0.4045 & 0.3651 & \textbf{0.2895} & 0.2616 & 0.6727 \\
            Text-Embedding-Ada-002 & 0.6622 & 0.6997 & 0.2909 & 0.4110 & 0.3454 & 0.3003 & 0.2701 & 0.6772 \\
            CodeBERT & 0.8111 & 0.4544 & 0.4410 & 0.4476 & 0.1167 & 0.5456 & 0.3337 & 0.6689\\
            LineVul & 0.8294 & 0.3474 & 0.4910 & 0.4069 & 0.0729 & 0.6526 & 0.317 & 0.6372\\
            UniXcoder & 0.8126 & \textbf{0.4789} & 0.4473 & \textbf{0.4626} & 0.1198 & 0.5211 & 0.3495 & 0.6795\\
            CodeT5 & 0.8151 & 0.4681 & 0.4526 & 0.4602 & 0.1147 & 0.5319 & 0.3487 & \textbf{0.6767}\\
            CodeT5P & \textbf{0.8528} & 0.2934 & 0.6362 & 0.4016 & 0.0340 & 0.7066 & \textbf{0.3627} & 0.6297\\
        \midrule
        \rowcolor{gray!40}\multicolumn{9}{l}{\textbf{LLMs (Zero-Shot Prompting)}} \\ 
        \midrule
            DeepSeek-Coder & 0.8232 & 0.0108 & 0.1507 & 0.0201 & 0.0123 & 0.9892 & -0.0052 & 0.4992 \\
            Code Llama & 0.7480 & 0.0402 & 0.0697 & 0.0510 & 0.1087 & 0.9598 & -0.0865 & 0.4658\\
            Llama 3 & 0.5343 & 0.4838 & 0.1770 & 0.2592 & 0.4555 & 0.5162 & 0.0213 & 0.5142\\
            GPT-3.5-Turbo & 0.5169 & 0.6212 & 0.1997 & 0.3022 & 0.5042 & 0.3788 & 0.0877 & 0.5585\\
            GPT-4o & 0.8022 & 0.1501 & 0.3161 & 0.2036 & 0.0658 & 0.8499 & 0.1164 & 0.5422\\
            GRACE & 0.7865 & 0.1581 & 0.2826 & 0.2027 & 0.0832 & 0.8419 & 0.0958 & 0.5374\\
        \midrule
        \rowcolor{gray!40}\multicolumn{9}{l}{\textbf{LLMs (Few-Shot Prompting)}} \\ 
        \midrule
            DeepSeek-Coder & 0.7871 & 0.0245 & 0.0784 & 0.0374 & 0.0584 & 0.9755 & -0.0568 & 0.4831\\
            Code Llama & 0.7638 & 0.0324 & 0.0693 & 0.0441 & 0.0880 & 0.9676 & -0.0774 & 0.4722\\
            Llama 3 & 0.5903 & 0.4171 & 0.1840 & 0.2553 & 0.3746 & 0.5829 & 0.0327 & 0.5212 \\
            GPT-3.5-Turbo & 0.6199 & 0.4426 & 0.2066 & 0.2817 & 0.3442 & 0.5574 & 0.0767 & 0.5492 \\
            GPT-4o & 0.7708 & 0.4446 & 0.3556 & 0.3951 & 0.1632 & 0.5554 & 0.2583 & 0.6407\\
        \midrule
        \rowcolor{gray!40}\multicolumn{9}{l}{\textbf{LLMs (Instruction Tuning + Zero-Shot Prompting)}} \\ 
        \midrule
            DeepSeek-Coder & 0.7382 & 0.1138 & 0.1455 & 0.1278 & 0.1353 & 0.8862 & -0.0238 & 0.4893 \\
            Code Llama & 0.2829 & 0.8901 & 0.1767 & 0.2948 & 0.84 & 0.1099 & 0.0522 & 0.5250\\
            Llama 3 & 0.7291 & 0.1590 & 0.1716 & 0.1651 & 0.1554 & 0.8410 & 0.0037 & 0.5018\\
            GPT-3.5-Turbo & 0.8463 & 0.1276 & 0.7602 & 0.2185 & 0.0081 & 0.8724 & 0.2697 & 0.5597 \\
            GPT-4o & 0.8451 & 0.0834 & \textbf{0.9659} & 0.1536 & \textbf{0.0006} & 0.9166 & 0.2589 & 0.5414\\
        \midrule
        \rowcolor{gray!40}\multicolumn{9}{l}{\textbf{LLMs (Instruction Tuning + Few-Shot Prompting)}} \\ 
        \midrule
            DeepSeek-Coder &0.7268 & 0.1570 & 0.1677 & 0.1622 & 0.1578 & 0.843 & -0.0008 & 0.4996 \\
            Code Llama & 0.3071 & 0.9254 & 0.1864 & 0.3102 & 0.8182 & 0.0746 & 0.1085 & 0.5536 \\
            Llama 3 & 0.7055 & 0.3474 & 0.2407 & 0.2843 & 0.2220 & 0.6526 & 0.1094 & 0.5627 \\
            GPT-3.5-Turbo & 0.8273 & 0.0353 & 0.3673 & 0.0645 & 0.0123 & 0.9647 & 0.0682 & 0.5115 \\
            GPT-4o & 0.7951 & 0.4073 & 0.3949 & 0.4010 & 0.1264 & 0.5927 & 0.2774 & 0.6404\\
        \bottomrule
        \bottomrule
    \end{tabular}
    \end{adjustbox}
\end{table}

Although our evaluation on balance dataset provide a controlled setting to compare models fairly, isolate intrinsic discriminative ability, and avoid conflating performance with class prevalence, we further evaluate the studied language models (See Table~10) on an imbalance dataset for establishing a deployment-faithful stress test that reveals how models behave when class skew affect calibration and the trade-off between false positives and false negatives. 
To construct the imbalanced function-level multilingual vulnerability dataset, we enlarge the benign class by mining version-control history. 
Specifically, we collect functions that do not exhibit commit changes (i.e., functions without security-relevant modifications across commits) and treat them as benign functions, then combine them with the labeled vulnerable functions under the same function-level labeling protocol. 
After filtering and deduplication across seven programming languages, the final imbalanced dataset contains 48,219 / 6,012 / 6,051 functions with corresponding labels for training / validation / testing, respectively. 
The splits are explicitly skewed: the training set contains 8,102 vulnerable and 40,117 benign functions; the validation set contains 1,012 vulnerable and 5,000 benign functions; and the test set contains 1,019 vulnerable and 5,032 benign functions. 
This construction better reflects practical scanning scenarios while still allowing results to be interpreted alongside the balanced setting for a more complete assessment of multilingual function-level vulnerability detection.

As shown on Table~\ref{tab:discuss_imbalanced}, the evaluation of PLMs reveals that models specifically architecture-optimized for code understanding generally handle the imbalanced nature of the dataset more effectively than general-purpose embeddings. 
UniXcoder achieves the highest F1-score (0.4626) among all models, demonstrating a superior balance between Precision and Recall. 
CodeT5P follows closely with the highest MCC (0.3627) and a strong AUC of 0.6297, indicating its robustness in distinguishing between vulnerable and non-vulnerable functions despite the majority class bias. 
These PLMs consistently outperform the Dummy classifier, which fail to achieve any meaningful MCC, highlighting that the specialized pre-training on code-specific tasks enables these models to capture the subtle semantic patterns associated with software vulnerabilities.

In contrast, LLMs in zero-shot and few-shot configurations show a wider variance in performance, often struggling with high FNR. 
In the zero-shot setting, models like DeepSeek-Coder and Code Llama exhibit extremely low F1-scores (below 0.1), with AUC values near 0.5, suggesting they are near-random in their ability to identify vulnerabilities without specific guidance. 
However, the introduction of few-shot prompting significantly boosts performance across the board. 
For instance, GPT-4o’s F1-score improves from 0.1264 in zero-shot to 0.3951 in few-shot, and its MCC jumps to 0.2583. 
This indicates that while LLMs possess vast general knowledge, they require in-context examples to calibrate to the specific nuances and imbalanced distribution of the vulnerability detection task.

Instruction tuning further refines the LLMs' capabilities, though it introduces distinct trade-offs between precision and recall. 
For GPT-4o, instruction tuning combined with few-shot prompting yields a high precision and the best LLM F1-score of 0.4010. 
Some instruction-tuned models like Code Llama show a dramatic surge in Recall (up to 0.9254) but at the cost of very high FPR, resulting in a low MCC. 
This suggests that instruction tuning can make certain models over-sensitive to potential threats, causing them to label most samples as vulnerable.
Conversely, GPT-4o maintains a very low FPR (0.0006 in zero-shot instruction tuning and 0.1264 in few-shot instruction tuning), making it highly reliable when it does flag a vulnerability, even if it misses a larger portion of the total vulnerable samples (high FNR).

Comparing the best-performing PLM (UniXcoder) with the best-performing LLM (GPT-4o under instruction tuning and few-shot), the PLM remains superior in overall classification balance for this imbalanced dataset. 
UniXcoder maintains a significantly higher F1-score (0.4626 vs. 0.4010) and MCC (0.3475 vs. 0.2774) than GPT-4o. While the LLM achieves a slightly better AUC in some configurations (0.6404), the PLM’s higher MCC indicates a more reliable correlation between its predictions and the actual ground truth in a skewed distribution. 
This suggests that for function-level detection multilingual vulnerability detection on imbalance scenario, the dense, specialized representations learned by PLMs like UniXcoder are currently more effective than the broad reasoning capabilities of LLMs, which may still struggle with the high precision requirements and class imbalance inherent in security-critical datasets.

\subsection{Deployment Cost of LLMs and PLMs on Multilingual Vulnerability Detection}
\begin{table}
\caption{Cost Estimation of LLMs}
\label{tab:cost}
\centering
\begin{adjustbox}{width=0.8\textwidth, center}
\begin{tabular}{lcc}
\toprule
\toprule
\textbf{Granularity} & \multicolumn{1}{c}{Function-level} & \multicolumn{1}{c}{Line-level} \\
\midrule
\textbf{\#Tokens of Training} & 3,442,643 & 2,376,633 \\
\textbf{\#Tokens of Input} & 439,680 & 253,697 \\
\textbf{\#Tokens of Output} & 2,026 & 39,726 \\
\midrule
\textbf{Cost of Training (GPT-3.5-Turbo / GPT-4o)} & \$27.54 / \$86.07 & \$19.01 / \$59.42 \\
\textbf{Cost of Input (GPT-3.5-Turbo / GPT-4o)} & \$1.32 / \$1.65 & \$0.76 / \$0.95 \\
\textbf{Cost of Output (GPT-3.5-Turbo / GPT-4o)} & \$0.01 / \$0.03 & \$0.24 / \$0.60 \\
\bottomrule
\bottomrule
\end{tabular}
\end{adjustbox}
\end{table}

Evaluating deployment cost is critical for assessing the practicality of multilingual vulnerability detection approaches, as real-world adoption depends not only on detection accuracy but also on economic feasibility.
While several recent studies on LLM-based vulnerability detection~\citep{lu2024grace, du2024vul} have reported promising performance, they often overlook deployment cost, leaving a gap in understanding the trade-off between accuracy and operational expenses.
In contrast, prior work on software engineering tasks such as SWE-Bench~\citep{yu2025utboost, yang2024swe} has established reporting API usage costs as a standard practice for evaluating LLM-based solutions, which we follow in this study.

For LLMs with more than 100B parameters, such as GPT-3.5-Turbo and GPT-4o, deployment typically requires data center–grade GPU infrastructure, making API-based usage a common and practical choice.
As shown in Table~\ref{tab:cost}, for function-level detection with instruction tuning, the training costs for GPT-3.5-Turbo and GPT-4o are approximately \$27.54 and \$86.07, respectively, with input costs of about \$1.32 and \$1.65, and output costs of about \$0.01 and \$0.03.
For line-level detection, the training costs are approximately \$19.01 and \$59.42, with input costs of about \$0.76 and \$0.95, and output costs of about \$0.24 and \$0.60.

In contrast, PLMs such as CodeBERT, UniXcoder, CodeT5, and CodeT5P typically have fewer than 0.3B parameters, allowing them to be trained and deployed on consumer-grade GPUs.
As such, their deployment costs are limited to compute resources (e.g., GPU/hour) without incurring per-token API charges.

To provide a rigorous comparison between API-based LLMs and locally deployed PLMs, we formalize the \textit{Total Cost of Ownership} (TCO)~\citep{stojkovic2025rearchitecting}, where the components of TCO consist of Capital Expenses (CapEx) and Operational Expenses (OpEx). While API costs are purely OpEx, PLM costs involve a blend of CapEx and OpEx.
The total cost of a single training or inference cycle for local PLM is defined as:
\begin{equation}
Cost_{PLM} = \left( \frac{P}{L} \times T \right) + \left( \frac{W \times E}{1000} \times T \right)
\end{equation}
where P is the purchase price of the hardware (CapEx), L is the useful lifespan of the hardware in hours, T is the total duration of the compute tasks in hours, W is the total system power consumption in Watts (TDP + system overhead), and E is Electricity cost per kilowatt-hour (kWh).
We ground our analysis using the NVIDIA RTX A6000, a standard professional-grade GPU. 
In this case, we assume that the P (price) is \$ 4,650, L (lifespan) is 3 years which is satandard depreciation for high-utilization AI hardware due to thermal wear~\citep{kshirsagar2025lifespan}, W (Power) is 400W (i.e., 300W GPU TDP and 100W System overhead), E (Elevtricity) is \$0.14/kWh which is projected by 2026 U.S. commercial energy rates (EIA)~\citep{eia2026steo}, and T (Total duration) is 24 hours for a PLM like CodeT5P.
The amortized hardware cost is \$4.25 and the energy cost is \$1.34. This results in a total training cost of \$5.59.
As shown in the comparison, while the A6000 requires a significant upfront investment, the cost per training run is approximately 15x lower than GPT-4o (\$86.07). The Break-Even Point, the moment the local GPU becomes cheaper than API calls, occurs after only 58 fine-tuning runs or a corresponding volume of inference tokens.

This distinction makes PLMs more cost-effective for large-scale or continuous deployment when in-house hardware is available, while LLMs offer greater scalability and accessibility at the expense of variable API costs.
For organizations with sufficient resources and no privacy concerns, using the API of closed-source LLMs for fine-tuning and inference offers a reliable option, particularly for line-level multilingual vulnerability detection.
Conversely, organizations with only consumer-level GPUs and/or strict privacy requirements should consider deploying PLMs for multilingual vulnerability detection as a promising solution.
A practical alternative is to implement a lightweight custom PLM (such as CodeT5P) that can be fine-tuned using the organization's own in-house hardware and dataset.

%% file: threats.tex
\section{Threats to Validity}
\label{sec:threats}
\textit{External Threats} primarily stem from our choice of multilingual vulnerability dataset and automated vulnerability detection approaches. 
Our study focused on the REEF dataset covering seven programming languages, which may limit the generalizability of our findings to other languages. 
To address this limitation, we plan to expand our vulnerability collection to include more programming languages and timelines using the REEF collection framework.
To minimize methodological threats, we performed a comprehensive literature review to ensure our selected PLMs and LLMs for automated vulnerability detection represent current state-of-the-art solutions.

\textit{Internal Threat} relates to our implementation and utilization of the studied approaches. 
To address this, we carefully implemented all existing approaches following their original replication packages, with two authors conducting thorough code reviews.
For PLMs, we followed the hyperparameter settings recommended by the original authors to ensure a fair reproduction of their approaches. However, the 512-token input limit of the smaller PLMs we used, which cannot be extended during fine-tuning, may restrict the models’ ability to capture context in longer code snippets. We acknowledge this architectural constraint as a limitation of our experimental setup.
For LLMs, we utilized official channels, accessing public models (DeepSeek-Coder, Code Llama, and Llama3) via HuggingFace and invoking APIs (GPT-3.5-Turbo and GPT-4o) according to documented guidelines.
Besides, this study excludes certain state-of-the-art LLMs fine-tuned for vulnerability detection, as many target single languages (mainly C/C++) or lack line-level localization. Since our focus is multilingual and dual-granularity performance, these models fall outside our scope. While this limits comparison with customized tools, our evaluation of 11 diverse models with multiple stretagies, from CodeBERT to GPT-4o, provides a robust baseline for multilingual vulnerability research. We include LineVul as a line-level baseline and will prioritize specialized models as they support broader languages in future studies.

\textit{Construct Threats} arise from our construction of vulnerable functions, clean functions, and metrics for evaluating PLMs and LLMs. 
We used Tree-sitter to parse commit data and collect before-change and after-change functions, which introduces a potential risk of incorrect function identification. 
To mitigate this threat, we conducted sanity checks on randomly selected samples to verify the tool's robustness. 
One potential threat to the validity of our performance metrics is data leakage between the training, validation, and test sets. This subtle form of leakage may occur when functions from the same project or developer appear in multiple sets. To mitigate this issue, we apply exact-match filtering, helping prevent potential overestimation.
A potential threat to the validity of our findings is the inherent noise in real-world vulnerability datasets. As noted by~\citet{ding2024vulnerability}, vulnerability-fixing commits are frequently 'tangled,' mixing the actual security patch with unrelated modifications such as refactoring or documentation. Consequently, the input data may contain code changes irrelevant to the vulnerability itself, potentially introducing noise into the training process. To mitigate this, we implemented preprocessing heuristics to filter out cosmetic and non-functional changes, specifically removing comments, blank lines, logging statements, and formatting updates. While these measures reduce the noise floor, future work incorporating fine-grained, line-level manual verification is necessary to fully isolate the security-relevant signals.
For evaluation, we employed established metrics from function-level and line-level vulnerability detection, including Accuracy, Precision, Recall, and F1-score.

%% file: related_work.tex
\section{Related Work}
\label{sec:background}

\subsection{Software Vulnerability}
\label{subsec:soft_vul}
Software vulnerabilities are security flaws in software systems that malicious actors can exploit.
These vulnerabilities can cause significant damage when left unpatched, particularly in widely-used software, leading to substantial financial losses~\citep{bilge12empirical}.
For example, the Log4Shell vulnerability (CVE-2021-44228)~\citep{luttwak2021log4shell} in the Log4j library, which allowed attackers to execute arbitrary Java code on servers and other systems, exposed approximately 93\% of enterprise cloud environments to significant security risks.
Two main classification systems in the community help categorize these issues: Common Weakness Enumeration (CWE) for identifying weakness types, and Common Vulnerability Exposure (CVE) for documenting specific vulnerability instances~\citep{cwe, cve}.

Recent studies highlight the prevalence of vulnerabilities in open-source software. 
~\citet{jia2022cargo} found that the Cargo ecosystem's security vulnerabilities are primarily memory-related, with 18\% of affected libraries still vulnerable in their latest versions, and 19.78\% of all versions in the ecosystem impacted by vulnerability propagation.
~\citet{zerouali2022impact} observed that vulnerabilities in npm packages affect a median of 30 package releases, compared to 59 releases in RubyGems packages.
Furthermore, a significant proportion of external GitHub projects is exposed to vulnerabilities originating from both direct and indirect dependencies.
~\citet{li2022vulnerability} presented a large-scale study of popular multilingual projects on GitHub and identified statistically significant associations between the vulnerability proneness of multilingual code (both overall and for specific categories) and its language selection.
\citet{alfadel2023empirical} analyzed 550 vulnerability reports related to 252 Python packages, providing insights into common security issues within the Python ecosystem.
A comprehensive analysis~\citep{mir2023effect} of 3 million Maven packages demonstrated that about one-third of the packages in the dataset are identified as vulnerable only when all the transitive dependencies are taken into consideration.
~\citet{hu2024empirical} investigated the prevalence and remediation delays of vulnerabilities in Go modules, finding that 66.10\% of modules are affected and identifying two types of delays that impede the timely resolution of vulnerabilities.
These empirical findings emphasize the widespread presence of vulnerabilities in open-source software across diverse programming languages, underscoring the urgent need for effective vulnerability detection in multilingual contexts.

\subsection{Automated Vulnerability Detection}
Automated vulnerability detection is a security assessment method that employs techniques such as static code analysis~\citep{10.1145/3576915.3624401} or dynamic execution testing~\citep{kim2019rvfuzzer} to automatically identify potential security weaknesses in software or systems.  
Recent breakthroughs in deep learning, especially PLMs, have revolutionized automated vulnerability detection, leading to significant advances in cybersecurity applications~\citep{zhou2024large,steenhoek2023empirical}.
These methods learn potential vulnerability patterns by constructing abstract representations of source code and establishing a nonlinear mapping relationship between source code characterization and vulnerability presence (i.e., whether a given code snippet contains security vulnerabilities).

Specifically, ~\citet{hanif2022vulberta} introduced VulBERTa, a deep learning model that leverages a custom tokenization pipeline to pre-train RoBERTa~\citep{liu2019roberta} on real-world C/C++ code, enabling enhanced vulnerability detection through learned code syntax and semantics.
LineVul~\citep{fu2022linevul} employs a BERT-based architecture~\citep{devlin2019bert} to create code representations for detecting vulnerabilities at the function level and uses its attention mechanism to identify specific vulnerable lines.
SVulD~\citep{ni2023distinguishing} trains a model to differentiate semantic representations of functions, irrespective of lexical similarity, for vulnerability detection, and provides natural language explanations to help developers intuitively understand the root causes of vulnerabilities.
~\citet{zhang2023vulnerability} decomposed the syntax-based Control Flow Graph (CFG) of a code snippet into multiple execution paths, extracted their representations through LLMs with intra- and inter-path attention mechanisms, and aggregated these representations to detect vulnerabilities.
~\citet{liu2024pre} pre-trains their PLM, PDBERT, by leveraging the Abstract Syntax Tree (AST) and Program Dependency Graph (PDG) of functions to predict statement-level control dependencies and token-level data dependencies.
They then fine-tune the PLM for the vulnerability detection task.
DeepDFA~\citep{steenhoek2024dataflow} introduces a graph-based learning framework that incorporates graph embeddings with PLMs to enhance vulnerability detection performance. In this study, they use general PLMs as vulnerability detection baselines.

However, PLMs require vast amounts of data to uncover patterns of vulnerability, and their generalization ability may be limited when confronted with unseen data.
Recently, LLMs that are pre-trained on extensive code corpora have demonstrated a remarkable ability to understand code semantics, showing significant performance improvements in code-related tasks~\citep{hou2023large}.
Several empirical studies~\citep{10479409, zhou2024large, yin2024multitask, yin2024pros, zhou2024largeworkshop} have explored using LLMs for vulnerability detection.
~\citet{10479409} examined ChatGPT's capability to detect vulnerabilities in C/C++ functions through a zero-shot prompt.
They found that GPT-3.5-Turbo and GPT-4 fail to detect single-language vulnerabilities at both function-level and line-level.
~\citet{yin2024pros} demonstrated that when relying solely on prompts, ChatGPT is highly susceptible to altering vulnerability classifications, reflecting low confidence in its assessments.
Although the inclusion of contextual information enhances its accuracy, the model still struggles to reliably predict severity ratings for certain categories of CWEs.
Vul-RAG~\citep{du2024vul}, a technique for vulnerability detection based on LLMs, constructs a comprehensive knowledge base of vulnerabilities, retrieves relevant information through semantic similarity, and utilizes reasoning to assess the presence of vulnerabilities in code.
MSIVD~\citep{yang2024security} integrates a multitask sequence-to-sequence language model with program control flow graphs, which are encoded as graph neural networks, to enable sequence-to-classification vulnerability detection.
GRACE~\citep{lu2024grace} leverages graph structural information in the code and in-context learning to enhance the effectiveness of LLM-based software vulnerability detection.

Some research has begun to systematically investigate the effectiveness of PLMs and LLMs on vulnerability detection.
For instance, a notable study by~\citet{ding2024vulnerability} investigated the performance of code LMs using the PrimeVul benchmark, highlighting significant limitations of current models in accurately identifying vulnerabilities in real-world C/C++ code. While our work shares a similar motivation in probing the limits of LLMs, we extend this line of inquiry in three key dimensions.
First, whereas~\citet{ding2024vulnerability} focuses primarily on single-language (C/C++) evaluation, our study (RQ1) explicitly investigates the multilingual capabilities of models across seven distinct programming languages, revealing how cross-language semantic commonalities and differences affect detection performance.
Second, while~\citet{ding2024vulnerability} focuses on function-level classification, we provide a dual-granularity analysis that includes line-level localization, offering a more fine-grained perspective on model interpretability.
Finally, we explore a broader array of prompting and tuning strategies specifically tailored for the multilingual scenario, providing a comprehensive assessment of how these models generalize beyond single-language constraints.
In the multilingual settings, \citet{cao2022mvd} propose a framework based on model distillation and pre-trained language models (i.e., CodeBERT) for multilingual vulnerability detection. 
However, their work focuses solely on function-level detection. In contrast, our study is a comprehensive empirical study by examining dual-granularity detection across 6 PLMs and 5 LLMs, employing different prompting strategies, zero-shot, few-shot, and instruction-tuning, on seven programming languages. We also incorporate line-level localization to provide finer-grained security insights.

Notably, researchers have observed significant variations in the detection difficulty of different vulnerability types.  
Studies by ~\citet{russell2018automated}, ~\citet{li2018vuldeepecker}, and ~\citet{xu2022acgdp} indicate that detecting certain types of vulnerabilities is more challenging.  
Furthermore, ~\citet{hin2022linevd} evaluated model performance in cross-project settings using a leave-one-out approach and found a slight decline in performance under such scenarios.  
However, as modern software development increasingly trends toward multilingual integration, the diversity of codebases introduces new challenges for software security. 
The effectiveness of existing vulnerability detection methods in multilingual scenarios remains unverified, making it difficult to meet the demands of real-world development.

%% file: declaration.tex
\section*{Statement and Declarations}
\textbf{Funding}\\
This work was supported by (1) National Key Research and Development Program of China (Grant No. 2024YFB4506300), (2) the National Natural Science Foundation of China (Grant No. 62322208), (3) JST under the Adopting Sustainable Partnerships for Innovative Research Ecosystem (ASPIRE) program, Grant Number JPMJAP2415, (4) JSPS for the KAKENHI grants (JP25K22845), (5) Bilateral Program grant JPJSBP120239929, and (6) the Inamori Research Institute for Science for supporting Yasutaka Kamei via the InaRIS Fellowship.\\

\noindent
\textbf{Ethical Approval}\\
Not applicable.\\

\noindent
\textbf{Informed Consent}\\
Not applicable.\\

\noindent
\textbf{Author Contributions}\\
Conceptualization: Honglin Shu and Dong Wang; 
Methodology: Honglin Shu, Michael Fu, and Dong Wang;  
Formal analysis and investigation: Honglin Shu and Michael Fu;
Writing original draft preparation: Honglin Shu and Junji Yu; 
Writing review \& editing: Honglin Shu, Dong Wang, and Michael Fu; 
Resources: Yasutaka Kamei and Junjie Chen; 
Supervision: Junjie Chen, Chakkrit Tantithamthavorn, and Yasutaka Kamei;
Replication of baseline technique: Honglin Shu, Junji Yu, and Michael Fu.\\

\noindent
\textbf{Data Availability Statement}\\
The data, model training and evaluation scripts that support the findings of this study are available at: (\url{https://github.com/SpanShu96/Large-Language-Model-for-Multilingual-Vulnerability-Detection/tree/main}).\\

\noindent
\textbf{Conflict of Interest Statement}\\
The authors of this article declared that they have no conflict of interest.\\

%% file: additionalref.bib
@inproceedings{alizadeh2025language,
  title={Language Models in Software Development Tasks: An Experimental Analysis of Energy and Accuracy},
  author={Alizadeh, Negar and Belchev, Boris and Saurabh, Nishant and Kelbert, Patricia and Castor, Fernando},
  booktitle={2025 IEEE/ACM 22nd International Conference on Mining Software Repositories (MSR)},
  pages={725--736},
  year={2025},
  organization={IEEE}
}

@article{zhong2025larger,
  title={Larger Is Not Always Better: Exploring Small Open-source Language Models in Logging Statement Generation},
  author={Zhong, Renyi and Li, Yichen and Yu, Guangba and Gu, Wenwei and Kuang, Jinxi and Huo, Yintong and Lyu, Michael R},
  journal={ACM Transactions on Software Engineering and Methodology},
  year={2025},
  publisher={ACM New York, NY}
}

@inproceedings{wang2023reef,
  title={Reef: A framework for collecting real-world vulnerabilities and fixes},
  author={Wang, Chaozheng and Li, Zongjie and Pena, Yun and Gao, Shuzheng and Chen, Sirong and Wang, Shuai and Gao, Cuiyun and Lyu, Michael R},
  booktitle={2023 38th IEEE/ACM International Conference on Automated Software Engineering (ASE)},
  pages={1952--1962},
  year={2023},
  organization={IEEE}
}

@article{ding2024vulnerability,
  title={Vulnerability detection with code language models: How far are we?},
  author={Ding, Yangruibo and Fu, Yanjun and Ibrahim, Omniyyah and Sitawarin, Chawin and Chen, Xinyun and Alomair, Basel and Wagner, David and Ray, Baishakhi and Chen, Yizheng},
  journal={arXiv preprint arXiv:2403.18624},
  year={2024}
}

@article{stojkovic2025rearchitecting,
  title={Rearchitecting Datacenter Lifecycle for AI: A TCO-Driven Framework},
  author={Stojkovic, Jovan and Zhang, Chaojie and Goiri, {\'I}{\~n}igo and Bianchini, Ricardo},
  journal={arXiv preprint arXiv:2509.26534},
  year={2025}
}

@techreport{eia2026steo,
  author      = {{U.S. Energy Information Administration}},
  title       = {Short-Term Energy Outlook (STEO)},
  institution = {U.S. Department of Energy},
  year        = {2025},
  month       = {December},
  note        = {Projections for 2026 Electricity and Natural Gas Markets},
  url         = {https://www.eia.gov/outlooks/steo/}
}

@article{mccabe1976complexity,
  title={A complexity measure},
  author={McCabe, Thomas J},
  journal={IEEE Transactions on software Engineering},
  number={4},
  pages={308--320},
  year={1976},
  publisher={IEEE}
}

@article{harrison1981complexity,
  title={A complexity measure based on nesting level},
  author={Harrison, Warren A and Magel, Kenneth I},
  journal={ACM Sigplan Notices},
  volume={16},
  number={3},
  pages={63--74},
  year={1981},
  publisher={ACM New York, NY, USA}
}

@book{halstead1977elements,
  title={Elements of Software Science (Operating and programming systems series)},
  author={Halstead, Maurice H},
  year={1977},
  publisher={Elsevier Science Inc.}
}

@article{steenhoek2024err,
  title={To err is machine: Vulnerability detection challenges llm reasoning},
  author={Steenhoek, Benjamin and Rahman, Md Mahbubur and Roy, Monoshi Kumar and Alam, Mirza Sanjida and Tong, Hengbo and Das, Swarna and Barr, Earl T and Le, Wei},
  journal={arXiv preprint arXiv:2403.17218},
  year={2024}
}

@inproceedings{10.1109/ICSE43902.2021.00056,
author = {Peitek, Norman and Apel, Sven and Parnin, Chris and Brechmann, Andr\'{e} and Siegmund, Janet},
title = {Program Comprehension and Code Complexity Metrics: An fMRI Study},
year = {2021},
isbn = {9781450390859},
publisher = {IEEE Press},
url = {https://doi.org/10.1109/ICSE43902.2021.00056},
doi = {10.1109/ICSE43902.2021.00056},
pages = {524–536},
numpages = {13},
series = {ICSE '21}
}

@article{chidamber1994metrics,
  title={A metrics suite for object oriented design},
  author={Chidamber, Shyam R and Kemerer, Chris F},
  journal={IEEE Transactions on software engineering},
  volume={20},
  number={6},
  pages={476--493},
  year={1994},
  publisher={IEEE}
}

@misc{kshirsagar2025lifespan,
  author       = {Kshirsagar, Mihir},
  title        = {Lifespan of {AI} Chips: The \$300 Billion Question},
  howpublished = {Center for Information Technology Policy (CITP) Blog, Princeton University},
  year         = {2025},
  month        = {October},
  day          = {15},
  url          = {https://blog.citp.princeton.edu/2025/10/15/lifespan-of-ai-chips-the-300-billion-question/},
  note         = {Accessed: January 2026}
}

@article{weissberg2025llm,
  title={LLM-based Vulnerability Discovery through the Lens of Code Metrics},
  author={Weissberg, Felix and Pirch, Lukas and Imgrund, Erik and M{\"o}ller, Jonas and Eisenhofer, Thorsten and Rieck, Konrad},
  journal={arXiv preprint arXiv:2509.19117},
  year={2025}
}

@article{uddin2025deep,
  title={Deep Learning Aided Software Vulnerability Detection: A Survey},
  author={Uddin, Md Nizam and Zhang, Yihe and Hei, Xiali},
  journal={arXiv preprint arXiv:2503.04002},
  year={2025}
}

@article{zerouali2022impact,
  title={On the impact of security vulnerabilities in the npm and RubyGems dependency networks},
  author={Zerouali, Ahmed and Mens, Tom and Decan, Alexandre and De Roover, Coen},
  journal={Empirical Software Engineering},
  volume={27},
  number={5},
  pages={107},
  year={2022},
  publisher={Springer}
}

@inproceedings{mir2023effect,
  title={On the effect of transitivity and granularity on vulnerability propagation in the maven ecosystem},
  author={Mir, Amir M and Keshani, Mehdi and Proksch, Sebastian},
  booktitle={2023 IEEE International Conference on Software Analysis, Evolution and Reengineering (SANER)},
  pages={201--211},
  year={2023},
  organization={IEEE}
}

@article{jia2022cargo,
  title={Cargo Ecosystem Dependency-Vulnerability Knowledge Graph Construction and Vulnerability Propagation Study},
  author={Jia, Peiyang and Liu, Chengwei and Sun, Hongyu and Sun, Chengyi and Gu, Mianxue and Liu, Yang and Zhang, Yuqing},
  journal={arXiv preprint arXiv:2210.07482},
  year={2022}
}

@misc{luttwak2021log4shell,
  author       = {Ami Luttwak and Alon Schindel},
  title        = {{Log4Shell 10 days later: Enterprises halfway through patching}},
  howpublished = {\url{https://www.wiz.io/blog/10-days-later-enterprises-halfway-through-patching-log4shell}},
  year         = {2021},
  month        = {December},
  note         = {Accessed: 2025-04-27}
}

@inproceedings{devlin2019bert,
  title={Bert: Pre-training of deep bidirectional transformers for language understanding},
  author={Devlin, Jacob and Chang, Ming-Wei and Lee, Kenton and Toutanova, Kristina},
  booktitle={Proceedings of the 2019 conference of the North American chapter of the association for computational linguistics: human language technologies, volume 1 (long and short papers)},
  pages={4171--4186},
  year={2019}
}

@inproceedings{ni2023distinguishing,
  title={Distinguishing look-alike innocent and vulnerable code by subtle semantic representation learning and explanation},
  author={Ni, Chao and Yin, Xin and Yang, Kaiwen and Zhao, Dehai and Xing, Zhenchang and Xia, Xin},
  booktitle={Proceedings of the 31st ACM Joint European Software Engineering Conference and Symposium on the Foundations of Software Engineering},
  pages={1611--1622},
  year={2023}
}

@article{liu2019roberta,
  title={Roberta: A robustly optimized bert pretraining approach},
  author={Liu, Yinhan and Ott, Myle and Goyal, Naman and Du, Jingfei and Joshi, Mandar and Chen, Danqi and Levy, Omer and Lewis, Mike and Zettlemoyer, Luke and Stoyanov, Veselin},
  journal={arXiv preprint arXiv:1907.11692},
  year={2019}
}

@inproceedings{hanif2022vulberta,
  title={Vulberta: Simplified source code pre-training for vulnerability detection},
  author={Hanif, Hazim and Maffeis, Sergio},
  booktitle={2022 International joint conference on neural networks (IJCNN)},
  pages={1--8},
  year={2022},
  organization={IEEE}
}

@article{zhang2023vulnerability,
  title={Vulnerability detection by learning from syntax-based execution paths of code},
  author={Zhang, Junwei and Liu, Zhongxin and Hu, Xing and Xia, Xin and Li, Shanping},
  journal={IEEE Transactions on Software Engineering},
  volume={49},
  number={8},
  pages={4196--4212},
  year={2023},
  publisher={IEEE}
}

@inproceedings{liu2024pre,
  title={Pre-training by predicting program dependencies for vulnerability analysis tasks},
  author={Liu, Zhongxin and Tang, Zhijie and Zhang, Junwei and Xia, Xin and Yang, Xiaohu},
  booktitle={Proceedings of the IEEE/ACM 46th International Conference on Software Engineering},
  pages={1--13},
  year={2024}
}

@inproceedings{nguyen2022regvd,
  title={Regvd: Revisiting graph neural networks for vulnerability detection},
  author={Nguyen, Van-Anh and Nguyen, Dai Quoc and Nguyen, Van and Le, Trung and Tran, Quan Hung and Phung, Dinh},
  booktitle={Proceedings of the ACM/IEEE 44th International Conference on Software Engineering: Companion Proceedings},
  pages={178--182},
  year={2022}
}

@inproceedings{hin2022linevd,
  title={LineVD: Statement-level vulnerability detection using graph neural networks},
  author={Hin, David and Kan, Andrey and Chen, Huaming and Babar, M Ali},
  booktitle={Proceedings of the 19th international conference on mining software repositories},
  pages={596--607},
  year={2022}
}

@article{zhou2019devign,
  title={Devign: Effective vulnerability identification by learning comprehensive program semantics via graph neural networks},
  author={Zhou, Yaqin and Liu, Shangqing and Siow, Jingkai and Du, Xiaoning and Liu, Yang},
  journal={Advances in neural information processing systems},
  volume={32},
  year={2019}
}

@article{xu2022acgdp,
  title={ACGDP: An augmented code graph-based system for software defect prediction},
  author={Xu, Jiaxi and Ai, Jun and Liu, Jingyu and Shi, Tao},
  journal={IEEE Transactions on Reliability},
  volume={71},
  number={2},
  pages={850--864},
  year={2022},
  publisher={IEEE}
}

@ARTICLE{li2018vuldeepecker,
  author={Zou, Deqing and Wang, Sujuan and Xu, Shouhuai and Li, Zhen and Jin, Hai},
  journal={IEEE Transactions on Dependable and Secure Computing}, 
  title={$\mu$VulDeePecker: A Deep Learning-Based System for Multiclass Vulnerability Detection}, 
  year={2021},
  volume={18},
  number={5},
  pages={2224-2236},
  doi={10.1109/TDSC.2019.2942930}}

@inproceedings{russell2018automated,
  title={Automated vulnerability detection in source code using deep representation learning},
  author={Russell, Rebecca and Kim, Louis and Hamilton, Lei and Lazovich, Tomo and Harer, Jacob and Ozdemir, Onur and Ellingwood, Paul and McConley, Marc},
  booktitle={2018 17th IEEE international conference on machine learning and applications (ICMLA)},
  pages={757--762},
  year={2018},
  organization={IEEE}
}

@article{li2021sysevr,
  title={Sysevr: A framework for using deep learning to detect software vulnerabilities},
  author={Li, Zhen and Zou, Deqing and Xu, Shouhuai and Jin, Hai and Zhu, Yawei and Chen, Zhaoxuan},
  journal={IEEE Transactions on Dependable and Secure Computing},
  volume={19},
  number={4},
  pages={2244--2258},
  year={2021},
  publisher={IEEE}
}

@inproceedings{steenhoek2024dataflow,
  title={Dataflow analysis-inspired deep learning for efficient vulnerability detection},
  author={Steenhoek, Benjamin and Gao, Hongyang and Le, Wei},
  booktitle={Proceedings of the 46th ieee/acm international conference on software engineering},
  pages={1--13},
  year={2024}
}

@inproceedings{fu2023chatgpt,
  title={Chatgpt for vulnerability detection, classification, and repair: How far are we?},
  author={Fu, Michael and Tantithamthavorn, Chakkrit Kla and Nguyen, Van and Le, Trung},
  booktitle={2023 30th Asia-Pacific Software Engineering Conference (APSEC)},
  pages={632--636},
  year={2023},
  organization={IEEE}
}

@inproceedings{fu2022linevul,
  title={Linevul: A transformer-based line-level vulnerability prediction},
  author={Fu, Michael and Tantithamthavorn, Chakkrit},
  booktitle={Proceedings of the 19th International Conference on Mining Software Repositories},
  pages={608--620},
  year={2022}
}

@article{li2021vuldeelocator,
  title={Vuldeelocator: a deep learning-based fine-grained vulnerability detector},
  author={Li, Zhen and Zou, Deqing and Xu, Shouhuai and Chen, Zhaoxuan and Zhu, Yawei and Jin, Hai},
  journal={IEEE Transactions on Dependable and Secure Computing},
  volume={19},
  number={4},
  pages={2821--2837},
  year={2021},
  publisher={IEEE}
}

@article{chakraborty2021deep,
  title={Deep learning based vulnerability detection: Are we there yet?},
  author={Chakraborty, Saikat and Krishna, Rahul and Ding, Yangruibo and Ray, Baishakhi},
  journal={IEEE Transactions on Software Engineering},
  volume={48},
  number={9},
  pages={3280--3296},
  year={2021},
  publisher={IEEE}
}

@misc{textemb2024,
  year = {2024},
  title = {New Generation of Embedding Model},
  author={OpenAI},
  howpublished ={\url{https://openai.com/blog/new-embedding-models-and-api-updates}}
}

@inproceedings{kim2019rvfuzzer,
  title={$\{$RVFuzzer$\}$: Finding input validation bugs in robotic vehicles through $\{$Control-Guided$\}$ testing},
  author={Kim, Taegyu and Kim, Chung Hwan and Rhee, Junghwan and Fei, Fan and Tu, Zhan and Walkup, Gregory and Zhang, Xiangyu and Deng, Xinyan and Xu, Dongyan},
  booktitle={28th USENIX Security Symposium (USENIX Security 19)},
  pages={425--442},
  year={2019}
}

@inproceedings{10.1145/3576915.3624401,
author = {Gobbi, Mat\'{\i}as F. and Kinder, Johannes},
title = {Poster: Using CodeQL to Detect Malware in npm},
year = {2023},
isbn = {9798400700507},
publisher = {Association for Computing Machinery},
address = {New York, NY, USA},
url = {https://doi.org/10.1145/3576915.3624401},
doi = {10.1145/3576915.3624401},
booktitle = {Proceedings of the 2023 ACM SIGSAC Conference on Computer and Communications Security},
pages = {3519–3521},
numpages = {3},
location = {Copenhagen, Denmark},
series = {CCS '23}
}

@misc{brunsfeld2024tree,
  author = {Max Brunsfeld},
  title = {tree-sitter/tree-sitter: v0.23.0},
  year = {2024},
  month = {Aug},
  publisher = {Zenodo},
  doi = {10.5281/zenodo.13375512}
}

@misc{whitesource2022mend,
  title = {Mend bolt},
  author = {{WhiteSource}},
  year = {2022},
  url = {https://www.mend.io/free-developer-tools/bolt}
}

@inproceedings{zhou2024out,
  title={Out of Sight, Out of Mind: Better Automatic Vulnerability Repair by Broadening Input Ranges and Sources},
  author={Zhou, Xin and Kim, Kisub and Xu, Bowen and Han, DongGyun and Lo, David},
  booktitle={Proceedings of the IEEE/ACM 46th International Conference on Software Engineering},
  pages={1--13},
  year={2024}
}

@article{wang2021codet5,
  title={Codet5: Identifier-aware unified pre-trained encoder-decoder models for code understanding and generation},
  author={Wang, Yue and Wang, Weishi and Joty, Shafiq and Hoi, Steven CH},
  journal={arXiv preprint arXiv:2109.00859},
  year={2021}
}

@article{feng2020codebert,
  title={Codebert: A pre-trained model for programming and natural languages},
  author={Feng, Zhangyin and Guo, Daya and Tang, Duyu and Duan, Nan and Feng, Xiaocheng and Gong, Ming and Shou, Linjun and Qin, Bing and Liu, Ting and Jiang, Daxin and others},
  journal={arXiv preprint arXiv:2002.08155},
  year={2020}
}

@article{roziere2023code,
  title={Code llama: Open foundation models for code},
  author={Roziere, Baptiste and Gehring, Jonas and Gloeckle, Fabian and Sootla, Sten and Gat, Itai and Tan, Xiaoqing Ellen and Adi, Yossi and Liu, Jingyu and Remez, Tal and Rapin, J{\'e}r{\'e}my and others},
  journal={arXiv preprint arXiv:2308.12950},
  year={2023}
}

@misc{chatgpt2022,
  year = {2022},
  title = {ChatGPT: Optimizing Language Models for Dialogue.},
  author={OpenAI},
  howpublished ={\url{https://openai.com/blog/chatgpt}}
}

@misc{openai2024gpt4o,
  author       = {OpenAI},
  title        = {GPT-4o: A Flagship Model by OpenAI},
  year         = {2024},
  howpublished = {\url{https://openai.com/index/gpt-4o-and-more-tools-to-chatgpt-free}},
  note         = {Accessed: 2024-10-19}
}

@article{zhang2023pre,
  title={Pre-trained model-based automated software vulnerability repair: How far are we?},
  author={Zhang, Quanjun and Fang, Chunrong and Yu, Bowen and Sun, Weisong and Zhang, Tongke and Chen, Zhenyu},
  journal={IEEE Transactions on Dependable and Secure Computing},
  year={2023},
  publisher={IEEE}
}

@article{tian2024large,
  title={Large Language Models for Equivalent Mutant Detection: How Far Are We?},
  author={Tian, Zhao and Shu, Honglin and Wang, Dong and Cao, Xuejie and Kamei, Yasutaka and Chen, Junjie},
  journal={arXiv preprint arXiv:2408.01760},
  year={2024}
}

@article{kojima2022large,
  title={Large language models are zero-shot reasoners},
  author={Kojima, Takeshi and Gu, Shixiang Shane and Reid, Machel and Matsuo, Yutaka and Iwasawa, Yusuke},
  journal={Advances in neural information processing systems},
  volume={35},
  pages={22199--22213},
  year={2022}
}

@misc{openai2024,
  year = {2024},
  author={OpenAI},
  howpublished ={\url{https://openai.com/}}
}

@article{wolf2019huggingface,
  title={Huggingface's transformers: State-of-the-art natural language processing},
  author={Wolf, Thomas and Debut, Lysandre and Sanh, Victor and Chaumond, Julien and Delangue, Clement and Moi, Anthony and Cistac, Pierric and Rault, Tim and Louf, R{\'e}mi and Funtowicz, Morgan and others},
  journal={arXiv preprint arXiv:1910.03771},
  year={2019}
}

@inproceedings{fan2020ac,
  title={AC/C++ code vulnerability dataset with code changes and CVE summaries},
  author={Fan, Jiahao and Li, Yi and Wang, Shaohua and Nguyen, Tien N},
  booktitle={Proceedings of the 17th International Conference on Mining Software Repositories},
  pages={508--512},
  year={2020}
}

@inproceedings{bhandari2021cvefixes,
  title={CVEfixes: automated collection of vulnerabilities and their fixes from open-source software},
  author={Bhandari, Guru and Naseer, Amara and Moonen, Leon},
  booktitle={Proceedings of the 17th International Conference on Predictive Models and Data Analytics in Software Engineering},
  pages={30--39},
  year={2021}
}

@article{dubey2024llama,
  title={The llama 3 herd of models},
  author={Dubey, Abhimanyu and Jauhri, Abhinav and Pandey, Abhinav and Kadian, Abhishek and Al-Dahle, Ahmad and Letman, Aiesha and Mathur, Akhil and Schelten, Alan and Yang, Amy and Fan, Angela and others},
  journal={arXiv preprint arXiv:2407.21783},
  year={2024}
}

@article{PORNPRASIT2024107523,
title = {Fine-tuning and prompt engineering for large language models-based code review automation},
journal = {Information and Software Technology},
volume = {175},
pages = {107523},
year = {2024},
issn = {0950-5849},
doi = {https://doi.org/10.1016/j.infsof.2024.107523},
url = {https://www.sciencedirect.com/science/article/pii/S0950584924001289},
author = {Chanathip Pornprasit and Chakkrit Tantithamthavorn},
}

@article{robertson2009probabilistic,
  title={The probabilistic relevance framework: BM25 and beyond},
  author={Robertson, Stephen and Zaragoza, Hugo and others},
  journal={Foundations and Trends{\textregistered} in Information Retrieval},
  volume={3},
  number={4},
  pages={333--389},
  year={2009},
  publisher={Now Publishers, Inc.}
}

@inproceedings{gao2023makes,
  title={What makes good in-context demonstrations for code intelligence tasks with llms?},
  author={Gao, Shuzheng and Wen, Xin-Cheng and Gao, Cuiyun and Wang, Wenxuan and Zhang, Hongyu and Lyu, Michael R},
  booktitle={2023 38th IEEE/ACM International Conference on Automated Software Engineering (ASE)},
  pages={761--773},
  year={2023},
  organization={IEEE}
}

@article{yuan2023evaluating,
  title={Evaluating instruction-tuned large language models on code comprehension and generation},
  author={Yuan, Zhiqiang and Liu, Junwei and Zi, Qiancheng and Liu, Mingwei and Peng, Xin and Lou, Yiling},
  journal={arXiv preprint arXiv:2308.01240},
  year={2023}
}

@INPROCEEDINGS {10479409,
author = {M. Fu and C. Tantithamthavorn and V. Nguyen and T. Le},
booktitle = {2023 30th Asia-Pacific Software Engineering Conference (APSEC)},
title = {ChatGPT for Vulnerability Detection, Classification, and Repair: How Far Are We?},
year = {2023},
volume = {},
issn = {},
pages = {632-636},
doi = {10.1109/APSEC60848.2023.00085},
url = {https://doi.ieeecomputersociety.org/10.1109/APSEC60848.2023.00085},
publisher = {IEEE Computer Society},
address = {Los Alamitos, CA, USA},
month = {dec}
}

@article{hu2021lora,
  title={Lora: Low-rank adaptation of large language models},
  author={Hu, Edward J and Shen, Yelong and Wallis, Phillip and Allen-Zhu, Zeyuan and Li, Yuanzhi and Wang, Shean and Wang, Lu and Chen, Weizhu},
  journal={arXiv preprint arXiv:2106.09685},
  year={2021}
}

@article{hou2023large,
  title={Large language models for software engineering: A systematic literature review},
  author={Hou, Xinyi and Zhao, Yanjie and Liu, Yue and Yang, Zhou and Wang, Kailong and Li, Li and Luo, Xiapu and Lo, David and Grundy, John and Wang, Haoyu},
  journal={ACM Transactions on Software Engineering and Methodology},
  year={2023},
  publisher={ACM New York, NY}
}

@inproceedings{hu2024empirical,
  title={Empirical Analysis of Vulnerabilities Life Cycle in Golang Ecosystem},
  author={Hu, Jinchang and Zhang, Lyuye and Liu, Chengwei and Yang, Sen and Huang, Song and Liu, Yang},
  booktitle={Proceedings of the IEEE/ACM 46th International Conference on Software Engineering},
  pages={1--13},
  year={2024}
}

@inproceedings{li2022vulnerability,
  title={On the vulnerability proneness of multilingual code},
  author={Li, Wen and Li, Li and Cai, Haipeng},
  booktitle={Proceedings of the 30th ACM Joint European Software Engineering Conference and Symposium on the Foundations of Software Engineering},
  pages={847--859},
  year={2022}
}

@article{bilge12empirical,
  title={An empirical study of zeroday attacks in the real world},
  author={Bilge, Leyla and Dumitras, Tudor},
  journal={CCS’12},
  pages={16--18},
year = {2012}
}

@article{alfadel2023empirical,
  title={Empirical analysis of security vulnerabilities in python packages},
  author={Alfadel, Mahmoud and Costa, Diego Elias and Shihab, Emad},
  journal={Empirical Software Engineering},
  volume={28},
  number={3},
  pages={59},
  year={2023}
}

@misc{CWE,
  year = {2024},
  author={CWE},
  howpublished ={\url{https://cwe.mitre.org/about/index.html}}
}

@misc{CVE,
  year = {2024},
  author={CVE},
  howpublished ={\url{https://www.cve.org/About/Overview}}
}

@article{guo2024deepseek,
  title={DeepSeek-Coder: When the Large Language Model Meets Programming--The Rise of Code Intelligence},
  author={Guo, Daya and Zhu, Qihao and Yang, Dejian and Xie, Zhenda and Dong, Kai and Zhang, Wentao and Chen, Guanting and Bi, Xiao and Wu, Yu and Li, YK and others},
  journal={arXiv preprint arXiv:2401.14196},
  year={2024}
}

@article{ouyang2022training,
  title={Training language models to follow instructions with human feedback},
  author={Ouyang, Long and Wu, Jeffrey and Jiang, Xu and Almeida, Diogo and Wainwright, Carroll and Mishkin, Pamela and Zhang, Chong and Agarwal, Sandhini and Slama, Katarina and Ray, Alex and others},
  journal={Advances in neural information processing systems},
  volume={35},
  pages={27730--27744},
  year={2022}
}

@inproceedings{cao2022mvd,
  title={MVD: memory-related vulnerability detection based on flow-sensitive graph neural networks},
  author={Cao, Sicong and Sun, Xiaobing and Bo, Lili and Wu, Rongxin and Li, Bin and Tao, Chuanqi},
  booktitle={Proceedings of the 44th international conference on software engineering},
  pages={1456--1468},
  year={2022}
}

@article{yin2024pros,
  title={Pros and cons! evaluating chatgpt on software vulnerability},
  author={Yin, Xin},
  journal={arXiv preprint arXiv:2404.03994},
  year={2024}
}

@inproceedings{zhou2024largeworkshop,
  title={Large language model for vulnerability detection: Emerging results and future directions},
  author={Zhou, Xin and Zhang, Ting and Lo, David},
  booktitle={Proceedings of the 2024 ACM/IEEE 44th International Conference on Software Engineering: New Ideas and Emerging Results},
  pages={47--51},
  year={2024}
}

@article{zhou2024large,
  title={Large language model for vulnerability detection and repair: Literature review and the road ahead},
  author={Zhou, Xin and Cao, Sicong and Sun, Xiaobing and Lo, David},
  journal={ACM Transactions on Software Engineering and Methodology},
  year={2024},
  publisher={ACM New York, NY}
}

@article{yin2024multitask,
  title={Multitask-based evaluation of open-source llm on software vulnerability},
  author={Yin, Xin and Ni, Chao and Wang, Shaohua},
  journal={IEEE Transactions on Software Engineering},
  year={2024},
  publisher={IEEE}
}

@article{yang2024security,
  title={Security vulnerability detection with multitask self-instructed fine-tuning of large language models},
  author={Yang, Aidan ZH and Tian, Haoye and Ye, He and Martins, Ruben and Goues, Claire Le},
  journal={arXiv preprint arXiv:2406.05892},
  year={2024}
}

@article{du2024vul,
  title={Vul-rag: Enhancing llm-based vulnerability detection via knowledge-level rag},
  author={Du, Xueying and Zheng, Geng and Wang, Kaixin and Feng, Jiayi and Deng, Wentai and Liu, Mingwei and Chen, Bihuan and Peng, Xin and Ma, Tao and Lou, Yiling},
  journal={arXiv preprint arXiv:2406.11147},
  year={2024}
}

@article{lu2024grace,
  title={GRACE: Empowering LLM-based software vulnerability detection with graph structure and in-context learning},
  author={Lu, Guilong and Ju, Xiaolin and Chen, Xiang and Pei, Wenlong and Cai, Zhilong},
  journal={Journal of Systems and Software},
  volume={212},
  pages={112031},
  year={2024},
  publisher={Elsevier}
}

@inproceedings{grieco2016toward,
  title={Toward large-scale vulnerability discovery using machine learning},
  author={Grieco, Gustavo and Grinblat, Guillermo Luis and Uzal, Lucas and Rawat, Sanjay and Feist, Josselin and Mounier, Laurent},
  booktitle={Proceedings of the sixth ACM conference on data and application security and privacy},
  pages={85--96},
  year={2016}
}

@article{scandariato2014predicting,
  title={Predicting vulnerable software components via text mining},
  author={Scandariato, Riccardo and Walden, James and Hovsepyan, Aram and Joosen, Wouter},
  journal={IEEE Transactions on Software Engineering},
  volume={40},
  number={10},
  pages={993--1006},
  year={2014},
  publisher={IEEE}
}

@article{wang2023codet5+,
  title={Codet5+: Open code large language models for code understanding and generation},
  author={Wang, Yue and Le, Hung and Gotmare, Akhilesh Deepak and Bui, Nghi DQ and Li, Junnan and Hoi, Steven CH},
  journal={arXiv preprint arXiv:2305.07922},
  year={2023}
}

@article{guo2022unixcoder,
  title={Unixcoder: Unified cross-modal pre-training for code representation},
  author={Guo, Daya and Lu, Shuai and Duan, Nan and Wang, Yanlin and Zhou, Ming and Yin, Jian},
  journal={arXiv preprint arXiv:2203.03850},
  year={2022}
}

@inproceedings{steenhoek2023empirical,
  title={An empirical study of deep learning models for vulnerability detection},
  author={Steenhoek, Benjamin and Rahman, Md Mahbubur and Jiles, Richard and Le, Wei},
  booktitle={2023 IEEE/ACM 45th International Conference on Software Engineering (ICSE)},
  pages={2237--2248},
  year={2023},
  organization={IEEE}
}

@inproceedings{ding2024traced,
  title={Traced: Execution-aware pre-training for source code},
  author={Ding, Yangruibo and Steenhoek, Benjamin and Pei, Kexin and Kaiser, Gail and Le, Wei and Ray, Baishakhi},
  booktitle={Proceedings of the 46th IEEE/ACM International Conference on Software Engineering},
  pages={1--12},
  year={2024}
}

@misc{difflib,
  year = {2024},
  author={Homepage},
  howpublished ={\url{https://docs.python.org/3/library/difflib.html}}
}

@article{guo2025deepseek,
  title={Deepseek-r1: Incentivizing reasoning capability in llms via reinforcement learning},
  author={Guo, Daya and Yang, Dejian and Zhang, Haowei and Song, Junxiao and Zhang, Ruoyu and Xu, Runxin and Zhu, Qihao and Ma, Shirong and Wang, Peiyi and Bi, Xiao and others},
  journal={arXiv preprint arXiv:2501.12948},
  year={2025}
}

@misc{aliyun_qwq_2024,
  author = {Alibaba Cloud},
  title = {QwQ Model Documentation},
  year = {2024},
  url = {https://www.alibabacloud.com/help/en/model-studio/user-guide/qwq},
  publisher = {Alibaba Cloud},
  note = {Accessed: March 29, 2025},
}

@article{yu2025preliminarystudylargelanguage,
      title={A Preliminary Study of Large Language Models for Multilingual Vulnerability Detection}, 
      author={Junji Yu and Honglin Shu and Michael Fu and Dong Wang and Chakkrit Tantithamthavorn and Yasutaka Kamei and Junjie Chen},
      year={2025},
      eprint={2505.07376},
      journal={arXiv}
}

@misc{replicationpackage,
  title = {Replication Package},
  author={Honglin Shu},
  year={2025},
  url = {https://github.com/SpanShu96/Large-Language-Model-for-Multilingual-Vulnerability-Detection/tree/main},
}

@article{treude2025interacting,
  title={Interacting with ai reasoning models: Harnessing" thoughts" for ai-driven software engineering},
  author={Treude, Christoph and Kula, Raula Gaikovina},
  journal={arXiv preprint arXiv:2503.00483},
  year={2025}
}

@article{yu2025utboost,
  title={Utboost: Rigorous evaluation of coding agents on swe-bench},
  author={Yu, Boxi and Zhu, Yuxuan and He, Pinjia and Kang, Daniel},
  journal={arXiv preprint arXiv:2506.09289},
  year={2025}
}

@article{yang2024swe,
  title={Swe-agent: Agent-computer interfaces enable automated software engineering},
  author={Yang, John and Jimenez, Carlos E and Wettig, Alexander and Lieret, Kilian and Yao, Shunyu and Narasimhan, Karthik and Press, Ofir},
  journal={Advances in Neural Information Processing Systems},
  volume={37},
  pages={50528--50652},
  year={2024}
}
